\pdfoutput=1
\documentclass[a4paper,UKenglish,cleveref,autoref,thm-restate]{lipics-v2021}

\hideLIPIcs

\newif\ifsubmit
\submittrue

\usepackage{csquotes}

\usepackage{mathtools}
\usepackage{amssymb}

\usepackage{siunitx}
\usepackage{float}
\usepackage{mdframed}
\usepackage{booktabs}
\usepackage{svg}
\usepackage{placeins}
\usepackage{pifont}
\usepackage{tikz}
\usetikzlibrary{arrows.meta, positioning, decorations.pathreplacing, decorations.pathmorphing}

\crefname{definition}{Definition}{Definitions}
\crefname{proposition}{Proposition}{Propositions}
\crefname{theorem}{Theorem}{Theorems}
\crefname{corollary}{Corollary}{Corollaries}
\crefname{example}{Example}{Examples}
\crefname{remark}{Remark}{Remarks}

\providecommand{\datalink}{https://github.com/ivan-vynyavskyy/modern-portfolio-theory-in-the-crypto-wilderness}

\ifsubmit
\else
\usepackage{silence} 
\usepackage[colorinlistoftodos,prependcaption,textsize=tiny,textwidth=\marginparwidth]{todonotes}

\fi

\usepackage[symbols, acronym, nonumberlist, nogroupskip, stylemods={mcols,longbooktabs}, section=subsection, numberedsection]{glossaries-extra}
\makenoidxglossaries

\glssetcategoryattribute{acronym}{nohyper}{true}
\setabbreviationstyle[acronym]{long-short}

\newacronym{DeFi}{DeFi}{decentralised finance}
\newacronym{MPT}{MPT}{modern portfolio theory}
\newacronym{MVO}{MVO}{mean-variance optimisation}
\newacronym{wETH}{wETH}{wrapped ETH}
\newacronym{WBTC}{WBTC}{wrapped BTC}
\newacronym{EOA}{EOA}{externally owned account}
\newacronym{CA}{CA}{contract account}
\newacronym{CAGR}{CAGR}{compound annual growth rate}
\newacronym{MRV}{MaxRet}{max return at target volatility}
\newacronym{MVR}{MinVar}{min variance at target return}
\newacronym{MSR}{MaxSR}{max Sharpe ratio}
\newacronym{TVL}{TVL}{total value locked}
\newacronym{ERC-20}{ERC-20}{Ethereum request for comments 20}
\newacronym{CAPM}{CAPM}{capital asset pricing model}
\newacronym{OLS}{OLS}{ordinary least squares}
\newacronym{HHI}{HHI}{Herfindahl-Hirschman index}

\newglossaryentry{token}{
    name={token},
    description={A digital asset issued on a blockchain platform, typically following a standard such as ERC-20 on Ethereum.}
}

\newglossaryentry{cryptoasset}{
    name={cryptoasset},
    description={Any cryptocurrency or digital token that can be transferred and traded on a blockchain.}
}

\newglossaryentry{on-chain}{
    name={on-chain},
    description={Refers to data or transactions that are recorded and publicly verifiable on a blockchain.}
}

\newglossaryentry{gini}{
    name={Gini coefficient},
    description={A measure of statistical dispersion ranging from 0 (perfect equality) to 1 (maximal inequality), used here to quantify wealth inequality among token holders.}
}

\newglossaryentry{efficient-frontier}{
    name={efficient frontier},
    description={The set of portfolios that achieve the highest expected return for a given level of volatility. Throughout the paper, the empirically relevant quantity is the \emph{constrained} efficient frontier: we optimise over each account's already-held tokens rather than over all available tokens, and native ETH enters only through its wrapped form (WETH).}
}

\newglossaryentry{sharpe-ratio}{
    name={Sharpe ratio},
    description={The ratio of a portfolio's expected excess return to its volatility, measuring risk-adjusted performance.}
}

\newglossaryentry{naive-allocation}{
    name={naive allocation},
    description={A baseline allocation that requires no estimation of returns or covariances. The $1/N$ (equal-weight) variant assigns weight $1/N$ to each of the $N$ held tokens; the market-cap variant weights each token by its market capitalisation. Used here as benchmarks against \gls{MPT}-optimal allocations.}
}

\newglossaryentry{covariance-matrix}{
    name={covariance matrix},
    description={Matrix $\Sigma$ whose entry $(i,j)$ is the covariance of returns between tokens $i$ and $j$; the diagonal contains per-token variances. Estimated empirically over a lookback window and used as input to \gls{MVO}.}
}

\glsxtrnewsymbol[description={A portfolio.}]{portfolio}{\ensuremath{\pi}}
\glsxtrnewsymbol[description={Market risk exposure of a portfolio, measuring co-movement with the market benchmark (\gls{CAPM}).}]{beta}{\ensuremath{\beta}}
\glsxtrnewsymbol[description={Risk-adjusted excess return, the residual beyond what market risk exposure alone explains (\gls{CAPM}).}]{alpha}{\ensuremath{\alpha}}
\glsxtrnewsymbol[description={USD closing price of asset $i$ at snapshot $t$.}]{p}{\ensuremath{p}}
\glsxtrnewsymbol[description={USD value of a position (per-asset) or portfolio (total, capitalised $V$), obtained by multiplying balances with prices.}]{v}{\ensuremath{v}}
\glsxtrnewsymbol[description={Portfolio weight vector, where $w_i$ is the fraction of total value allocated to asset $i$.}]{w}{\ensuremath{w}}
\glsxtrnewsymbol[description={Asset log-return (lookback window) or portfolio forward-realised return ($t \to t'$, arithmetic).}]{r}{\ensuremath{r}}
\glsxtrnewsymbol[description={Expected return of an asset or portfolio, estimated as the sample mean over the lookback window.}]{mu}{\ensuremath{\mu}}
\glsxtrnewsymbol[description={Standard deviation (volatility) of an asset or portfolio.}]{sigma}{\ensuremath{\sigma}}
\glsxtrnewsymbol[description={Variance (or volatility squared) of an asset or portfolio.}]{sigma2}{\ensuremath{\sigma^2}}
\glsxtrnewsymbol[description={Empirical covariance matrix of asset returns.}]{Sigma}{\ensuremath{\Sigma}}
\glsxtrnewsymbol[description={$\ell_1$ weight distance between the actual portfolio and the optimum under strategy $s$.}]{ds}{\ensuremath{d_s}}
\glsxtrnewsymbol[description={Market benchmark return, based on an equally-weighted basket of \gls{wETH} and \gls{WBTC}.}]{rm}{\ensuremath{r_{(m)}}}
\glsxtrnewsymbol[description={Asset index (i.e., token), $i = 1, \dots, N$.}]{idxi}{\ensuremath{i}}
\glsxtrnewsymbol[description={Account index (Ethereum address).}]{idxa}{\ensuremath{a}}
\glsxtrnewsymbol[description={Time index (monthly snapshot) at which the portfolio is observed and $\beta$ is estimated.}]{idxt}{\ensuremath{t}}
\glsxtrnewsymbol[description={Forward evaluation time, $t' = t + T^{(+)}$, at which realised returns and $\alpha$ are measured.}]{idxtprime}{\ensuremath{t'}}
\glsxtrnewsymbol[description={Optimisation-strategy label, $s \in \{\text{MinVar},\text{MaxRet},\text{MaxSR}\}$.}]{idxs}{\ensuremath{s}}
\glsxtrnewsymbol[description={Number of assets held by the account at snapshot $t$.}]{idxN}{\ensuremath{N}}
\glsxtrnewsymbol[description={Lookback-window length (days) used for estimating asset moments and covariances up to snapshot $t$.}]{Tminus}{\ensuremath{T^{(-)}}}
\glsxtrnewsymbol[description={Forward holding-period length (days) over which realised returns and $\alpha$ are evaluated starting from snapshot $t$.}]{Tplus}{\ensuremath{T^{(+)}}}

\def\CoingeckoInitTokensN/{\num{5394}}
\def\CoingeckoERCTokensN/{\num{5359}}
\def\CoingeckoCleanedTokensN/{\num{4562}}
\def\CoingeckoRemovedTokensN/{\num{830}}

\def\TotalAccountsRaw/{116182156}
\def\TotalAccountsRounded/{\the\numexpr\TotalAccountsRaw//1000000\relax\,M}

\def\EOAAccountsEnd/{\num{109007240}}
\def\CAAccountsEnd/{\num{7174916}}

\def\EOACAGR/{\SI{50.53}{\percent}}
\def\EOAAvgIncreaseYear/{\num{16785778}}

\def\CACAGR/{\SI{56.93}{\percent}}
\def\CAAvgIncreaseYear/{\num{1128397}}

\def\TotalAccountsEnd/{\num{\TotalAccountsRaw/}}
\def\CAShareEndPct/{\SI{6.18}{\percent}}

\def\HoldersAvgYtwenty{\num{27088}}
\def\HoldersMedYtwenty{\num{3312}}

\def\HoldersAvgYtwentyfive{\num{12429}}
\def\HoldersMedYtwentyfive{\num{1746}}

\def\HoldersPctGtTenKYtwenty{\num{28.3}\%}
\def\HoldersPctGtTenKYtwentyfive{\num{15.6}\%}

\def\HoldersPctOneKTenKYtwenty{\num{39.7}\%}

\def\HoldersPctOneKTenKYtwentyfour{\num{50.4}\%}

\def\HoldersPctHundredOneKYtwenty{\num{17.7}\%}
\def\HoldersPctHundredOneKYtwentyfive{\num{27.5}\%}

\def\WealthBucketCAZeroOneP/{\num{51.63}}
\def\WealthBucketCAOneHundredP/{\num{28.86}}
\def\WealthBucketCAHundredOneKP/{\num{10.75}}
\def\WealthBucketCAOneKTenKP/{\num{5.08}}
\def\WealthBucketCATenKHundredKP/{\num{2.00}}
\def\WealthBucketCAOverHundredKP/{\num{1.67}}

\def\WealthBucketEOAZeroOneP/{\num{29.80}}
\def\WealthBucketEOAOneHundredP/{\num{49.43}}
\def\WealthBucketEOAHundredOneKP/{\num{13.89}}
\def\WealthBucketEOAOneKTenKP/{\num{5.02}}
\def\WealthBucketEOATenKHundredKP/{\num{1.44}}
\def\WealthBucketEOAOverHundredKP/{\num{0.42}}

\def\WealthShareCAMeanP/{\num{31.80}}

\def\WealthShareEOAMeanP/{\num{68.20}}

\def\LOneObservationsRaw/{239634756}
\def\LOneObservationsN/{\num{\LOneObservationsRaw/}}
\def\LOneObservationsM/{\the\numexpr\LOneObservationsRaw//1000000\relax\,M}

\def\LOneMeanSaferRisk/{12.59}
\def\LOneMeanBetterReturn/{28.30}
\def\LOneMeanMaxSharpe/{45.09}
\def\LOneMedianBetterReturn/{14.52}
\def\LOneMedianMaxSharpe/{48.79}
\def\LOneMonthsN/{72}

\def\LOneSrCombinedLeOnePct/{63.35}
\def\LOneBrCombinedLeOnePct/{40.09}
\def\LOneMsCombinedLeOnePct/{17.87}
\def\LOneMsHighBinPct/{19.49}
\def\LOneMs60to80Pct/{24.66}

\def\LOneCorrBrSr/{0.452}
\def\LOneCorrBrMs/{0.360}
\def\LOneCorrSrMs/{0.385}

\def\LOneJointAllThreeN/{\num{3099}}
\def\LOneJointAllThreePct/{0.001}
\def\LOneJointBrSrPct/{0.26}
\def\LOneJointBrMsPct/{0.05}
\def\LOneJointSrMsPct/{0.03}

\def\LOneNearOptTwoTokenPctLo/{92}
\def\LOneNearOptTwoTokenPctHi/{95}
\def\LOneNearOptMedianValueLo/{27}
\def\LOneNearOptMedianValueHi/{33}
\def\LOneNearOptBrN/{96.1}
\def\LOneNearOptBrPct/{40.1}
\def\LOneNearOptSrN/{151.8}
\def\LOneNearOptSrPct/{63.4}
\def\LOneNearOptMsN/{42.8}
\def\LOneNearOptMsPct/{17.9}

\def\LOneOlsBrRsq/{0.064}
\def\LOneOlsSrRsq/{0.358}
\def\LOneOlsMsRsq/{0.080}
\def\LOneOlsExtBrRsq/{0.104}
\def\LOneOlsExtSrRsq/{0.632}
\def\LOneOlsExtMsRsq/{0.132}

\def\LOneRfBrRsq/{0.376}
\def\LOneRfBrMae/{19.55}
\def\LOneRfSrRsq/{0.676}
\def\LOneRfSrMae/{5.35}
\def\LOneRfMsRsq/{0.412}
\def\LOneRfMsMae/{19.35}
\def\LOneRfSrTokensImp/{95.5}
\def\LOneRfBrMsValueImp/{45}
\def\LOneRfBrMsTokensImpLo/{27}
\def\LOneRfBrMsTokensImpHi/{31}
\def\LOneRfBrMsMonthImpLo/{25}
\def\LOneRfBrMsMonthImpHi/{28}

\def\LOneSpearmanSrTokens/{0.811}
\def\LOnePearsonSrTokens/{0.576}
\def\LOneSpearmanBrTokens/{0.451}
\def\LOneSpearmanMsTokens/{0.399}

\def\LOneFivePlusBrLeOnePct/{0.77}
\def\LOneFivePlusSrLeOnePct/{0.49}
\def\LOneFivePlusMsLeOnePct/{0.16}
\def\LOneFivePlusMsHighBinPct/{50.6}
\def\LOneFivePlusRfSrRsq/{0.173}
\def\LOneFivePlusMsMonthImp/{42.7}
\def\LOneFivePlusSpearmanSrTokens/{0.365}
\def\LOneFivePlusCorrBrMs/{0.596}

\def\WealthPeakTwentyone/{\$1\,T}
\def\WealthPeakTwentyfiveB/{\$1.47\,T}

\def\TopOneWealthTwenty/{97.25}
\def\TopOneWealthTwentyfive/{99.03}
\def\TopFiveTenWealthMin/{99.3}

\def\PortSizeOneTokenPct/{83.35}
\def\PortSizeTwoTokenPct/{10.14}
\def\PortSizeThreeFourTokenPct/{4.26}
\def\PortSizeFivePlusPct/{2.17}
\def\PortSizeTwoOfMultiPct/{61.4}

\def\LOneNaiveEqualMean/{37.7}
\def\LOneNaiveEqualMedian/{40.0}
\def\LOneNaiveMcapMean/{38.4}
\def\LOneNaiveMcapMedian/{31.2}

\def\LOneDeltaBrEqualMean/{-0.042}
\def\LOneDeltaBrEqualCloserPct/{32.9}
\def\LOneDeltaBrEqualFartherPct/{20.1}
\def\LOneDeltaBrEqualUnchangedPct/{46.9}
\def\LOneDeltaSrEqualMean/{-0.074}
\def\LOneDeltaSrEqualCloserPct/{27.4}
\def\LOneDeltaSrEqualFartherPct/{9.1}
\def\LOneDeltaSrEqualUnchangedPct/{63.5}
\def\LOneDeltaMsEqualMean/{+0.173}
\def\LOneDeltaMsEqualCloserPct/{10.2}
\def\LOneDeltaMsEqualFartherPct/{50.8}
\def\LOneDeltaMsEqualUnchangedPct/{39.1}

\def\LOneDeltaBrMcapMean/{+0.130}
\def\LOneDeltaBrMcapCloserPct/{22.7}
\def\LOneDeltaBrMcapFartherPct/{37.1}
\def\LOneDeltaBrMcapUnchangedPct/{39.9}
\def\LOneDeltaSrMcapMean/{-0.008}
\def\LOneDeltaSrMcapCloserPct/{18.8}
\def\LOneDeltaSrMcapFartherPct/{17.9}
\def\LOneDeltaSrMcapUnchangedPct/{63.0}
\def\LOneDeltaMsMcapMean/{+0.205}
\def\LOneDeltaMsMcapCloserPct/{29.8}
\def\LOneDeltaMsMcapFartherPct/{52.0}
\def\LOneDeltaMsMcapUnchangedPct/{17.9}

\def\LOneDeltaSrEqualRatio/{3}

\def\CAPMBetaBaselineMedian/{0.989}
\def\CAPMBetaBaselinePtwentyfive/{0.58}
\def\CAPMBetaBaselinePseventyfive/{1.28}

\def\CAPMAlphaBaselineMedian/{-1.91}
\def\CAPMAlphaMcapMedian/{-0.30}
\def\CAPMAlphaEqualMedian/{-2.78}
\def\CAPMAlphaBetterMedian/{-2.31}
\def\CAPMAlphaSaferMedian/{-1.85}
\def\CAPMAlphaMaxSharpeMedian/{-1.92}

\def\CAPMFracPositiveBaseline/{38.9}
\def\CAPMFracPositiveMcap/{39.7}
\def\CAPMFracPositiveEqual/{37.9}
\def\CAPMFracPositiveBetter/{37.8}
\def\CAPMFracPositiveSafer/{38.5}
\def\CAPMFracPositiveMaxSharpe/{38.0}

\def\CAPMRetBaselineMedian/{-1.87}
\def\CAPMRetBetterMedian/{-2.33}
\def\CAPMRetSaferMedian/{-2.00}
\def\CAPMRetMaxSharpeMedian/{-1.92}
\def\CAPMRetEqualMedian/{-2.82}
\def\CAPMRetMcapMedian/{-0.24}
\def\CAPMRetMarketMedian/{1.89}

\def\CAPMRetHitBetter/{48.5}
\def\CAPMRetHitSafer/{51.0}
\def\CAPMRetHitMaxSharpe/{49.8}
\def\CAPMRetHitEqual/{50.4}
\def\CAPMRetHitMcap/{53.3}

\def\CAPMSpearmanSrAlpha/{+0.026}
\def\CAPMSpearmanMsAlpha/{+0.101}
\def\CAPMSpearmanBetaAlpha/{-0.143}

\def\CAPMDeltaBetaMcapMean/{-0.102}

\def\CAPMObsRaw/{235868007}
\def\CAPMBlocksN/{72}
\def\CAPMHorizonDays/{20}

\def\PctBeatRetBaselineBlockMedian/{42.98}
\def\PctBeatRetBaselineBlockMean/{41.55}
\def\PctPosAlphaBaselineBlockMedian/{40.23}
\def\PctPosAlphaBaselineBlockMean/{41.69}
\def\PctBeatRetBetterBlockMedian/{36.94}
\def\PctBeatRetBetterBlockMean/{40.75}
\def\PctPosAlphaBetterBlockMedian/{36.07}
\def\PctPosAlphaBetterBlockMean/{40.10}
\def\PctBeatRetSaferBlockMedian/{40.80}
\def\PctBeatRetSaferBlockMean/{42.21}
\def\PctPosAlphaSaferBlockMedian/{38.23}
\def\PctPosAlphaSaferBlockMean/{41.65}
\def\PctBeatRetMaxSharpeBlockMedian/{38.61}
\def\PctBeatRetMaxSharpeBlockMean/{41.36}
\def\PctPosAlphaMaxSharpeBlockMedian/{36.69}
\def\PctPosAlphaMaxSharpeBlockMean/{40.61}
\def\PctBeatRetEqualBlockMedian/{36.05}
\def\PctBeatRetEqualBlockMean/{41.79}
\def\PctPosAlphaEqualBlockMedian/{37.20}
\def\PctPosAlphaEqualBlockMean/{41.24}
\def\PctBeatRetMcapBlockMedian/{42.78}
\def\PctBeatRetMcapBlockMean/{43.91}
\def\PctPosAlphaMcapBlockMedian/{39.58}
\def\PctPosAlphaMcapBlockMean/{42.74}

\def\PctBeatRetBaselineBlockMin/{6.69}
\def\PctBeatRetBaselineBlockMax/{95.92}

\def\PctBeatRetMedianStrategySpread/{6.7}
\def\PctPosAlphaMedianStrategySpread/{4.2}

\def\PctBeatCrossStratCorrLo/{0.918}

\def\PctBlocksBelowFifty/{69.4}

\def\BeatExcRetBaselineBotOne/{$-$39.87}
\def\BeatExcRetBaselineMed/{$-$2.71}
\def\BeatExcRetBaselineTopOne/{67.31}
\def\BeatExcRetBaselineSkew/{27.44}
\def\BeatExcRetBetterBotOne/{$-$38.12}
\def\BeatExcRetBetterMed/{$-$4.09}
\def\BeatExcRetBetterTopOne/{71.28}
\def\BeatExcRetBetterSkew/{33.15}
\def\BeatExcRetSaferBotOne/{$-$36.64}
\def\BeatExcRetSaferMed/{$-$3.18}
\def\BeatExcRetSaferTopOne/{66.15}
\def\BeatExcRetSaferSkew/{29.52}
\def\BeatExcRetMaxSharpeBotOne/{$-$42.39}
\def\BeatExcRetMaxSharpeMed/{$-$3.19}
\def\BeatExcRetMaxSharpeTopOne/{76.60}
\def\BeatExcRetMaxSharpeSkew/{34.20}
\def\BeatExcRetEqualBotOne/{$-$31.41}
\def\BeatExcRetEqualMed/{$-$3.12}
\def\BeatExcRetEqualTopOne/{65.65}
\def\BeatExcRetEqualSkew/{34.25}
\def\BeatExcRetMcapBotOne/{$-$34.00}
\def\BeatExcRetMcapMed/{$-$2.32}
\def\BeatExcRetMcapTopOne/{55.04}
\def\BeatExcRetMcapSkew/{21.04}
\def\BeatAlphaBaselineBotOne/{$-$40.84}
\def\BeatAlphaBaselineMed/{$-$2.28}
\def\BeatAlphaBaselineTopOne/{72.35}
\def\BeatAlphaBaselineSkew/{31.50}
\def\BeatAlphaBetterBotOne/{$-$40.13}
\def\BeatAlphaBetterMed/{$-$2.86}
\def\BeatAlphaBetterTopOne/{75.31}
\def\BeatAlphaBetterSkew/{35.17}
\def\BeatAlphaSaferBotOne/{$-$37.23}
\def\BeatAlphaSaferMed/{$-$1.84}
\def\BeatAlphaSaferTopOne/{66.89}
\def\BeatAlphaSaferSkew/{29.66}
\def\BeatAlphaMaxSharpeBotOne/{$-$45.14}
\def\BeatAlphaMaxSharpeMed/{$-$2.20}
\def\BeatAlphaMaxSharpeTopOne/{80.76}
\def\BeatAlphaMaxSharpeSkew/{35.62}
\def\BeatAlphaEqualBotOne/{$-$31.46}
\def\BeatAlphaEqualMed/{$-$2.60}
\def\BeatAlphaEqualTopOne/{68.25}
\def\BeatAlphaEqualSkew/{36.80}
\def\BeatAlphaMcapBotOne/{$-$35.06}
\def\BeatAlphaMcapMed/{$-$0.63}
\def\BeatAlphaMcapTopOne/{55.99}
\def\BeatAlphaMcapSkew/{20.94}
\def\BeatExcRetBaselineBotOneAbs/{39.87}

\def\RetDistZeroMedianBaseline/{-0.05}
\def\RetDistHighMedianBaseline/{-6.09}

\def\RetNegativePctLo/{57}
\def\RetNegativePctHi/{59}

\def\FitPsiMRV/{1.1319}
\def\FitGammaMRV/{0.4803}
\def\FitDinfMRV/{100.00}
\def\FitRsqMRV/{0.9937}
\def\FitMaeMRV/{1.12}
\def\FitPsiMVR/{1.7139}
\def\FitGammaMVR/{0.7774}
\def\FitDinfMVR/{80.69}
\def\FitRsqMVR/{0.9999}
\def\FitMaeMVR/{1.56}
\def\FitPsiMSR/{1.1211}
\def\FitGammaMSR/{0.9022}
\def\FitDinfMSR/{88.62}
\def\FitRsqMSR/{0.9999}
\def\FitMaeMSR/{1.34}

\def\TopOneCAPctWallets/{6.98}
\def\TopOneEOAPctWallets/{93.02}
\def\TopOneCAPctValue/{32.26}
\def\TopOneEOAPctValue/{67.74}

\def\TopPointOneCAPctWallets/{18.95}
\def\TopPointOneCAPctValue/{33.70}

\def\TagCoveragePctMean/{23.25}
\def\TagCoveragePctMin/{10.70}
\def\TagCoveragePctMax/{35.14}
\def\TagUnlabeledPct/{76.20}

\def\TagCatUserPct/{33.83}
\def\TagCatDexPairPct/{19.32}
\def\TagCatServicePct/{15.66}
\def\TagCatExchangePct/{7.97}
\def\TagCatDefiTokenPct/{7.40}

\def\TagConDefiPct/{63.07}
\def\TagConCustodyPct/{31.73}
\def\TagConExchangePct/{17.09}
\def\TagConDexPct/{13.88}

\def\RfExtRetRsqLo/{0.56}
\def\RfExtRetRsqHi/{0.65}
\def\RfExtRetMonthImpLo/{70}
\def\RfExtRetMonthImpHi/{79}
\def\RfExtRetBetaImpLo/{14}
\def\RfExtRetBetaImpHi/{22}
\def\RfExtRetValueImpLo/{5}
\def\RfExtRetValueImpHi/{8}
\def\RfExtAlphaMonthImpLo/{55}
\def\RfExtAlphaMonthImpHi/{62}
\def\RfExtAlphaBetaImpLo/{25}
\def\RfExtAlphaBetaImpHi/{31}
\def\RfDeltaRsqRetLo/{+0.07}
\def\RfDeltaRsqRetHi/{+0.10}
\def\RfDeltaRsqAlphaLo/{+0.08}
\def\RfDeltaRsqAlphaHi/{+0.12}

\def\ShapMonthPp/{11.4}
\def\ShapBetaPp/{2.7}

\def\RfDefault/{5}

\def\StablecoinUSDCAvgAPY/{5.51}
\def\StablecoinUSDCMedianAPY/{4.77}
\def\StablecoinUSDCMonths/{44}
\def\StablecoinUSDTAvgAPY/{4.47}
\def\StablecoinUSDTMedianAPY/{3.48}
\def\StablecoinUSDTMonths/{52}
\def\StablecoinPooledAvgAPY/{4.94}
\def\StablecoinPooledMedianAPY/{3.88}
\def\StablecoinPooledMonths/{52}

\def\RfRobustBlocks/{72}
\def\RfRobustWalletsPerBlock/{5{,}000}
\def\RfRobustTotalWallets/{360{,}000}

\def\RfRobustWeightLOneMedianPct/{0}
\def\RfRobustWeightLOneMeanPct/{4.5}
\def\RfRobustWeightLOneTopFivePct/{37.1}

\def\RfRobustGapDiffMedianPP/{0}
\def\RfRobustGapDiffMeanPP/{-0.4}
\def\RfRobustGapDiffTopFivePP/{13.2}
\def\RfRobustVsEqualDiffMedianPP/{0}
\def\RfRobustVsEqualDiffMeanPP/{-0.7}
\def\RfRobustVsEqualDiffTopFivePP/{2.4}
\def\RfRobustVsMcapDiffMedianPP/{0}
\def\RfRobustVsMcapDiffMeanPP/{-2.6}
\def\RfRobustVsMcapDiffTopFivePP/{18.5}

\def\ReconTransferEvents/{1,977,546,094}
\def\ReconTokensDiscovered/{4,847}
\def\ReconParquetSizeGB/{42.4}
\def\ReconTopTokenTransfers/{698}
\def\ReconTopTokenSharePct/{35.3}
\def\ReconMedianTokenTransfers/{25,030}
\def\ReconTokensAfterFilter/{4,562}
\def\ReconPipelineWorkers/{48}
\def\ReconBatchSize/{20,000}
\def\ReconOutputRowsFirst/{477.3}
\def\ReconOutputRowsSecond/{982.2}
\def\ReconOutputRows/{1,459.5}
\def\ReconOutputSizeGB/{196.6}
\def\ReconSnapshotBlocks/{72}

\def\MptObservations/{239,634,756}
\def\MptObservationsM/{239.6}
\def\MptBlocks/{72}
\def\MptWorkers/{64}
\def\MptCores/{128}
\def\MptOutputSizeGB/{144}
\def\MptStrategies/{6}
\def\MptHorizonDays/{20}
\def\MptMedianWalletsPerBlock/{3,348,198}
\def\MptMinWalletsPerBlock/{879,029}
\def\MptMaxWalletsPerBlock/{7,111,049}

\def\TotalBlocks/{72}

\def\LOneSmallMRVMean/{24.2}
\def\LOneLargeMRVMean/{56.8}

\def\LOneSmallMVRMean/{7.5}
\def\LOneLargeMVRMean/{48.5}

\def\LOneSmallMSRMean/{41.5}
\def\LOneLargeMSRMean/{70.2}

\def\LOneDeltaMRV/{32.6}
\def\LOneCohenMRV/{4.93}
\def\LOneDeltaMVR/{41.1}
\def\LOneCohenMVR/{13.77}
\def\LOneDeltaMSR/{28.7}
\def\LOneCohenMSR/{5.48}

\title{Modern Portfolio Theory in the Crypto-Wilderness}

\titlerunning{Modern Portfolio Theory in the Crypto-Wilderness}

\author{Ivan Vynyavskyy}{Complexity Science Hub \& TU Wien, Austria}{ivan.vynyavskyy@gmail.com}{}{}
\author{Stefan Kitzler}{Complexity Science Hub \& Austrian Institute of Technology, Austria}{kitzler@csh.ac.at}{}{}
\author{Bernhard Haslhofer}{Complexity Science Hub, Austria}{haslhofer@csh.ac.at}{}{}
\author{Aviv Yaish}{Yale University \& IC3, USA\\Complexity Science Hub, Austria}{a@yai.sh}{}{}

\authorrunning{I. Vynyavskyy}

\Copyright{Ivan Vynyavskyy, Stefan Kitzler, Bernhard Haslhofer, and Aviv Yaish}

\ccsdesc[500]{Applied computing~Digital cash}
\ccsdesc[300]{Mathematics of computing~Mathematical optimization}
\ccsdesc[100]{Applied computing~Economics}

\keywords{blockchain, portfolio theory, Ethereum, DeFi, on-chain data}

\category{}

\relatedversion{}

\ifdefined\anonymousAck
\acknowledgements{Acknowledgements anonymised for double-blind review. \vspace{1\baselineskip}}
\else
\acknowledgements{This work is partially funded by the Austrian security research program KIRAS of the Federal Ministry of Finance (BMF) under the project LLEA (grant agreement 926183).
The computational results have been achieved using the Austrian Scientific Computing (ASC) infrastructure. \vspace{1\baselineskip}}
\fi

\EventEditors{Aggelos Kiayias and Maria Kyropoulou}
\EventNoEds{2}
\EventLongTitle{8th Conference on Advances in Financial Technologies (AFT 2026)}
\EventShortTitle{AFT 2026}
\EventAcronym{AFT}
\EventYear{2026}
\EventDate{October 6--9, 2026}
\EventLocation{London, United Kingdom}
\EventLogo{}
\SeriesVolume{8}
\ArticleNo{1}

\ifdefined\anonymousAck
\hypersetup{pdfauthor={},pdftitle={},pdfsubject={},pdfkeywords={},pdfcreator={},pdfproducer={}}
\fi

\begin{document}

\ifdefined\anonymousAck\else\nolinenumbers\fi

\maketitle

\ifdefined\anonymousAck
\vspace{2\baselineskip}
\else
\vspace{1\baselineskip}
\fi

\begin{abstract}

\Gls{MPT} prescribes how to maximise the return of an asset portfolio for a given level of risk. The optimal trade-off between return and variance defines the \gls{efficient-frontier}.
Whether actual cryptoasset portfolios approximate this prescription and whether proximity to the frontier translates into realised performance remain difficult to test at large scale in traditional markets due to their opaque nature and the inaccessibility of data.
As we show, public blockchains make these questions measurable: every \gls{token} transfer is recorded, thus enabling complete portfolio reconstruction for every account at any point in time.
We leverage this transparency to reconstruct \gls{cryptoasset} portfolios for over \TotalAccountsRounded/ Ethereum accounts across the full chain history (2015--2025), measure their distance to the constrained \gls{efficient-frontier}, and quantify how deviations translate into realised performance.
Here we show that market entry timing, not allocation choice, is the dominant predictor of realised cryptoasset returns.
On-chain wealth is highly concentrated and portfolios are pervasively under-diversified, with single-asset holdings accounting for \PortSizeOneTokenPct/\% of accounts.
Two-asset portfolios sit closest to the efficient frontier defined by their held assets, a proximity that reflects the narrowness of their opportunity set rather than deliberate optimisation.
Passive market-capitalisation weighting outperforms every \gls{MPT} optimisation strategy in median realised return, and entry month alone explains \RfExtRetMonthImpLo/--\RfExtRetMonthImpHi/\% of the variance in returns, far exceeding the contribution of allocation choice.
Mean-variance optimisation therefore appears neither descriptive of observed behaviour nor prescriptively useful in the cryptoasset domain, even if \gls{MPT} retains its value as a normative benchmark.

\keywords{Modern Portfolio Theory, Mean-Variance Optimisation, DeFi, Blockchain}

\end{abstract}

\ifdefined\anonymousAck
\vspace{4\baselineskip}
\fi

\setcounter{totalnumber}{10}
\setcounter{topnumber}{10}
\setcounter{bottomnumber}{10}
\renewcommand{\topfraction}{.99}
\renewcommand{\bottomfraction}{.99}
\renewcommand{\textfraction}{.01}

\section{Introduction}
\label{sec:introduction}

Modern portfolio theory (\gls{MPT}), for which Markowitz received the Nobel Prize in Economics~\cite{markowitz1991foundations}, prescribes that portfolios should lie on the mean-variance \gls{efficient-frontier}, that is, maximise expected return for a given level of risk.
MPT's core assumptions are that investors are \emph{perfectly rational}, actively optimising their allocation, and \emph{risk-averse}, preferring lower variance for equal return.
These assumptions are idealisations challenged in this paper.

Cryptoasset markets are a uniquely suitable empirical test bed for \gls{MPT}.
First, blockchains record every token transfer in a publicly auditable ledger, enabling the reconstruction of complete portfolio histories for every account, unlike traditional markets where holdings are observable only through selective brokerage disclosures.
Second, tokens are accessible permissionlessly and adhere to standards such as ERC-20 so on-chain portfolio choices reflect actual allocation decisions rather than institutional constraints.
Third, the Ethereum ecosystem is particularly well-suited for this examination, spanning a decade of multiple market cycles and hosting the largest decentralised finance ecosystem.
This setting enables us to examine both assumptions empirically at population scale.

In traditional markets, investors are systematically under-diversified and often underperform passive benchmarks~\cite{Kumar2008,calvet2007down,Barber2000}, while on-chain studies reveal highly concentrated token distributions and centralised governance~\cite{victor2022measuring,nadler2022decentralized}.
Existing crypto-portfolio work applies \gls{MPT} only to hypothetical asset baskets~\cite{briere2015virtual, platanakis2019portfolio, trimborn2020investing, petukhina2021investing} rather than to actual on-chain holdings, and each strand relies on sampled or selected data over limited windows.
Yet no study has measured, at scale, how actual on-chain portfolios compare to MPT-optimal allocations, or whether deviating from the constrained efficient frontier has performance consequences.

We fill this gap with the first large-scale empirical study comparing \emph{actual} on-chain cryptoasset portfolios to rebalanced \emph{theoretical optimal} allocations under \gls{MPT}.
We reconstruct on-chain cryptoasset portfolios at scale and pursue three objectives: First (\textbf{Obj A}), we characterise the composition and diversification patterns of actual on-chain cryptoasset portfolios. Second (\textbf{Obj B}), we quantify the deviation of actual portfolios from the mean-variance efficient frontier under optimisation strategies and held-asset constraints. Third (\textbf{Obj C}), we examine how such deviations affect portfolio performance relative to the market.

To summarise, we contribute the following:
\begin{enumerate}

  \item \textbf{Portfolio reconstruction method.} We propose a method for reconstructing actual on-chain token portfolios from transfer events, processing \CoingeckoCleanedTokensN/ tokens across the full Ethereum history (2015--2025) and covering \TotalAccountsRounded/ accounts via an incremental ledger-based approach validated against on-chain balances (\cref{sec:data_and_methods}). We contribute to open science by making the code (and the full dataset upon publication) publicly available\footnote{\url{\datalink}}.

  \item \textbf{Actual portfolio composition and diversification.} Aggregate on-chain wealth peaks at \WealthPeakTwentyfiveB/, but is highly concentrated: the vast majority of accounts hold less than \$100, and \TopOneWealthTwentyfive/\% of token wealth is controlled by the top 1\% of holders, predominantly centralised cryptoasset exchanges and decentralised finance (DeFi) protocols. Single-asset portfolios account for \PortSizeOneTokenPct/\% of all accounts, and only \PortSizeFivePlusPct/\% hold five or more assets. Multi-asset portfolios become more prevalent as account wealth increases, a pattern inconsistent with rational, risk-averse portfolio optimisation (\cref{sec:descriptive_statistics}).

  \item \textbf{Deviation from efficient frontier.} We measure the $\ell_1$ weight distance between actual portfolios and three frontier strategies (minimum variance, maximum return, maximum \gls{sharpe-ratio}). Portfolio size and suboptimality follow a power-law decay: small portfolios ($<5$ assets) lie close to the efficient frontier by construction, while larger portfolios deviate substantially, with distances reaching \LOneLargeMVRMean/\%–\LOneLargeMSRMean/\% depending on the strategy (\cref{sec:mpt_optimal_portfolio}).

  \item \textbf{Deviation effects.} We compare realised returns of actual portfolios against MPT-optimal counterfactuals. Proximity to the constrained efficient frontier does not predict better risk-adjusted performance. Instead, market entry timing, rather than allocation, is the dominant predictive feature of realised returns (\cref{sec:returns_and_capm}).

\end{enumerate}

The results have implications for both theory and practice.
For theory, pervasive under-diversification and the dominance of market entry timing over frontier proximity challenge the descriptive adequacy of \gls{MPT} in the cryptoasset domain. The mismatch points to behavioural factors and a dominant common market factor, rather than mean-variance optimisation, as the organising principles of actual cryptoasset portfolios.

For practice, our findings caution against importing mean-variance frameworks wholesale into cryptoasset portfolio construction: rebalancing toward \gls{MPT} optima does not reliably improve risk-adjusted performance, and passive market-capitalisation weighting offers a more robust default. The persistence of single-asset portfolios and the extreme wealth concentration we document further suggest that frictions, behavioural factors, and investor-protection concerns deserve greater attention than allocation theory in shaping interventions across retail, advisory, platform, and regulatory contexts.

The remainder of the paper reviews related work, presents the method and findings for each objective, and concludes with a discussion of implications and limitations.

\section{Background and Related Work}
\label{sec:background}

This section reviews the \gls{MPT} framework and efficient frontier, empirical portfolio studies in traditional finance, organised around our three objectives, empirical studies in blockchain markets, and closes with the synthesis and research gap that motivates the present work.

\subsection{Modern Portfolio Theory}
\label{sec:portfolio_theory}

The \gls{MPT} framework was introduced by Markowitz~\cite{markowitz1952portfolio}, who proposed a method for allocating assets to maximise expected returns for a given level of risk, known as mean-variance optimisation.
\gls{MPT} relies on several assumptions: investors are perfectly rational and risk-averse; markets are frictionless (no transaction costs, taxes, or borrowing constraints); and prices are efficient, instantaneously reflecting all available information.
Building on \gls{MPT}, Sharpe~\cite{sharpe1964capital} introduced the \gls{CAPM}, which models an asset's expected return as a function of its systematic risk relative to the market.

Central to \gls{MPT} is the concept of the \emph{efficient frontier}, the upper boundary of the minimum-variance set in expected-return/standard-deviation space: portfolios on the frontier achieve the highest possible expected return for a given level of risk, while portfolios below it are suboptimal. Any actual portfolio can thus be evaluated against the frontier, quantifying the gap between what an account holds and what mean-variance optimisation prescribes. For a comprehensive introduction to \gls{MPT} and relevant literature, see Elton et al.~\cite{elton2014modern}.

\subsection{Empirical Studies in Traditional Finance}
\label{sec:traditional_portfolio_measurements}

A large body of empirical research has studied the portfolios of investors in traditional assets. Campbell~\cite{campbell2006household} establishes household finance as a field, surveying evidence that while many households invest effectively, a significant minority (typically poorer and less educated) make portfolio mistakes. The findings summarised below provide context for each of our objectives.

\textbf{Composition and diversification.} A central finding is that investors hold fewer assets than theory prescribes. Kumar and Goetzmann~\cite{Kumar2008} analyse more than 40,000 equity accounts at a U.S.\ discount brokerage and find that the average investor holds only four stocks (median of three), with greater under-diversification among younger, low-income, and less-educated investors. This concentration often takes the form of familiarity bias: French and Poterba~\cite{french1991investor} document strikingly high domestic ownership shares despite substantial diversification gains from international investing, and Huberman~\cite{huberman2001familiarity} shows that shareholders concentrate holdings in geographically familiar companies. Benartzi and Thaler~\cite{benartzi2001naive} further show that 401(k) participants tend to divide contributions equally across available options, the so-called \emph{1/n heuristic}, another departure from mean-variance optimisation; DeMiguel et al.~\cite{demiguel2009optimal} find that this naive rule is hard to beat by mean-variance strategies out-of-sample, while Hellum et al.~\cite{hellum2026complex} show that this gap closes once the number of assets exceeds the number of training observations. Together, these studies characterise actual portfolios as simpler and less diversified than \gls{MPT} prescribes.

\textbf{Deviation from optimal allocations.} Quantifying how far actual portfolios lie from theoretical optima requires both a benchmark and a distance measure. Using comprehensive administrative data on all Swedish households, Calvet et al.~\cite{calvet2007down} measure the welfare costs of investment mistakes: while most households actually outperform their domestic stock index through international diversification, a minority incurs substantial losses from under-diversification. Massa and Simonov~\cite{Massa2006} further show that Swedish investors concentrate holdings in stocks related to their employment income and geographic proximity rather than hedging against non-financial income risk, a systematic deviation from theoretically optimal allocations. At the fund level, Didier et al.~\cite{Didier2013} find that even mutual funds exhibit limited international diversification, with global funds potentially achieving 2.6--5.5\% higher annual risk-adjusted returns by broadening their allocations. These studies typically rely on welfare proxies or return differentials rather than portfolio-level distance metrics, leaving open how directly observable allocations compare to the efficient frontier.

\textbf{Performance consequences.} Individual investors generally underperform passive benchmarks. Barber and Odean~\cite{Barber2000} find that among 66,465 household accounts, the most active traders earned 11.4\% annually versus 17.9\% for the market, with overconfidence driving excessive trading. Odean~\cite{Odean1998} documents the disposition effect: investors realise gains more readily than losses, reducing annual returns by 4.4 percentage points. Classically, Brinson et al.~\cite{brinson1986determinants} show that asset allocation policy explains 93.6\% of the variation in quarterly portfolio returns across 91 large U.S.\ pension funds, with security selection and market timing playing minor roles, a benchmark our cryptoasset findings invert.

\subsection{Empirical Studies in Blockchain Markets}
\label{sec:blockchain_measurements}

Blockchain ledgers provide a uniquely complete empirical record of asset holdings; this subsection describes the Ethereum setting that makes portfolio reconstruction possible and reviews prior measurement work on token distributions and crypto-portfolio optimisation.

\textbf{Ethereum as an empirical setting.} Fungible tokens (ERC-20) on Ethereum are digital assets managed by smart contracts (\glspl{CA}) that track balances for each account. Ethereum distinguishes two account types: \glspl{EOA}, controlled by private keys, hold tokens directly; \glspl{CA} implement programmable logic and hold tokens on behalf of protocols such as \gls{DeFi} applications for lending, trading, and liquidity provision. Every token transfer is recorded as an event in the blockchain logs, providing a complete and auditable history of all token movements and forming the empirical basis for portfolio reconstruction from public data.

\textbf{On-chain measurement studies.} Prior work has leveraged this data to characterise token distributions and network structure. Nadler and Sch\"ar~\cite{nadler2022decentralized} analyse the token distribution of \gls{DeFi} governance tokens and find that protocol governance is heavily centralised. Victor and L\"uders~\cite{victor2022measuring} conduct an early empirical study of Ethereum ERC-20 tokens over the first 6 million blocks, analysing token distributions and trade networks. These studies establish that on-chain holdings are concentrated, but stop short of reconstructing portfolios at the account level or evaluating them against a portfolio-theoretic benchmark.

\textbf{\Gls{MPT} applied to cryptoassets.}
A separate strand applies \gls{MPT} to cryptoassets, typically on curated baskets rather than actual holdings. Bri\`ere et al.~\cite{briere2015virtual} examine Bitcoin's role in diversified portfolios; Platanakis and Urquhart~\cite{platanakis2019portfolio}, Trimborn et al.~\cite{trimborn2020investing}, and Petukhina et al.~\cite{petukhina2021investing} apply mean-variance optimisation to small baskets of major cryptoassets. Borri et al.~\cite{borri2025cryptocurrency} apply an empirical asset pricing approach to tokens and find that their risk-adjusted performance is comparable to traditional markets. None of these studies compares actual account holdings to MPT-optimal allocations at population scale.

\subsection{Synthesis and Research Gap}
\label{sec:research_gap}

Together, these three strands establish that actual investors in traditional markets systematically deviate from \gls{MPT} prescriptions, that on-chain portfolios can be measured at scale from public blockchain data, and that prior work on crypto-portfolio optimisation has focused on hypothetical baskets rather than actual holdings. Yet no study has characterised actual on-chain portfolio composition at population scale, measured how these portfolios compare to \gls{MPT}-optimal allocations, or tested whether deviating from the efficient frontier has measurable performance consequences. Blockchain data enables this analysis at a completeness and scale impossible with traditional brokerage data; the present work fills this gap.

\section{Data and Methods}
\label{sec:data_and_methods}

\subsection{Overall Approach}
\label{sec:approach}

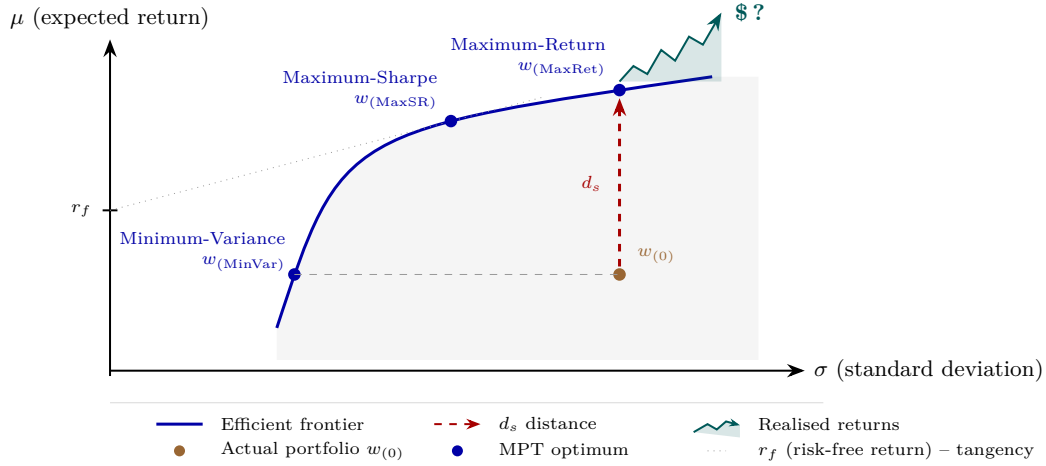
\begin{figure}
  \centering
  \resizebox{\linewidth}{!}{\begin{tikzpicture}[
  arr/.style={-{Stealth[length=2.5mm]}, thick},
  xscale=1.3, yscale=0.75,
]

\fill[gray!8]
    (1.3, 0.8) .. controls (2.0, 4.5) .. (6.0, 5.5)
    -- (6.5, 5.5) -- (6.5, 0.2) -- (1.3, 0.2) -- cycle;

\draw[blue!65!black, very thick]
    (1.3, 0.8) .. controls (2.0, 4.5) .. (6.0, 5.5);

\draw[arr] (-0.5, 0) -- (7.0, 0)
    node[right, font=\small] {$\sigma$ (standard deviation)};
\draw[arr] (-0.5, -0.1) -- (-0.5, 6.2)
    node[above, font=\small] {$\mu$ (expected return)};

\draw[thick] (-0.58, 3.0) -- (-0.42, 3.0);
\node[font=\scriptsize, anchor=east] at (-0.6, 3.0) {$r_f$};

\draw[gray!80, thin, dotted] (-0.5, 3.0) -- (4.2, 5.13);

\fill[brown!80!black] (5.0, 1.80) ellipse [x radius=1.92pt, y radius=3.33pt];
\node[font=\scriptsize\bfseries, text=brown!80!black, anchor=south west]
    at (5.1, 1.85) {$w_{(0)}$};

\fill[blue!65!black] (1.49, 1.80) ellipse [x radius=1.92pt, y radius=3.33pt];
\node[font=\scriptsize, text=blue!65!black, anchor=south east, align=right]
    at (1.49, 1.70) {Minimum-Variance\\$w_{(\text{MinVar})}$};

\fill[blue!65!black] (3.18, 4.67) ellipse [x radius=1.92pt, y radius=3.33pt];
\node[font=\scriptsize, text=blue!65!black, anchor=south east, align=right]
    at (3.13, 4.72) {Maximum-Sharpe\\$w_{(\text{MaxSR})}$};

\fill[blue!65!black] (5.0, 5.25) ellipse [x radius=1.92pt, y radius=3.33pt];
\node[font=\scriptsize, text=blue!65!black, anchor=south east, align=right]
    at (4.95, 5.34) {Maximum-Return\\$w_{(\text{MaxRet})}$};

\draw[gray!80, thin, dashed] (1.49, 1.80) -- (5.0, 1.80);

\draw[red!65!black, very thick, dashed, -{Stealth[length=2.5mm]}]
    (5.0, 1.95) -- (5.0, 5.10);
\node[font=\scriptsize\bfseries, text=red!65!black, anchor=east]
    at (4.90, 3.50) {$d_s$};

\fill[teal!70!black, opacity=0.15]
    (5.0, 5.41) -- (5.15, 5.70) -- (5.30, 5.55) -- (5.45, 6.00)
    -- (5.60, 5.80) -- (5.75, 6.25) -- (5.90, 6.10) -- (6.10, 6.70)
    -- (6.10, 5.41) -- cycle;
\draw[teal!70!black, thick, -{Stealth[length=2.5mm]}]
    (5.0, 5.41) -- (5.15, 5.70) -- (5.30, 5.55) -- (5.45, 6.00)
    -- (5.60, 5.80) -- (5.75, 6.25) -- (5.90, 6.10) -- (6.10, 6.70);
\node[font=\small\bfseries, text=teal!70!black, anchor=west]
    at (6.15, 6.70) {\$\,?};

\draw[gray!30] (-0.5, -0.6) -- (7.2, -0.6);

\draw[blue!65!black, very thick] (0.0, -1.0) -- (0.5, -1.0);
\node[font=\scriptsize, anchor=west] at (0.6, -1.0) {Efficient frontier};

\draw[red!65!black, thick, dashed, -{Stealth[length=2mm]}] (3.0, -1.0) -- (3.5, -1.0);
\node[font=\scriptsize, anchor=west] at (3.6, -1.0) {$d_s$~distance};

\fill[brown!80!black] (0.25, -1.5) ellipse [x radius=1.69pt, y radius=2.93pt];
\node[font=\scriptsize, anchor=west] at (0.6, -1.5) {Actual portfolio $w_{(0)}$};

\fill[blue!65!black] (3.25, -1.5) ellipse [x radius=1.69pt, y radius=2.93pt];
\node[font=\scriptsize, anchor=west] at (3.6, -1.5) {MPT optimum};

\draw[gray!80, thin, dotted] (5.95, -1.5) -- (6.15, -1.5);
\node[font=\scriptsize, anchor=west] at (6.4, -1.5) {$r_f$ (risk-free return) -- tangency};

\fill[teal!70!black, opacity=0.15] (5.8, -1.15) -- (5.95, -1.0) -- (6.05, -1.1) -- (6.15, -0.95) -- (6.3, -1.02) -- (6.3, -1.25) -- cycle;
\draw[teal!70!black, thick, -{Stealth[length=1.5mm]}] (5.8, -1.15) -- (5.95, -1.0) -- (6.05, -1.1) -- (6.15, -0.95) -- (6.3, -1.02);
\node[font=\scriptsize, anchor=west] at (6.4, -1.0) {Realised returns};

\end{tikzpicture}}
  \caption{Overall analytical approach in expected-return/standard-deviation space. Each actual portfolio
    \tikz[baseline=-0.5ex]\fill[brown!80!black] (0,0) circle (1.8pt);\,$w_{(0)}$
    is characterised (\textbf{Obj~A}), compared to three frontier optima
    \tikz[baseline=-0.5ex]\fill[blue!65!black] (0,0) circle (1.8pt);
    on the efficient frontier
    \tikz[baseline=-0.5ex]\draw[blue!65!black, very thick] (0,0) -- (0.4,0);
    via the weight-space distance
    \tikz[baseline=-0.5ex]\draw[red!65!black, thick, dashed, -{Stealth[length=1.5mm]}] (0,0) -- (0.4,0);\,$d_{s}$
    (\textbf{Obj~B}), and evaluated on realised returns
    \tikz[baseline=-0.5ex]{\fill[teal!70!black, opacity=0.15] (0,-0.06) -- (0.1,0.06) -- (0.2,-0.02) -- (0.3,0.08) -- (0.4,0.04) -- (0.4,-0.06) -- cycle; \draw[teal!70!black, thick, -{Stealth[length=1.5mm]}] (0,-0.06) -- (0.1,0.06) -- (0.2,-0.02) -- (0.3,0.08) -- (0.4,0.04);}
    (\textbf{Obj~C}).}
  \label{fig:approach}
\end{figure}

Our analysis reconstructs actual on-chain portfolios and evaluates them against \gls{MPT}-optimal allocations along the three objectives. \Cref{fig:approach} provides a schematic overview.

Each actual portfolio is represented by its asset-weight vector $w_{(0)}$, corresponding to expected return $\mu_{(0)}$ and standard deviation $\sigma_{(0)}$.
We place $w_{(0)}$ in the feasible $(\mu,\sigma)$ region and characterise its composition and diversification (Obj~A). We compute its weight-space distance $d_{s}$ to three reference portfolios on the efficient frontier: the minimum-variance (MinVar), maximum-return (MaxRet), and maximum-Sharpe-ratio (MaxSR) optima. This quantifies deviation from \gls{MPT} prescriptions (Obj~B). We compare the realised returns of $w_{(0)}$ against those references over matched holding periods, testing whether proximity to the frontier predicts risk-adjusted performance (Obj~C).

The remainder of this section describes the data sources underlying our analysis, the method for reconstructing on-chain portfolios from transfer events, the computation of \gls{MPT}-optimal reference portfolios, and the distance and performance metrics used to evaluate actual portfolios against these references.

\subsection{Data Collection}
\label{sec:data_collection}

We collect data from two sources: the Ethereum blockchain, which provides transactions and event logs, and the Coingecko platform~\cite{coingecko}, which provides token metadata and prices.

\begin{itemize}

  \item \textbf{Ethereum blockchain:} We run a full node~\cite{erigon} and extract all transactions and event logs from the genesis block (30~July 2015, block~\#0) to 5~December 2025 (block~\#23,943,069). Smart contracts emit information to the blockchain \textit{logs} when on-chain events occur. In particular, \texttt{Transfer} events following the ERC-20 standard record token movements between accounts and form the empirical basis for portfolio reconstruction.

  \item \textbf{Coingecko:} We collect metadata for the top tokens on Coingecko, ranked by market capitalisation, including the Ethereum contract address, name, and symbol. For each token we additionally collect daily USD closing prices, market capitalisation, fully diluted valuation ($\mathrm{FDV}=\text{Price}\times\text{Total Supply}$), total circulating supply, and trading volume.

\end{itemize}

\begin{figure}
  \centering
  \resizebox{\textwidth}{!}{\begin{tikzpicture}[x=1cm, y=1cm]

\pgfmathsetmacro{\sy}{2.1/5394}

\pgfmathsetmacro{\hS}{5394*\sy}
\pgfmathsetmacro{\hA}{5359*\sy}
\pgfmathsetmacro{\hB}{5006*\sy}
\pgfmathsetmacro{\hC}{4832*\sy}
\pgfmathsetmacro{\hD}{4812*\sy}
\pgfmathsetmacro{\hE}{4562*\sy}

\def\xA{0}     \def\xB{2.5}   \def\xC{5.0}
\def\xD{7.5}   \def\xE{10.0}  \def\xF{12.5}

\def\bw{0.14}

\definecolor{cFlow1}{HTML}{B0B0B0}
\definecolor{cFlow2}{HTML}{9DB89A}
\definecolor{cFlow3}{HTML}{8AC085}
\definecolor{cFlow4}{HTML}{70C870}
\definecolor{cFlow5}{HTML}{50D050}
\definecolor{cDrop}{HTML}{A84040}
\definecolor{cBar1}{HTML}{808080}
\definecolor{cBar2}{HTML}{6B8B68}
\definecolor{cBar3}{HTML}{559055}
\definecolor{cBar4}{HTML}{409540}
\definecolor{cBar5}{HTML}{2B9B2B}
\definecolor{cBarF}{HTML}{208020}

\fill[cFlow1] (\xA+\bw,0) rectangle (\xB-\bw,\hA);
\fill[cFlow2] (\xB+\bw,0) rectangle (\xC-\bw,\hB);
\fill[cFlow3] (\xC+\bw,0) rectangle (\xD-\bw,\hC);
\fill[cFlow4] (\xD+\bw,0) rectangle (\xE-\bw,\hD);
\fill[cFlow5] (\xE+\bw,0) rectangle (\xF-\bw,\hE);

\def\dyA{3.4}  \def\dyB{3.1}  \def\dyC{2.8}  \def\dyD{2.5}  \def\dyE{2.2}
\def\dxoff{0.7}

\pgfmathsetmacro{\dxA}{\xA+\dxoff}
\fill[cDrop, opacity=0.4, draw=cDrop, draw opacity=0.6, line width=0.4pt]
  (\xA+\bw, \hA)
  .. controls (\xA+\bw+0.25, \hA) and (\dxA-0.25, \dyA) ..
  (\dxA, \dyA)
  -- (\dxA, \dyA+0.04)
  .. controls (\dxA-0.25, \dyA+0.04) and (\xA+\bw+0.25, \hS) ..
  (\xA+\bw, \hS) -- cycle;
\fill[cDrop] (\dxA-0.04,\dyA-0.04) rectangle (\dxA+0.04,\dyA+0.08);

\pgfmathsetmacro{\dxB}{\xB+\dxoff}
\fill[cDrop, opacity=0.4, draw=cDrop, draw opacity=0.6, line width=0.4pt]
  (\xB+\bw, \hB)
  .. controls (\xB+\bw+0.25, \hB) and (\dxB-0.25, \dyB) ..
  (\dxB, \dyB)
  -- (\dxB, \dyB+0.08)
  .. controls (\dxB-0.25, \dyB+0.08) and (\xB+\bw+0.25, \hA) ..
  (\xB+\bw, \hA) -- cycle;
\fill[cDrop] (\dxB-0.08,\dyB-0.06) rectangle (\dxB+0.08,\dyB+0.14);

\pgfmathsetmacro{\dxC}{\xC+\dxoff}
\fill[cDrop, opacity=0.4, draw=cDrop, draw opacity=0.6, line width=0.4pt]
  (\xC+\bw, \hC)
  .. controls (\xC+\bw+0.25, \hC) and (\dxC-0.25, \dyC) ..
  (\dxC, \dyC)
  -- (\dxC, \dyC+0.06)
  .. controls (\dxC-0.25, \dyC+0.06) and (\xC+\bw+0.25, \hB) ..
  (\xC+\bw, \hB) -- cycle;
\fill[cDrop] (\dxC-0.06,\dyC-0.05) rectangle (\dxC+0.06,\dyC+0.11);

\pgfmathsetmacro{\dxD}{\xD+\dxoff}
\fill[cDrop, opacity=0.4, draw=cDrop, draw opacity=0.6, line width=0.4pt]
  (\xD+\bw, \hD)
  .. controls (\xD+\bw+0.25, \hD) and (\dxD-0.25, \dyD) ..
  (\dxD, \dyD)
  -- (\dxD, \dyD+0.04)
  .. controls (\dxD-0.25, \dyD+0.04) and (\xD+\bw+0.25, \hC) ..
  (\xD+\bw, \hC) -- cycle;
\fill[cDrop] (\dxD-0.04,\dyD-0.04) rectangle (\dxD+0.04,\dyD+0.08);

\pgfmathsetmacro{\dxE}{\xE+\dxoff}
\fill[cDrop, opacity=0.4, draw=cDrop, draw opacity=0.6, line width=0.4pt]
  (\xE+\bw, \hE)
  .. controls (\xE+\bw+0.25, \hE) and (\dxE-0.25, \dyE) ..
  (\dxE, \dyE)
  -- (\dxE, \dyE+0.06)
  .. controls (\dxE-0.25, \dyE+0.06) and (\xE+\bw+0.25, \hD) ..
  (\xE+\bw, \hD) -- cycle;
\fill[cDrop] (\dxE-0.06,\dyE-0.05) rectangle (\dxE+0.06,\dyE+0.11);

\fill[cBar1] (\xA-\bw,0) rectangle (\xA+\bw,\hS);
\fill[cBar2] (\xB-\bw,0) rectangle (\xB+\bw,\hA);
\fill[cBar3] (\xC-\bw,0) rectangle (\xC+\bw,\hB);
\fill[cBar4] (\xD-\bw,0) rectangle (\xD+\bw,\hC);
\fill[cBar5] (\xE-\bw,0) rectangle (\xE+\bw,\hD);
\fill[cBarF] (\xF-\bw,0) rectangle (\xF+\bw,\hE);

\pgfmathsetmacro{\yLbl}{\hE*0.5}

\node[left, align=center, font=\scriptsize]
  at (\xA-\bw-0.15, \yLbl)
  {Start:\\\num{5394}\\Tokens};

\node[align=center, font=\scriptsize,
      fill=white, fill opacity=0.7, text opacity=1,
      rounded corners=2pt, inner sep=3pt]
  at ({(\xA+\bw+\xB-\bw)/2}, \yLbl)
  {Step~1:\\ERC-20~$\checkmark$\\\num{5359}};

\node[align=center, font=\scriptsize,
      fill=white, fill opacity=0.7, text opacity=1,
      rounded corners=2pt, inner sep=3pt]
  at ({(\xB+\bw+\xC-\bw)/2}, \yLbl)
  {Step~2:\\Price data~$\checkmark$\\\num{5006}};

\node[align=center, font=\scriptsize,
      fill=white, fill opacity=0.7, text opacity=1,
      rounded corners=2pt, inner sep=3pt]
  at ({(\xC+\bw+\xD-\bw)/2}, \yLbl)
  {Step~3:\\Volume~$\checkmark$\\\num{4832}};

\node[align=center, font=\scriptsize,
      fill=white, fill opacity=0.7, text opacity=1,
      rounded corners=2pt, inner sep=3pt]
  at ({(\xD+\bw+\xE-\bw)/2}, \yLbl)
  {Step~4:\\Supply \& mcap~$\checkmark$\\\num{4812}};

\node[align=center, font=\scriptsize,
      fill=white, fill opacity=0.7, text opacity=1,
      rounded corners=2pt, inner sep=3pt]
  at ({(\xE+\bw+\xF-\bw)/2}, \yLbl)
  {Step~5:\\Validation~$\checkmark$\\\num{4562}};

\node[right, align=center, font=\scriptsize\bfseries,
      fill=white, fill opacity=0.7, text opacity=1,
      rounded corners=2pt, inner sep=3pt]
  at (\xF+\bw+0.15, \yLbl)
  {Final:\\\textbf{4562}\\Tokens};

\node[right, font=\scriptsize] at (\dxA+0.06, \dyA+0.02)
  {non-compliant\quad(\ding{55}\,35)};
\node[right, font=\scriptsize] at (\dxB+0.10, \dyB+0.04)
  {insufficient pricing\quad(\ding{55}\,318)};
\node[right, font=\scriptsize] at (\dxC+0.08, \dyC+0.03)
  {negligible volume\quad(\ding{55}\,174)};
\node[right, font=\scriptsize] at (\dxD+0.06, \dyD+0.02)
  {invalid supply/mcap\quad(\ding{55}\,55)};
\node[right, font=\scriptsize] at (\dxE+0.08, \dyE+0.03)
  {inconsistent balance\quad(\ding{55}\,250)};

\end{tikzpicture}}
  \caption{Token cleaning pipeline. Out of \CoingeckoInitTokensN/~initial tokens, \CoingeckoCleanedTokensN/ pass all filtering stages (ERC-20 compliance; price history, trading volume, and supply/market-cap validation; reconstruction validation) and are retained for subsequent analysis.}
  \label{fig:tokens_per_category}
\end{figure}
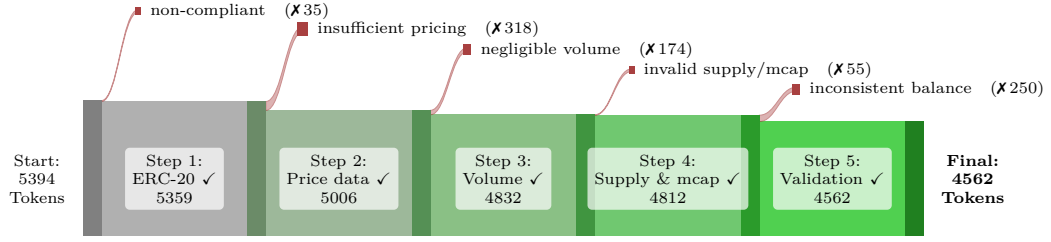

\Cref{fig:tokens_per_category} summarises the multi-stage cleaning pipeline described below.
Applied to the \CoingeckoInitTokensN/ initial tokens, \CoingeckoCleanedTokensN/ (${\approx}84.5\%$) pass all stages and are retained for subsequent analysis.

\begin{enumerate}
  \item \textbf{Standard compatibility.} We verify ERC-20 compliance by extracting the runtime bytecode from \texttt{CREATE} and \texttt{CREATE2} traces and checking for ERC-20 function and event selectors~\cite{victor2022measuring}. Proxy contracts are resolved to their underlying implementation.

  \item \textbf{Market activity.} We apply three filters for sufficient trading activity and reliable price data. First, we exclude tokens with fewer than 15~days of price history, which is insufficient for return estimation. Second, we exclude tokens with negligible historical trading volume. Third, we validate supply data, excluding tokens whose market capitalisation or FDV exceeds Ethereum's (an indicator of misconfigured supply).

  \item \textbf{On-chain consistency.} Finally, we validate that our event-based reconstruction matches the on-chain state. For each candidate token, we reconstruct balances at randomly sampled (account, block) pairs and compare them to the on-chain balance retrieved via RPC. Tokens with any disagreement between reconstructed and on-chain balances are excluded, since discrepancies indicate non-standard token implementations (e.g., rebasing tokens, tokens with transfer hooks, or tokens that modify balances outside \texttt{Transfer} events).
\end{enumerate}

\subsection{Portfolio Reconstruction}
\label{sec:portfolio_reconstruction}

\begin{figure}
  \centering
  \resizebox{\linewidth}{!}{\begin{tikzpicture}[
    xscale=0.75,
    arr/.style={-{Stealth[length=2.5mm]}, thick},
    arrA/.style={arr, brown!70!black},
    arrB/.style={arr, blue!70!black},
    flowarr/.style={-{Stealth[length=3mm]}, very thick, gray!50},
    transferlabel/.style={font=\footnotesize, inner sep=1.5pt},
    deltabox/.style={draw, minimum width=0.9cm, minimum height=0.4cm,
                     align=center, font=\small, rounded corners=1.5pt},
    deltaboxA/.style={deltabox, brown!70!black, fill=brown!8},
    deltaboxB/.style={deltabox, blue!70!black, fill=blue!5},
    balbox/.style={draw, minimum width=1.0cm, minimum height=0.4cm,
                   text width=0.85cm, align=right, font=\small, fill=white,
                   inner sep=2pt},
    balboxA/.style={balbox, draw=brown!70!black, text=brown!70!black},
    balboxB/.style={balbox, draw=blue!70!black, text=blue!70!black},
    stepbox/.style={draw, rounded corners=3pt, minimum width=1.2cm,
                    minimum height=0.5cm, font=\normalsize\bfseries, align=center},
    smalltransfer/.style={draw, minimum width=0.6cm, minimum height=0.35cm,
                          align=center, font=\footnotesize, rounded corners=1.5pt},
    smalltransferA/.style={smalltransfer, brown!70!black, fill=brown!8},
    smalltransferB/.style={smalltransfer, blue!70!black, fill=blue!5},
    dotnode/.style={circle, fill=gray!40, inner sep=1.5pt},
]

\fill[gray!5, rounded corners=5pt] (-4.7, 1.8) rectangle (-0.1, -6.6);
\node[stepbox, fill=gray!12, draw=gray!50, text=gray!60!black]
    at (-2.4, 1.5) {Step 1: Token-level};

\draw[black, thick, -{Stealth[length=2.5mm]}] (-2.4, 0.15) -- node[fill=white, font=\footnotesize, inner sep=2pt] {\textit{MapReduce}} (-2.4, -1.1);

\node[font=\scriptsize, text=gray!50!black, anchor=south, align=center] at (-2.4, 0.5)
    {\textit{Transfer events} \\ \textit{across all accounts}};

\node[font=\footnotesize, text=gray!60] at (-3.6, -1.25) {\textit{From}};
\node[font=\footnotesize, text=gray!60] at (-1.2, -1.25) {\textit{To}};

\node[dotnode, label={[font=\tiny]below:A}] (f1) at (-3.4, -1.7) {};
\node[smalltransferA] (t1) at (-2.4, -1.7) {100 \textsc{x}};
\node[dotnode, label={[font=\tiny]below:B}] (to1) at (-1.4, -1.7) {};
\draw[brown!50, thick] (f1) -- (t1);
\draw[brown!50, thick] (t1) -- (to1);

\node[dotnode, label={[font=\tiny]below:B}] (f2) at (-3.4, -2.7) {};
\node[smalltransferA] (t2) at (-2.4, -2.7) {50 \textsc{x}};
\node[dotnode, label={[font=\tiny]below:C}] (to2) at (-1.4, -2.7) {};
\draw[brown!50, thick] (f2) -- (t2);
\draw[brown!50, thick] (t2) -- (to2);

\node[dotnode, label={[font=\tiny]below:C}] (f3) at (-3.4, -3.7) {};
\node[smalltransferB] (t3) at (-2.4, -3.7) {200 \textsc{y}};
\node[dotnode, label={[font=\tiny]below:A}] (to3) at (-1.4, -3.7) {};
\draw[blue!40, thick] (f3) -- (t3);
\draw[blue!40, thick] (t3) -- (to3);

\node[dotnode, label={[font=\tiny]below:A}] (f4) at (-3.4, -4.7) {};
\node[smalltransferA] (t4) at (-2.4, -4.7) {30 \textsc{x}};
\node[dotnode, label={[font=\tiny]below:C}] (to4) at (-1.4, -4.7) {};
\draw[brown!50, thick] (f4) -- (t4);
\draw[brown!50, thick] (t4) -- (to4);

\draw[flowarr] (-0.5, -1.5) -- (1.0, -1.5);

\fill[gray!5, rounded corners=5pt] (1.6, 1.8) rectangle (14.0, -6.6);
\node[stepbox, fill=gray!12, draw=gray!50, text=gray!60!black]
    at (7.5, 1.5) {Step 2: Address-level Ledger};

\def\accA{0}
\def\accB{-2.4}
\def\accC{-4.8}

\def\xStart{1.6}
\def\tZero{2.4}
\def\tOneA{4.6}
\def\tOneMid{5.6}
\def\tOneB{6.6}
\def\tOneC{8.6}
\def\tTwoMid{9.6}
\def\tTwo{10.6}
\def\tEnd{12.6}
\def\xRight{13.6}

\draw[thick] (\xStart, \accA) -- (\xRight, \accA);
\node[font=\normalsize\bfseries, anchor=east] at (\xStart-0.15, \accA) {Alice};
\node[font=\scriptsize, anchor=east] at (\xStart-0.15, \accA-0.35) {(A)};

\draw[thick] (\xStart, \accB) -- (\xRight, \accB);
\node[font=\normalsize\bfseries, anchor=east] at (\xStart-0.15, \accB) {Bob};
\node[font=\scriptsize, anchor=east] at (\xStart-0.15, \accB-0.35) {(B)};

\draw[thick] (\xStart, \accC) -- (\xRight, \accC);
\node[font=\normalsize\bfseries, anchor=east] at (\xStart-0.15, \accC) {Carol};
\node[font=\scriptsize, anchor=east] at (\xStart-0.15, \accC-0.35) {(C)};

\fill[gray!12, rounded corners=2pt] (\tZero-0.85, \accA+0.8) rectangle (\tZero+0.85, \accC-0.8);
\fill[gray!12, rounded corners=2pt] (\tEnd-0.85, \accA+0.8) rectangle (\tEnd+0.85, \accC-0.8);

\foreach \x/\lab in {\tOneMid/{$t\!+\!1$}, \tTwoMid/{$t\!+\!2$}} {
    \draw[gray!40, dashed] (\x, \accA+0.8) -- (\x, \accC-0.8);
    \node[font=\scriptsize\bfseries, anchor=south] at (\x, \accA+0.8) {\lab};
}
\node[font=\scriptsize\bfseries, anchor=south] at (\tZero, \accA+0.8) {$t$};
\node[font=\scriptsize\bfseries, anchor=south] at (\tEnd, \accA+0.8) {$t\!+\!\Delta t$};

\node[balboxA, anchor=south] at (\tZero, \accA) {500 \textsc{x}};
\node[balboxB, anchor=north] at (\tZero, \accA) {0 \textsc{y}};

\node[balboxA, anchor=south] at (\tZero, \accB) {0 \textsc{x}};
\node[balboxB, anchor=north] at (\tZero, \accB) {0 \textsc{y}};

\node[balboxA, anchor=south] at (\tZero, \accC) {0 \textsc{x}};
\node[balboxB, anchor=north] at (\tZero, \accC) {300 \textsc{y}};

\node[deltaboxA] (t1s) at (\tOneA, \accA) {$-100$};
\node[deltaboxA] (t1r) at (\tOneA, \accB) {$+100$};
\draw[arrA] (t1s.south) -- node[transferlabel, right, text=brown!70!black] {100 \textsc{x}} (t1r.north);

\node[deltaboxA] (t2s) at (\tOneB, \accB) {$-50$};
\node[deltaboxA] (t2r) at (\tOneB, \accC) {$+50$};
\draw[arrA] (t2s.south) -- node[transferlabel, right, text=brown!70!black] {50 \textsc{x}} (t2r.north);

\node[deltaboxB] (t3s) at (\tOneC, \accC) {$-200$};
\node[deltaboxB] (t3r) at (\tOneC, \accA) {$+200$};
\draw[arrB] (t3s.north) -- node[transferlabel, right, pos=0.35, text=blue!70!black] {200 \textsc{y}} (t3r.south);

\node[deltaboxA] (t4s) at (\tTwo, \accA) {$-30$};
\node[deltaboxA] (t4r) at (\tTwo, \accC) {$+30$};
\draw[arrA] (t4s.south) -- node[transferlabel, right, pos=0.35, text=brown!70!black] {30 \textsc{x}} (t4r.north);

\node[balboxA, anchor=south] at (\tEnd, \accA) {370 \textsc{x}};
\node[balboxB, anchor=north] at (\tEnd, \accA) {200 \textsc{y}};

\node[balboxA, anchor=south] at (\tEnd, \accB) {50 \textsc{x}};
\node[balboxB, anchor=north] at (\tEnd, \accB) {0 \textsc{y}};

\node[balboxA, anchor=south] at (\tEnd, \accC) {80 \textsc{x}};
\node[balboxB, anchor=north] at (\tEnd, \accC) {100 \textsc{y}};

\draw[decorate, decoration={brace, mirror, amplitude=5pt}, thick, gray!60]
    (\tZero, \accC-1.0) -- (\tEnd, \accC-1.0)
    node[midway, below=6pt, font=\normalsize, text=gray!60!black] {$\Delta t$};

\fill[gray!5, rounded corners=5pt] (14.7, 1.8) rectangle (19.3, -6.6);
\node[stepbox, fill=gray!12, draw=gray!50, text=gray!60!black]
    at (17.0, 1.5) {Step 3: Portfolio-level};

\draw[flowarr] (14.1, -1.5) -- (15.3, -1.5);

\node[font=\footnotesize] at (17.0, \accA)
    {$w^{(A,\, t+\Delta t)} \propto P \begin{pmatrix} 370 \\ 200 \end{pmatrix}$};

\node[font=\footnotesize] at (17.0, \accB)
    {$w^{(B,\, t+\Delta t)} \propto P \begin{pmatrix} 50 \\ 0 \end{pmatrix}$};

\node[font=\footnotesize] at (17.0, \accC)
    {$w^{(C,\, t+\Delta t)} \propto P \begin{pmatrix} 80 \\ 100 \end{pmatrix}$};

\node[font=\scriptsize\bfseries, anchor=west] at (-4.2, -7.2) {Legend:};

\fill[brown!8, draw=brown!70!black, rounded corners=1.5pt] (-1.8, -7.05) rectangle (-1.3, -7.35);
\node[font=\footnotesize, anchor=west] at (-1.2, -7.2) {Token \textsc{x} (brown)};

\fill[blue!5, draw=blue!70!black, rounded corners=1.5pt] (2.9, -7.05) rectangle (3.4, -7.35);
\node[font=\footnotesize, anchor=west] at (3.5, -7.2) {Token \textsc{y} (blue)};

\node[font=\footnotesize, anchor=west] at (7.6, -7.2)
    {$t = (\text{block}, \text{log index})$};

\node[font=\footnotesize, anchor=west] at (12.0, -7.2)
    {$P$: diagonal USD price matrix at time $t\!+\!\Delta t$};

\end{tikzpicture}}
  \caption{Three-step portfolio reconstruction pipeline for two example tokens (X, \textcolor{brown!70!black}{brown}; Y, \textcolor{blue!70!black}{blue}) and three example accounts (Alice, Bob, Carol). \textbf{Step~1} (left): a MapReduce procedure extracts \texttt{Transfer} events from blockchain logs and produces per-token transfer files. \textbf{Step~2} (centre): for each token, transfers are recorded as signed ledger entries (debit $-$ for senders, credit $+$ for receivers), ordered by block height and log index. Cumulative sums over the interval $\Delta t$ yield per-account balances at the snapshot time $t$. \textbf{Step~3} (right): per-account balances at snapshot $t$ are combined with the asset prices $p^{(t)}_i$ to yield position values $v^{(a,t)}_i$, which are normalised into portfolio weight vectors $w^{(a,t)}$. In our analysis, $\Delta t$ corresponds to one month.}
  \label{fig:reconstruction_pipeline}
\end{figure}
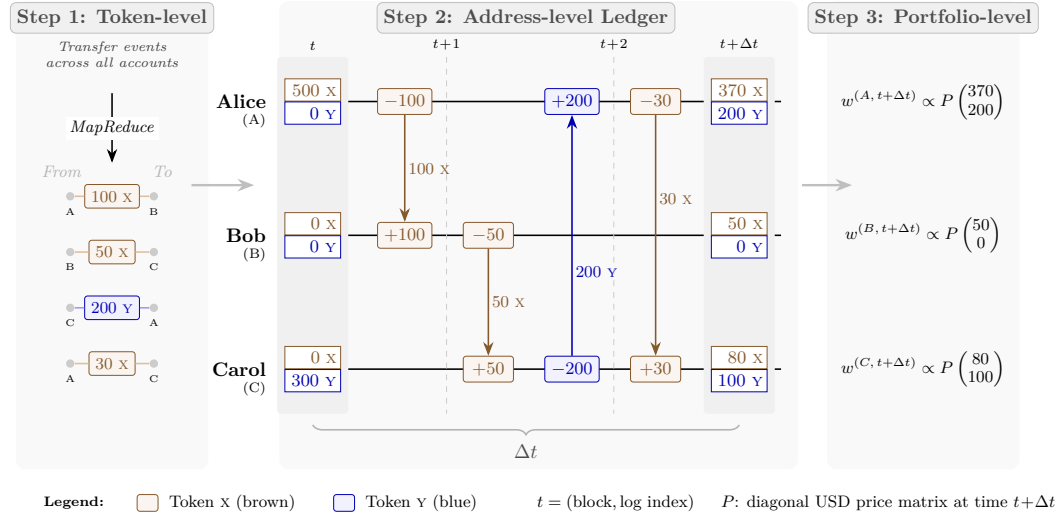

As \Cref{fig:reconstruction_pipeline} shows, starting from raw Ethereum \textit{logs} and \textit{traces}, we reconstruct the token holdings of each account at any point in time (block height) in three steps: extracting token-level transfer events, aggregating transfers into per-address balance ledgers, and computing portfolio snapshots at target time points. Implementation details on the underlying storage infrastructure are provided in Appendix~\ref{sec:computational_details}.

\textbf{Step 1: Token-level extraction.} We extract the logs for the ERC-20 transfer-signature topic and write per-token files of (sender, recipient, value) tuples, parallelised across tokens via a MapReduce pipeline~\cite{dean2004mapreduce}. Some assets require additional token-specific handling: Wrapped ETH (WETH), for instance, exposes \texttt{Deposit} and \texttt{Withdrawal} events, which we map to synthetic \texttt{Transfer} entries from or to the zero address to unify the event stream.

\textbf{Step 2: Account-level ledger.} We transform the \texttt{Transfer} stream into per-(account, token) ledgers, where each account corresponds to an Ethereum address. Each event produces a signed entry (debit for sender, credit for recipient), ordered by block height and log index. The ledger supports balance aggregation over any block range from genesis to a target height.

\textbf{Step 3: Portfolio snapshots.} A portfolio is the collection of tokens held by an account at a given point in time, with Ethereum block heights converted to datetimes via an internal mapping. For each account and target timestamp, we sum the corresponding ledger entries, price each position with forward-filled daily USD closes, and normalise the position values into a fully-invested weight vector. We sample \emph{monthly snapshots} at the first day of each calendar month, matching the literature~\cite{Kumar2008} and balancing resolution against compute.

For each account $a$ and time $t$, we obtain the USD value of each held position $v^{(a,t)}_i$ for asset $i$, the total portfolio value $V^{(a,t)} = \sum_i v^{(a,t)}_i$, the number of held tokens $N^{(a,t)}$, and the fully-invested weight vector\footnote{\emph{Notation convention.} Throughout, when the $(a,t)$ indices are suppressed (e.g.\ $w$, $V$, $N$, $\mu$, $\sigma$), the quantity refers to the portfolio of account $a$ at time $t$.}
\begin{equation}
w^{(a,t)} = \bigl(w^{(a,t)}_1, \dots, w^{(a,t)}_N\bigr), \qquad w^{(a,t)}_i = \frac{v^{(a,t)}_i}{V^{(a,t)}},
\end{equation}
where $\sum_{i=1}^N w^{(a,t)}_i = 1$. Reconstruction correctness is validated against on-chain balances as described in \cref{sec:data_collection} (step~3).

\subsection{Efficient Frontier Computation}
\label{sec:frontier_computation}

Building on the \gls{MPT} framework introduced in \cref{sec:portfolio_theory}, we estimate the efficient frontier for each account and snapshot date $t$ from empirical return statistics over a historical lookback window, constrained to the account's already-held assets. Token-specific functionalities (e.g., underlying lending positions, staking yields) are disregarded: tokens are treated as simple assets characterised by prices.

\paragraph*{Return and risk estimation.}

\begin{definition}[Expected return]
\label{eq:asset-mean}
The expected asset-return of asset $i$ over a $T^{(-)}$-day lookback window ending at $t$ is
\begin{equation*}
  \mu^{(t)}_i \;=\; \frac{1}{T^{(-)}}\sum_{\tau = 0}^{T^{(-)}-1} r^{(t-\tau)}_i, \qquad \text{with} \qquad
  r^{(t)}_i \;=\; \log\!\left(\frac{p^{(t)}_i}{p^{(t-1)}_i}\right)
\end{equation*}
the log-return of asset $i$ at time $t$ and $p^{(t)}_i$ the USD closing price of token $i$ at time $t$.
\end{definition}

\begin{definition}[Portfolio mean and variance]
\label{eq:portfolio-moments}
The expected return and variance of account $a$'s portfolio at time $t$ are
\begin{equation*}
  \mu^{(a,t)} \;=\; \sum_{i=1}^{N} w^{(a,t)}_i \mu^{(t)}_i, \qquad
  \bigl(\sigma^{(a,t)}\bigr)^2 \;=\; \bigl(w^{(a,t)}\bigr)^\top \Sigma^{(t)} \, w^{(a,t)},
\end{equation*}
where $\Sigma^{(t)} \in \mathbb{R}^{N \times N}$ is the empirical \gls{covariance-matrix} over the lookback ending at $t$, and $N$ is the number of assets held at $t$.
\end{definition}

\paragraph*{Reference portfolios under optimisation strategies.}
For each account $a$ and time $t$, we apply three optimisation strategies, $s \in \{\text{MinVar}, \text{MaxRet}, \text{MaxSR}\}$, each producing a reference portfolio $w_{(s)}$ on the constrained efficient frontier.

\begin{definition}[Frontier portfolios]
\label{def:reference_portfolios}
Let $w_{(0)} \equiv w^{(a,t)}$ denote the actual portfolio with moments $\mu_{(0)}, \sigma_{(0)}$, and $w_{(s)}$ the portfolio optimised under strategy $s$ with moments $\mu_{(s)}, \sigma_{(s)}$:
\begin{description}
  \item[\Gls{MVR}:] $\min_{w}\, \sigma_{(s)}$ s.t.\ $\mu_{(s)} = \mu_{(0)} \;\longrightarrow\; w_{(\text{MinVar})}$.
  \item[\Gls{MRV}:] $\max_{w}\, \mu_{(s)}$ s.t.\ $\sigma_{(s)} = \sigma_{(0)} \;\longrightarrow\; w_{(\text{MaxRet})}$.
  \item[\Gls{MSR}:] $\max_{w}\, (\mu_{(s)} - r_f) / \sigma_{(s)} \;\longrightarrow\; w_{(\text{MaxSR})}$.
\end{description}
\end{definition}

The \gls{MVR} and \gls{MRV} anchor to the actual portfolio's expected return and standard deviation, respectively, corresponding to the horizontal and vertical projections of $w_{(0)}$ onto the frontier (\cref{fig:approach}). The \gls{MSR} is the \emph{tangency portfolio}, the point where the line from $r_f$ touches the frontier, and it depends on the assets held but not on their initial weights.
We set $r_f = \RfDefault/\%$ per annum for the \gls{MSR} optimisation, close to on-chain stablecoin lending yields and consistent with~\cite{platanakis2019portfolio}, who proxy $r_f$ by the U.S.\ Treasury bill rate (2023--2024~\cite{fred_tb3ms}). \Cref{app:risk-free-rate} motivates the calibration and shows that results are insensitive to $r_f$.

\paragraph*{Shared constraints.}
All three optimisation problems share the following constraints, which together with the held-asset restriction define the \emph{constrained efficient frontier} used throughout the paper:
\begin{itemize}
  \item \emph{Fully invested}: $\sum_i w_{(s),i} = 1$, with $w_{(s),i} = 0$ whenever $w_{(0),i} = 0$ (pure reallocation across held assets, no external funds).
  \item \emph{Long-only}: $w_{(s),i} \ge 0$ (no spot-level short-selling available for most ERC-20 tokens).
  \item \emph{Single-asset cap}: $w_{(s),i} \le w_{\max} = 0.9$ (prevents collapse to a near-single-asset solution).
\end{itemize}

\paragraph*{Configuration.}
We solve all three optimisation problems with sequential least-squares programming (SLSQP). The following parameters are shared across all optimisation instances:

\begin{itemize}
  \item \emph{Lookback window:} $T^{(-)}=60$ daily observations (minimum 45 per asset required, with tokens below the threshold excluded at that time point). The window balances stable covariance estimation against the frequent regime changes of cryptoasset markets.
  \item \emph{Covariance shrinkage:} We apply Ledoit--Wolf shrinkage~\cite{ledoit2004honey} to stabilise the empirical covariance matrix when $N$ is large relative to $T^{(-)}$.
  \item \emph{Return regularisation:} Sample mean returns are highly noisy over the $T^{(-)}$-day lookback in crypto markets. We shrink the sample mean toward a neutral prior of equal expected returns across assets:
  \[
  \mu_i^{\text{shrunk}} = \lambda\, \bar\mu + (1-\lambda)\, \mu_i,
  \]
  where $\bar\mu = \frac{1}{N}\sum_{j=1}^N \mu_j$ is the cross-asset mean and $\lambda = 0.5$ (further details in \cref{sec:mean_reversion_details}).
\end{itemize}

\subsection{Distance and Performance Metrics}
\label{sec:metrics}

We evaluate the optimisation outcomes along two dimensions: how far each optimised portfolio deviates from the actual allocation (Obj~B), and whether the resulting shift improves risk-adjusted returns (Obj~C).

\paragraph*{Weight $\ell_1$ distance.}

\begin{definition}[Weight $\ell_1$ distance]
\label{eq:l1_weight_distance}
The $\ell_1$ distance between the actual allocation $w_{(0)}$ and an optimised reference $w_{(s)}$ is
\begin{equation*}
  d_{s}
  =
  \tfrac{1}{2}\left\lVert w_{(0)} - w_{(s)} \right\rVert_{1}
  =
  \frac{1}{2}\sum_{i=1}^{N} \left|\, w_{(0),i} - w_{(s),i} \,\right|.
\end{equation*}
\end{definition}

We use $\ell_1$ rather than Euclidean ($\ell_2$) because $d_{s}$ admits a direct financial interpretation: it equals the minimum one-way turnover required to rebalance from $w_{(0)}$ to $w_{(s)}$, and is bounded in $[0,1]$ for long-only, fully-invested portfolios, with $0$ indicating identical allocations and $1$ disjoint ones. See \cref{sec:l1_distance} for the formal derivation.

\paragraph*{Risk-adjusted performance.}
\label{sec:capm}

\begin{definition}[Forward-realised return]
\label{eq:portfolio_forward_return}

The forward-realised return of account $a$'s allocation $w^{(a,t)}$ between consecutive snapshots $t$ and $t' = t + T^{(+)}$ is
\begin{equation*}
  r^{(a,t)}_{t'} \;=\; \sum_{i=1}^{N} w^{(a,t)}_i \left(\frac{p^{(t')}_i}{p^{(t)}_i} - 1\right),
\end{equation*}
where $p^{(t)}_i$ and $p^{(t')}_i$ are the asset prices at the start and end of the holding window; the expression is evaluated at the actual $w_{(0)}$ or any theoretical $w_{(s)}$.
\end{definition}

Raw returns conflate allocation with market movement: in a bull market, even naive allocations are positive. To isolate the allocation's contribution, we apply the \acrlong{CAPM} (\acrshort{CAPM}) and decompose the return into a component proportional to the portfolio's market exposure ($\beta$) and an excess residual ($\alpha$), the risk-adjusted return.
As the market benchmark $r^{(t)}_{(m)}$, we take the equally-weighted average of the daily log-returns of \gls{wETH} and \gls{WBTC}. Equal weighting prevents structural overlap with the market-cap-weight strategy.

\begin{definition}[Market exposure]
\label{def:market_exposure}
The market exposure of account $a$'s portfolio at time $t$ is
\begin{equation*}
  \beta^{(a,t)} \;=\; \sum_{i} w^{(a,t)}_i \, \beta^{(t)}_i, \qquad \text{with} \qquad
  \beta^{(t)}_i \;=\; \frac{\mathrm{Cov}_{\tau}\bigl(r^{(t-\tau)}_i,\, r^{(t-\tau)}_{(m)}\bigr)}{\mathrm{Var}_{\tau}\bigl(r^{(t-\tau)}_{(m)}\bigr)}
\end{equation*}
the sensitivity of asset $i$ to the market at time $t$, estimated over the same $T^{(-)}$-day lookback, and $w_{(s)}$ admissible in place of $w^{(a,t)}$.
\end{definition}

A value $\beta^{(a,t)} = 1$ implies lockstep movement with the market, $\beta^{(a,t)} > 1$ amplifies market moves, and $\beta^{(a,t)} < 1$ dampens them.

\begin{definition}[Risk-adjusted excess return]
\label{eq:alpha}
The risk-adjusted excess return of account $a$'s portfolio at $t'$ is
\begin{equation*}
  \alpha^{(a,t)}_{t'} \;=\; r^{(a,t)}_{t'} - \beta^{(a,t)} \cdot r^{(t')}_{(m)}.
\end{equation*}
\end{definition}

Here, we set $r_f = 0$ to avoid an off-chain rate assumption. Positive $\alpha^{(a,t)}_{t'}$ indicates outperformance beyond market exposure, and negative values indicate underperformance.

\paragraph*{Ethics and data handling.}
The study uses public Ethereum logs, public token-price data, and aggregate address tags. We do not deanonymise addresses, and address tags are used only for aggregate contract/category-level analysis. The released artifacts do not expose private address-level information, only public data already publicly available on-chain.

\section{Portfolio Characteristics}
\label{sec:descriptive_statistics}

This section addresses \textbf{Obj A} by characterising the wealth, concentration, and diversification of on-chain portfolios.
The analysis covers \CoingeckoCleanedTokensN/ ERC-20 tokens and \TotalAccountsEnd/ (\TotalAccountsRounded/) accounts at monthly snapshots across Ethereum's history (July 2015 -- December 2025).

\subsection{Evolution of Wealth}
\label{sec:evolution_of_wealth}

We measure account wealth $V$ as the USD value of assets held at each monthly snapshot.
We start with an aggregate view by summing these values across all accounts, providing an overview of the total ERC-20 wealth in the ecosystem and its evolution over time.

\begin{figure}[htbp]
  \centering
  \includegraphics[width=1\linewidth]{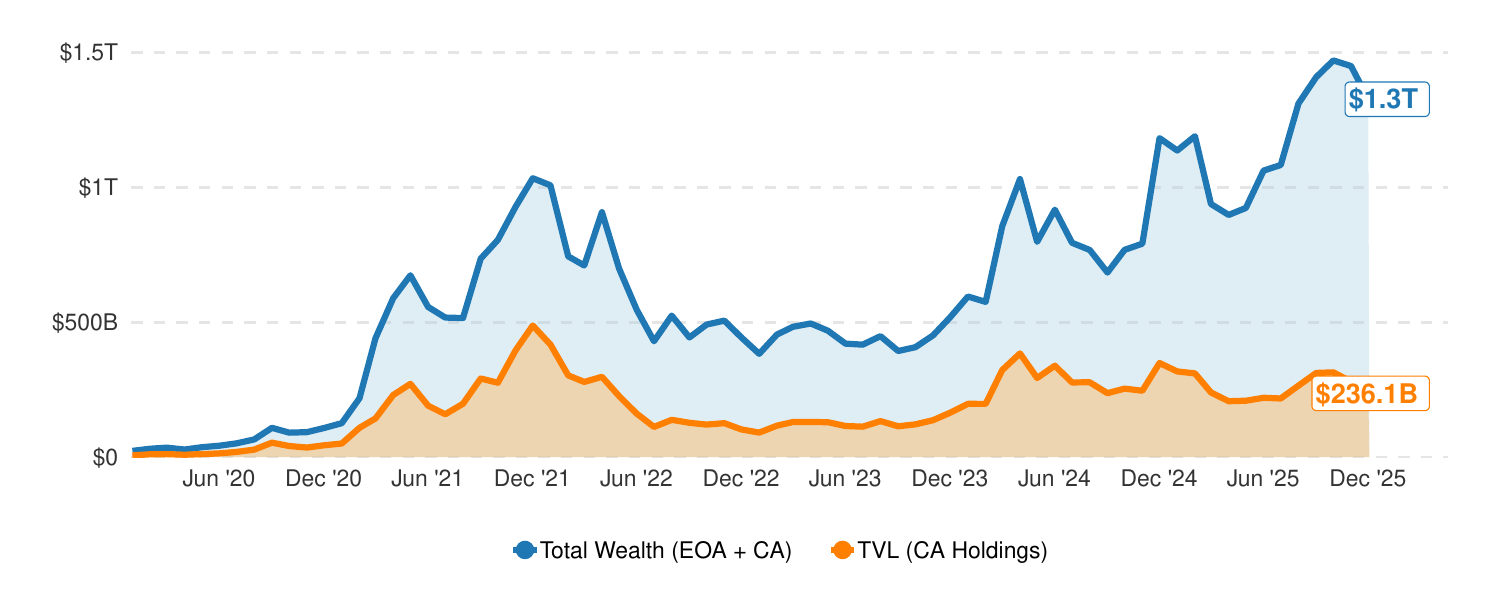}
  \caption{Total reconstructed portfolio value for all accounts over time, contrasted with \gls{TVL} (total value locked) in \gls{DeFi} protocols.}
  \label{fig:wealth_composition}
\end{figure}

\cref{fig:wealth_composition} traces the total USD value of reconstructed portfolios over time and contrasts it with \gls{TVL}~\cite{defillama2025}.
Total wealth follows the well-known boom--bust--recovery cycle of the broader crypto market,
peaking near \WealthPeakTwentyone/ in late 2021 and tracking the \gls{TVL} trajectory at consistently higher levels, before contracting sharply through 2022 as a sequence of market shocks unfolded.\footnote{Terra/LUNA (May 2022), Celsius and Three Arrows Capital (June--July 2022), FTX (November 2022).}
From 2024 onwards the two series decouple, with portfolio wealth exhibiting an additional growth trend beyond \gls{TVL}, potentially due to the rise of meme coins held outside \gls{DeFi} protocols, and peaking at \WealthPeakTwentyfiveB/ during 2025.

Disaggregating wealth by account type reveals a marked asymmetry:
although \glspl{CA} represent only \CAShareEndPct/ of all asset-holding accounts,
they hold on average \WealthShareCAMeanP/\% of total wealth,
while \glspl{EOA} account for the remaining \WealthShareEOAMeanP/\%.
The \gls{CA} share is highest during the 2020--2021 \gls{DeFi} expansion and declines toward end-2025 as wealth shifts to \gls{EOA} accounts.
This split is informative when comparing portfolio wealth with the \gls{TVL} metric:
\gls{TVL} captures only value locked in \gls{DeFi} protocols and therefore systematically underestimates total on-chain wealth by ignoring the large share held in \gls{EOA} accounts.
Although our reconstruction does not cover native ETH, the reconstructed account wealth still substantially exceeds \gls{TVL}, indicating the magnitude of wealth not captured by the metric.
Both series are also subject to double-counting of wrapped or derivative tokens~\cite{saggese2025tvl}.

\subsection{Wealth Distribution and Concentration}
\label{sec:wealth_concentration}

We characterise the wealth distribution across accounts from both ends: first via shares of accounts by wealth bin and account type, then via the top $1\%$ and $10\%$ wealth shares as our primary measure of concentration, complemented by standard inequality indices.

\begin{figure}[htbp]
  \centering
  \includegraphics[width=1\linewidth]{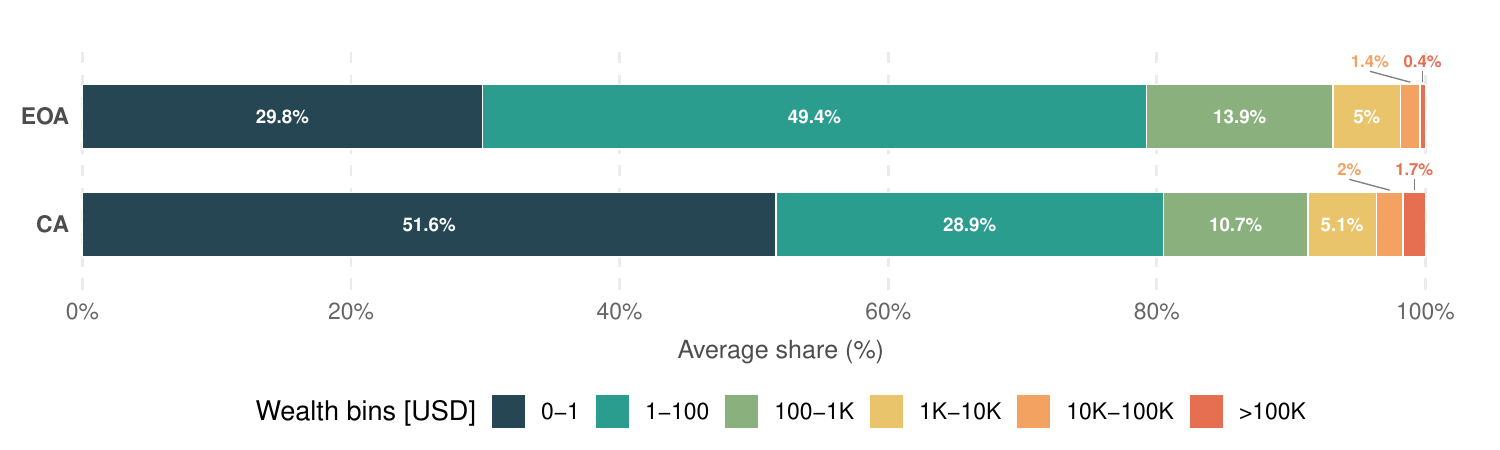}
  \caption{Share of \gls{EOA} and \gls{CA} accounts by wealth bucket (USD), computed across all \ReconSnapshotBlocks/ monthly snapshots.}
  \label{fig:wealth-composition-bar}
\end{figure}

As \cref{fig:wealth-composition-bar} shows, the distribution of accounts
across wealth bins differs systematically by account type but converges
on the same headline: the majority of both \glspl{CA} and \glspl{EOA}
consistently hold less than \$100, with \glspl{CA} concentrated in the
dust\footnote{Millions of accounts can hold negligible residual balances after a trade, which can inflate the denominator of the distribution and mechanically increase measured concentration.} bin ($\leq\$1$, $\WealthBucketCAZeroOneP/\%$ share over the entire observation period)
and \glspl{EOA} in the \$1--\$100 bin ($\WealthBucketEOAOneHundredP/\%$).
The two types reach their low-wealth-share minima at different times,
the \gls{CA} minimum around mid-2021 (``DeFi Summer'') and the \gls{EOA}
minimum around late 2021 at the broader market peak, jointly producing
the two wealth peaks visible in \cref{fig:wealth_composition}. Account
growth trajectories, \gls{CA}/\gls{EOA} wealth shares over time, and
the full bin-level distributions are reported in
Appendix~\ref{sec:account_wealth}.

Turning to the top of the distribution, we restrict the analysis to accounts whose portfolio value exceeds \$1, to mitigate the ``dust'' problem noted above.

We find that the top $1\%$ of accounts consistently controls the vast majority of total portfolio value,
increasing from $\TopOneWealthTwenty/\%$ to $\TopOneWealthTwentyfive/\%$ over 2020--2025.
The top $10\%$ likewise holds nearly the entire ecosystem value, always above $\TopFiveTenWealthMin/\%$,
while the number of accounts quadruples over the same period.
These findings are corroborated by standard inequality measures (Gini, HHI), which remain uniformly high throughout the observation period.
We treat these indices as supplementary rather than primary because address-level inequality measures are debated in pseudonymous blockchain settings, where one entity can control many addresses or one address can serve many entities~\cite{yaish2026inequality,chemaya2025quantifying}.

Narrowing to the top $0.1\%$, account type composition departs from the ecosystem average:
\glspl{CA} make up $\TopPointOneCAPctWallets/\%$ of these addresses, far above their \CAShareEndPct/ population share, with DeFi protocols, exchanges, and custodial wallets dominating the tagged categories.
However, the aggregate \gls{CA} value share within this slice ($\TopPointOneCAPctValue/\%$) is essentially unchanged from the ecosystem-wide \WealthShareCAMeanP/\%, meaning the same fraction of the pie is simply held by a proportionally larger group.
The \glspl{EOA} that reach the top $0.1\%$ are correspondingly whales with highly concentrated individual wealth.

Yearly concentration statistics and account type breakdowns are provided in Appendix~\ref{sec:appendix_characteristics}, as well as the full Gini and HHI analysis and tag-based attribution details.

\subsection{Portfolio Size}
\label{sec:portfolio_characteristics}

Having characterised wealth levels and concentration, we now examine portfolio size $N$, the number of distinct assets held, as a measure of diversification.
Portfolio size serves as a proxy for diversification, as holding more assets reduces overall portfolio variance.
We focus on accounts holding at least one asset at the snapshot.

\begin{figure}[htbp]
  \centering
  \includegraphics[width=1\linewidth]{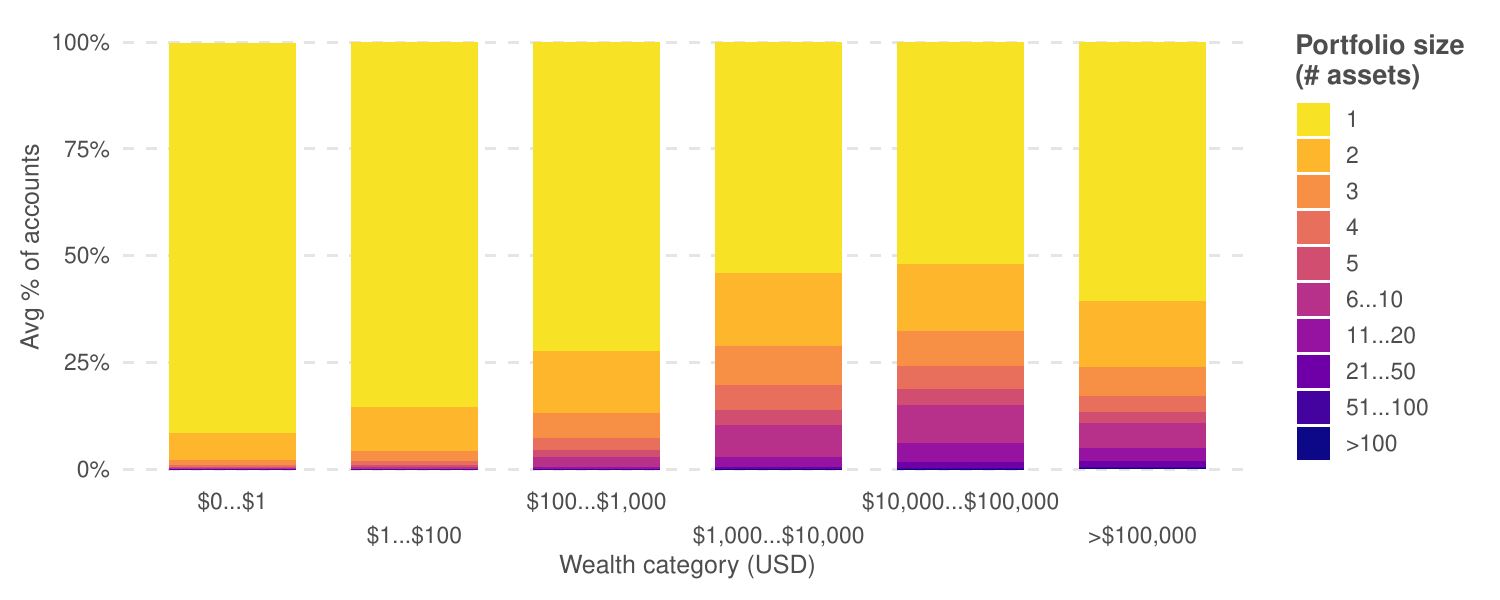}
  \caption{Portfolio size distribution: share of accounts holding a given number of distinct assets, by wealth category, averaged across all monthly snapshots.}
  \label{fig:portfolio_size}
\end{figure}

\cref{fig:portfolio_size} shows the distribution of portfolio sizes across wealth categories, averaged across all monthly snapshots.
Overall, \PortSizeOneTokenPct/\% of accounts hold only a single asset, \PortSizeTwoTokenPct/\% hold two, \PortSizeThreeFourTokenPct/\% hold three to four, and only \PortSizeFivePlusPct/\% hold five or more distinct assets.
Single-asset accounts dominate in every wealth category, but the share of multi-asset portfolios rises steadily with wealth, with the exception of the highest wealth category, where the trend reverses, likely reflecting the prevalence of \glspl{CA} such as exchanges and liquidity pools that concentrate holdings in a small number of high-value tokens.

\subsection*{Takeaways}
Total on-chain wealth tracks the \gls{TVL} trajectory at consistently higher levels, peaking at \WealthPeakTwentyfiveB/ during 2025.
\glspl{CA} account for only \CAShareEndPct/ of asset-holding addresses but hold \WealthShareCAMeanP/\% of aggregate wealth, so \gls{TVL} systematically undercounts on-chain wealth by ignoring the \gls{EOA}-held share.
On-chain wealth is highly concentrated: the top $1\%$ of accounts controls $\TopOneWealthTwentyfive/\%$ in 2025, with the top $0.1\%$ dominated by DeFi protocols, exchanges, and custodial wallets.
While \PortSizeOneTokenPct/\% of accounts hold only a single asset, the multi-asset share rises with wealth except at the very top.
\emph{Overall, on-chain portfolios are small along both dimensions: most accounts hold modest wealth (<\$100) in only one or two tokens.}

\section{Distance to the Efficient Frontier}
\label{sec:mpt_optimal_portfolio}

This section tackles \textbf{Obj B} by measuring how far portfolios sit from the constrained efficient frontier and identifying what predicts their suboptimality.
For every account holding at least two tokens, we solve the three \gls{MVO} problems by following strategies (\gls{MRV}, \gls{MVR}, \gls{MSR}) at \LOneMonthsN/ monthly snapshots (Jan 2020 -- Dec 2025), yielding \LOneObservationsN/ (\LOneObservationsM/) account-month optimisations.
We provide supplementary analyses in Appendix~\ref{sec:appendix_distance} and computational details in Appendix~\ref{sec:computational_details}.

\subsection{Measuring Portfolio Distance}
\label{sec:measuring_distance}

We quantify the gap between actual ($w_{(0)}$) and optimal ($w_{(s)}$) allocations using the weight distance $d$ defined in Def.~\eqref{eq:l1_weight_distance}, examining overall means, distributional patterns across the three strategies, and temporal stability.
We rescale the raw distance from $[0,1]$ to $[0,100]\%$ throughout, where $0\%$ denotes coincidence with the optimum and $100\%$ zero weight overlap.

\begin{figure}[htb]
    \centering
    \includegraphics[width=1\linewidth]{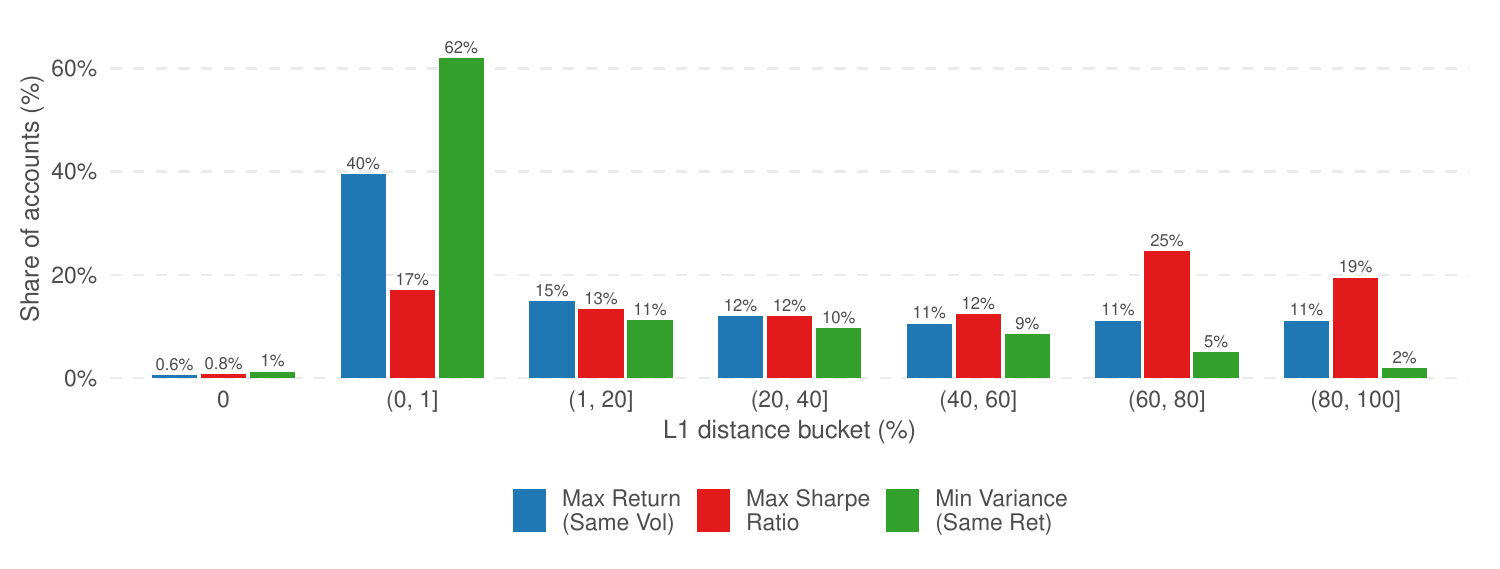}
    \caption{Distribution of portfolio distance to the constrained efficient frontier for the three optimisation strategies. Share of \LOneObservationsM/ account-month optimisations across seven distance bins.}
    \label{fig:l1_distrubution}
\end{figure}

\textbf{Distance prevalences.}
\cref{fig:l1_distrubution} groups the distances into buckets for each of the three optimisation strategies, reporting average shares over the entire observation period.
The resulting distances reveal a clear ordering across strategies:
on-chain portfolios are closest to the \gls{MVR} optimal solution ($\LOneMeanSaferRisk/\%$ mean), followed by \gls{MRV} ($\LOneMeanBetterReturn/\%$), and farthest from the \gls{MSR} optimum ($\LOneMeanMaxSharpe/\%$).
However, these means mask important distributional differences: the three strategies show strikingly different profiles.
\gls{MVR} and \gls{MRV} are left-skewed, with the bulk of account-observations concentrated near the optimum ($\LOneSrCombinedLeOnePct/\%$ and $\LOneBrCombinedLeOnePct/\%$ within $1\%$, respectively).
In contrast, \gls{MSR} is right-skewed, with $\LOneMs60to80Pct/\%$ of observations in the $(60,80]\%$ bin.

\begin{figure}[htbp]
    \centering
    \includegraphics[width=1\linewidth]{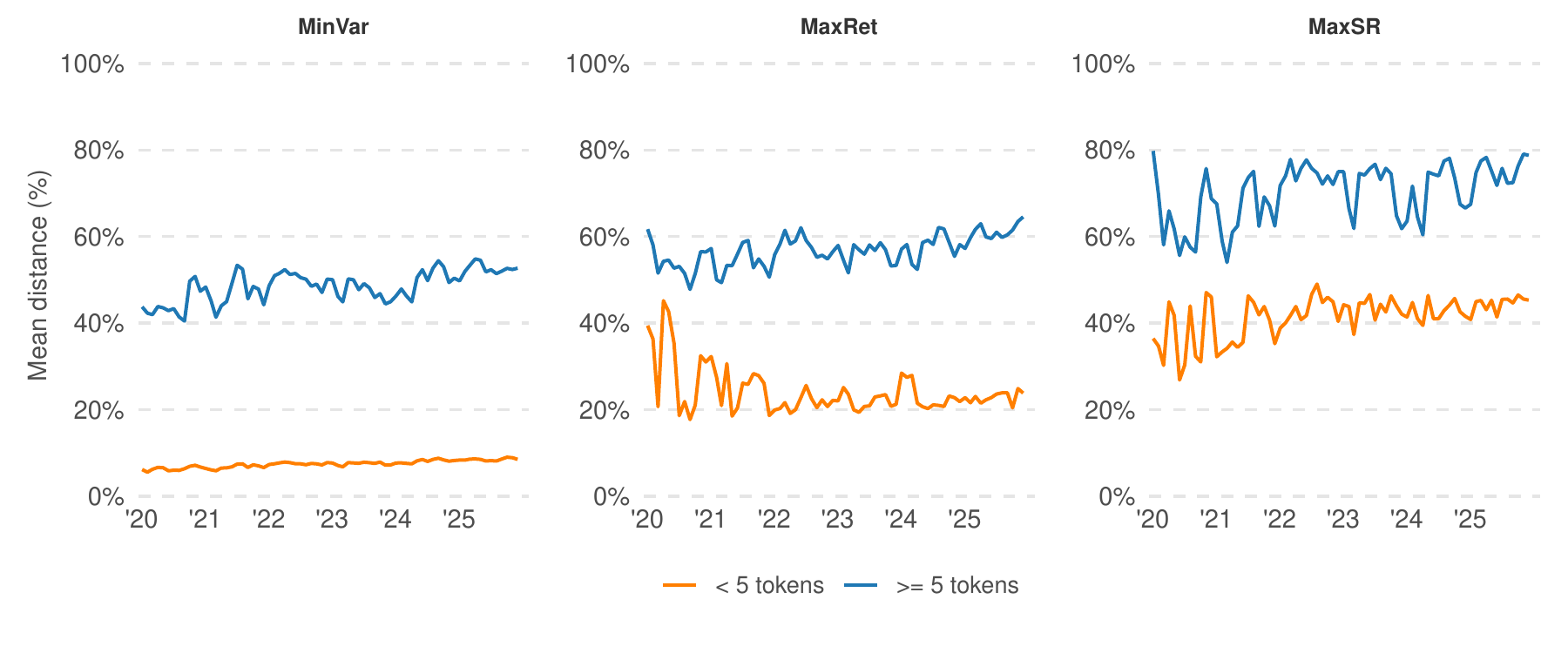}
    \caption{Mean distance over time by portfolio size. Trajectories for portfolios holding fewer than five assets versus five or more, with standard-error bands; one panel per optimisation strategy.}
    \label{fig:l1_mean_token_group}
\end{figure}

\textbf{Temporal dynamics.}
Distance may vary with portfolio size and evolve with market conditions over time.
To investigate, we stratify accounts at an illustrative threshold of $N=5$ for portfolio size and track the mean distance over time in \cref{fig:l1_mean_token_group}, with one panel per optimisation strategy and standard-error bands around each trajectory.
All three distances exhibit a slight upward trend over time.

Small portfolios ($N<5$) have a consistently small average distance to \gls{MVR} (\LOneSmallMVRMean/\,\%), indicating near-structural alignment, while \gls{MSR} shows the largest (\LOneSmallMSRMean/\,\%).
Large portfolios ($N \geq 5$) show a sharp increase to \LOneLargeMVRMean/\,\% for \gls{MVR}, and \LOneLargeMSRMean/\,\% for \gls{MSR}.
A paired Wilcoxon signed-rank test across the \TotalBlocks/ monthly snapshots confirms this small--large gap is significant with large effect sizes.

Notably, this near-optimal share is almost entirely due to two-asset portfolios, which have little room to deviate from the minimum-variance weights and therefore sit close to the optimum by construction: once we restrict to $N \geq 5$, the share within $1\%$ of the optimum drops to near zero.
Test statistics, Cohen's $d$ values, per-size-bucket plots, threshold sensitivity, cross-strategy correlations, and robustness analysis are in Appendix~\ref{sec:appendix_distance}.

\subsection{Determinants and Modeling of Distance}
\label{sec:determinants_modeling}

Having characterised how distance is distributed and evolves over time, we now ask what predicts suboptimality and whether it admits a parsimonious model.
We first identify the dominant determinants, then formalise the distance--size relationship ($d \sim N$) with a parametric model, and finally benchmark actual allocations against naive strategies.

\begin{figure}[htbp]
    \centering
    \includegraphics[width=1\linewidth]{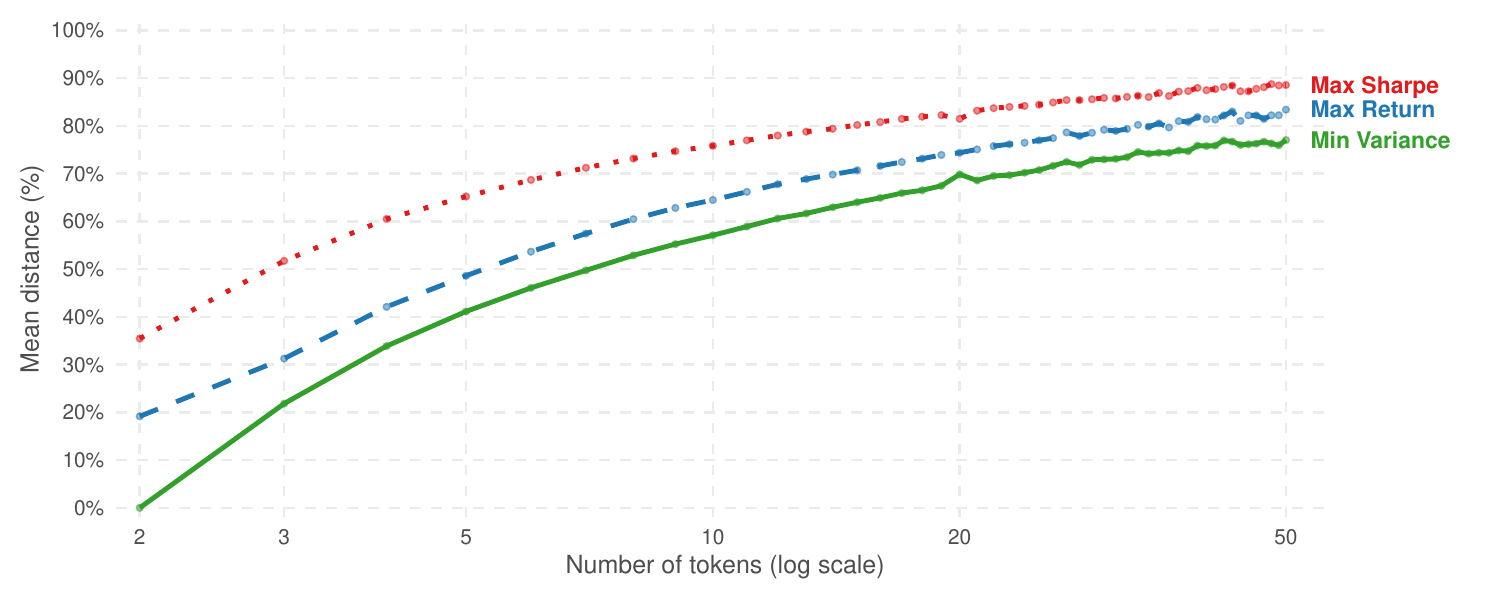}
    \caption{Mean distance versus portfolio size. Distance $d$ as a function of size $N$ on a log-scaled axis for all three optimisation strategies.}
    \label{fig:tokens_vs_distance}
\end{figure}

On a sampled subset, we regress distance ($d$) on three features: portfolio size ($N$), the log portfolio value ($\log V$), and the snapshot month ($t$).
As shown in \cref{fig:tokens_vs_distance}, $N$ is the dominant determinant of \emph{mean} distance across all three strategies.
Random Forest models show that $N$ also dominates the \emph{variance} of \gls{MVR} distance across individual accounts ($\LOneRfSrTokensImp/\%$), driven by a near-optimal cluster of almost exclusively two-asset portfolios, while for \gls{MRV} and \gls{MSR} portfolio value becomes the strongest individual-level predictor ($\LOneRfBrMsValueImp/\%$).

\textbf{Parametric model of distance.}
As \cref{fig:tokens_vs_distance} shows, the $d \sim N$ relationship is monotone and concave across all three strategies: distance rises steeply from $N=2$ to $N=5$ and plateaus thereafter, consistent with diminishing marginal deviation as portfolio size grows. This plateau also motivates the $N \geq 5$ stratification used in \cref{sec:measuring_distance}.
We therefore formalise it with a parsimonious three-parameter power-decay model:
\begin{equation}\label{eq:l1_fit}
  \hat{d}_s(N) \;=\; \delta_s^{\infty} \,\bigl(1 \;-\; \psi_s \cdot N^{-\gamma_s}\bigr),
\end{equation}
where $\hat{d}_s$ is the fitted mean distance in percent, $N \geq 2$ the portfolio size, $\delta_s^{\infty} \leq 100\%$ the asymptotic ceiling, $\psi_s > 0$ a shape parameter setting the depth of the dip from the asymptote, and $\gamma_s > 0$ the exponent governing how quickly distance saturates as $N$ grows.
All three fits achieve $R^2 \geq 0.99$ (\cref{tab:l1_fit}).
The asymptote shows that \gls{MRV} reaches $100\%$, while \gls{MVR} and \gls{MSR} plateau at $\FitDinfMVR/\%$ and $\FitDinfMSR/\%$, implying persistent structural overlap with the minimum-variance and Sharpe-optimal allocations.
Estimation details and fitted curves are in Appendix~\ref{sec:appendix_distance}.

\begin{table*}[htbp]
\centering
\caption{Power-decay distance model. Fitted parameters and goodness-of-fit for \cref{eq:l1_fit}, by optimisation strategy.}
\label{tab:l1_fit}
\begin{tabular*}{\textwidth}{l @{\extracolsep{\fill}} rrr rr}
\toprule
 & \multicolumn{3}{c}{Model Parameters}
 & \multicolumn{2}{c}{Model Quality} \\
\cmidrule(lr){2-4} \cmidrule(lr){5-6}
Strategy $s$ & $\psi_s$ & $\gamma_s$ & $\delta_s^{\infty}$ (\%) & $R^2$ & MAE (\%) \\
\midrule
\gls{MRV} & \FitPsiMRV/ & \FitGammaMRV/ & \FitDinfMRV/ & \FitRsqMRV/ & \FitMaeMRV/ \\
\gls{MVR} & \FitPsiMVR/ & \FitGammaMVR/ & \FitDinfMVR/ & \FitRsqMVR/ & \FitMaeMVR/ \\
\gls{MSR} & \FitPsiMSR/ & \FitGammaMSR/ & \FitDinfMSR/ & \FitRsqMSR/ & \FitMaeMSR/ \\
\bottomrule
\end{tabular*}

\end{table*}

\textbf{Comparison with naive allocation benchmarks.}
To put these distance results into perspective, we compare the optimisation strategies against two \gls{naive-allocation} benchmarks that require no estimation of returns or covariances: an \emph{equal-weight} portfolio ($w_i = 1/N$) and a \emph{market-capitalisation-weight} portfolio.
The mean distance to equal-weight is \LOneNaiveEqualMean/\% and to market-capitalisation-weight \LOneNaiveMcapMean/\%, both substantially larger than the \LOneMeanSaferRisk/\% distance to \gls{MVR}.
While both means are similar, the medians reveal a distributional difference: for equal-weight the median (\LOneNaiveEqualMedian/\%) exceeds the mean (\LOneNaiveEqualMean/\%), whereas for market-capitalisation-weighted (\LOneNaiveMcapMedian/\% median, \LOneNaiveMcapMean/\% mean) the reverse holds.
The typical account is thus noticeably closer to market-capitalisation-weight than to equal-weight portfolios, which is intuitive: concentrated holdings in one or two large-cap tokens mechanically resemble a market-cap-weighted portfolio, even without deliberately targeting the weighting.

This benchmark comparison examines the structural effect of \gls{MPT} optimisation, revealing that the three strategies shift accounts in qualitatively distinct ways.
\gls{MRV} and \gls{MVR} both produce a negative mean distance-delta against equal-weight, meaning optimisation on average moves accounts \emph{closer} to an equal-weight allocation;
\gls{MVR} additionally leaves \LOneDeltaSrEqualUnchangedPct/\% of accounts unchanged, consistent with many portfolios already lying near the constrained minimum-variance solution.
\gls{MSR}, by contrast, pushes the majority of accounts farther from both equal-weight and cap-weight (full decomposition in Appendix~\ref{sec:appendix_distance}).

\subsection*{Takeaways}

Across the three strategies, portfolios are closest to the minimum-variance optimum (\gls{MVR}, mean $\LOneMeanSaferRisk/\%$) and farthest from the Sharpe-optimal frontier (\gls{MSR}, $\LOneMeanMaxSharpe/\%$).
The large near-\gls{MVR} share (distance $\leq 1\%$: \LOneSrCombinedLeOnePct/\%) is structural: two-asset portfolios mechanically approximate the minimum-variance solution, and the share collapses for larger portfolios ($N \geq 5$).
Portfolio size $N$ is the dominant determinant of the mean distance to the constrained efficient frontier across all three strategies.
Larger portfolios exhibit substantial distance to all three frontiers, indicating considerable room for rebalancing.
Relative to naive benchmarks, actual portfolios are nearer to market-cap-weighting than to equal-weighting, reflecting concentrated holdings in large-cap tokens.
\gls{MRV} and \gls{MVR} optimisation moves accounts \emph{closer} to equal-weight, whereas \gls{MSR} pushes them farther from both naive benchmarks.
\emph{Overall, on-chain portfolios sit closest to the minimum-variance frontier,
but this proximity is largely explained by the dominant two-asset portfolio size.}

\section{Realised Portfolio Performance}
\label{sec:returns_and_capm}

Finally, we address \textbf{Obj C} by quantifying how deviations from the constrained efficient frontier translate into realised performance relative to the market. For each account snapshot, we compute the $T^{(+)}$-day realised return under the actual, optimised, and naive allocations.

\subsection{Forward Returns and Risk-Adjusted Returns}
\label{sec:forward_returns}

We set $T^{(+)} = \CAPMHorizonDays/$ days, keeping adjacent forward returns approximately non-overlapping~\cite{hansen1980forward} and yielding a 3:1 $T^{(-)}/T^{(+)}$ ratio standard in rolling mean--variance evaluation~\cite{demiguel2009optimal, kan2007optimal}. The actual allocation $w_{(0)}$ serves as baseline against the optimised \gls{MPT} strategies (\gls{MRV}, \gls{MVR}, \gls{MSR}) and naive allocations (equal-weight, market-cap-weight).
\cref{tab:ret_alpha_summary} reports, for the forward-realised return $r^{(a,t)}_{t'}$, the median and the hit rate, which is the fraction of accounts beating the baseline.
Alongside, the risk-adjusted return~$\alpha$ is documented with the median and the fraction of accounts with positive results for each strategy.
Throughout, we take the median within each monthly snapshot and weight snapshots equally, so that cohort-size growth does not dominate the statistics.

\begin{table}[htbp]
\centering
\small
\caption{Realised returns and risk-adjusted performance by strategy. Median forward return over the $T^{(+)}$-day horizon, hit rate versus the baseline allocation, and \gls{CAPM} $\alpha$.}
\label{tab:ret_alpha_summary}
\begin{tabular*}{\textwidth}{l @{\extracolsep{\fill}} rrrr}
\toprule
Strategy & Median Return (\%) & Hit Rate (\%) & Median $\alpha$ (\%) & Positive $\alpha$ (\%) \\
\midrule
Baseline (actual account)      & \CAPMRetBaselineMedian/    & ---                    & \CAPMAlphaBaselineMedian/    & \CAPMFracPositiveBaseline/ \\
\gls{MRV}                     & \CAPMRetBetterMedian/      & \CAPMRetHitBetter/     & \CAPMAlphaBetterMedian/      & \CAPMFracPositiveBetter/ \\
\gls{MVR}                     & \CAPMRetSaferMedian/       & \CAPMRetHitSafer/      & \CAPMAlphaSaferMedian/       & \CAPMFracPositiveSafer/ \\
\gls{MSR}                     & \CAPMRetMaxSharpeMedian/   & \CAPMRetHitMaxSharpe/  & \CAPMAlphaMaxSharpeMedian/   & \CAPMFracPositiveMaxSharpe/ \\
Equal-weight ($1/N$)          & \CAPMRetEqualMedian/       & \CAPMRetHitEqual/      & \CAPMAlphaEqualMedian/       & \CAPMFracPositiveEqual/ \\
Market-cap-weight             & \CAPMRetMcapMedian/        & \CAPMRetHitMcap/       & \CAPMAlphaMcapMedian/        & \CAPMFracPositiveMcap/ \\
\midrule
Market benchmark              & $+$\CAPMRetMarketMedian/ & ---                    & ---                          & --- \\
\bottomrule
\end{tabular*}

\end{table}

While the market benchmark $r_m$ reaches \CAPMRetMarketMedian/, the baseline realised-return is \CAPMRetBaselineMedian/\% for the actual allocation.
Market-cap-weighting achieves the highest hit rate (\CAPMRetHitMcap/\%) and the least negative median return (\CAPMRetMcapMedian/\%).
\gls{MVR} and equal-weight also exceed a 50\% hit rate (\CAPMRetHitSafer/\% and \CAPMRetHitEqual/\%, respectively), with \gls{MSR} and \gls{MRV} only marginally below.

The risk-adjusted return is negative for every allocation.
Market-cap-weighting has the least negative median~$\alpha$ (\CAPMAlphaMcapMedian/\%) and the highest positive-$\alpha$ fraction (\CAPMFracPositiveMcap/\%), consistent with the \gls{CAPM} prediction that a cap-weighted allocation is hardest to beat. Among the \gls{MPT} strategies, only \gls{MVR} is marginally less negative ($\alpha = \CAPMAlphaSaferMedian/\%$ vs.\ baseline \CAPMAlphaBaselineMedian/\%).

Aggregating per snapshot rather than averaging across all accounts in the observation period gives the same picture: across the six strategies, the median across snapshots of the share of accounts beating the wETH--WBTC market index varies only from \PctBeatRetEqualBlockMedian/\% to \PctBeatRetBaselineBlockMedian/\%, and the share with positive~$\alpha$ from \PctPosAlphaBetterBlockMedian/\% to \PctPosAlphaBaselineBlockMedian/\%.

\subsection{Predictors of Performance}
\label{sec:l1_vs_performance}

To identify which account characteristics predict performance, we train Random Forest models on six features: entry month, portfolio value, portfolio size $N$, portfolio $\beta$, and binary indicators for holding ETH or BTC wrappers.

\begin{figure}[htbp]
  \centering
  \includegraphics[width=1\linewidth]{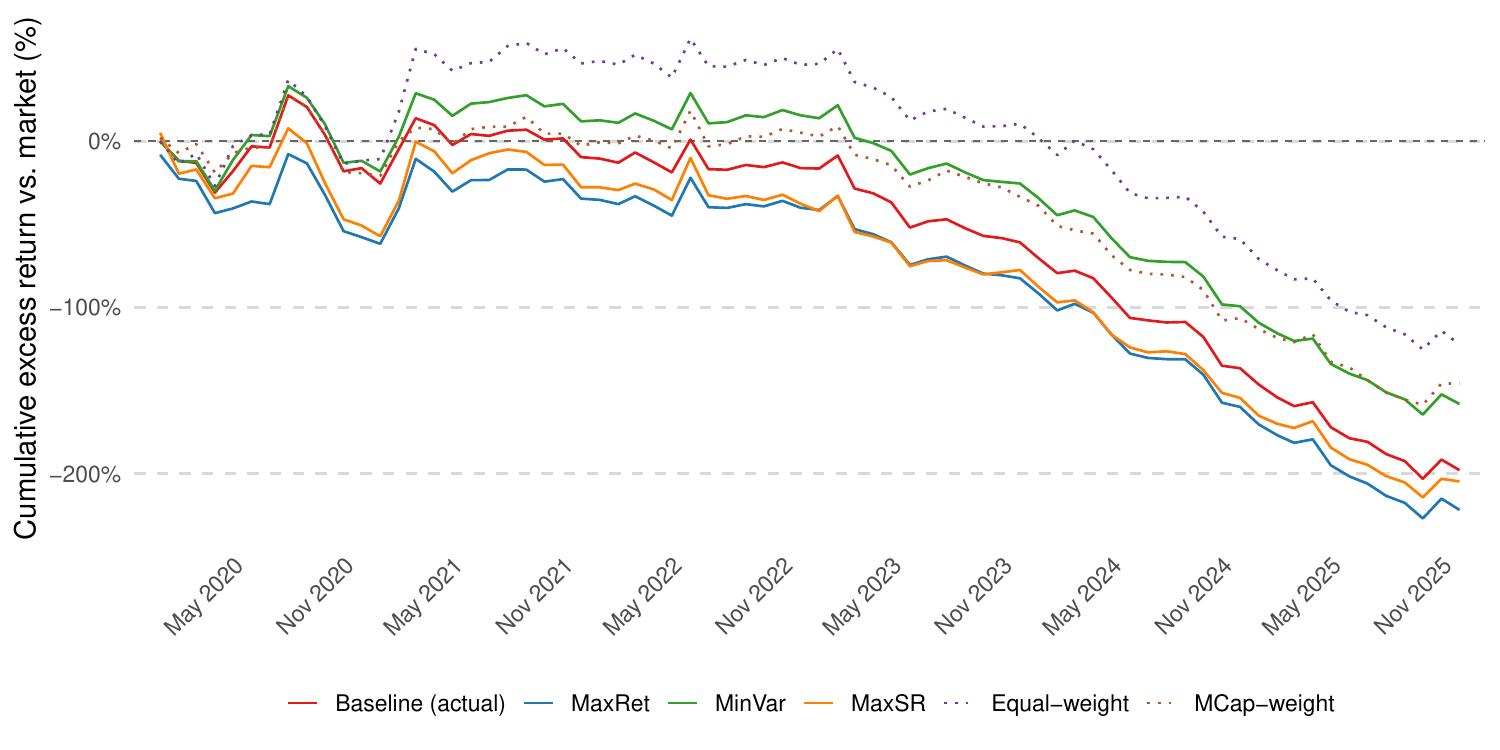}
  \caption{Cumulative excess return over the market benchmark. Running sum of monthly median return differentials by strategy, with allocations re-optimised independently at each snapshot.}
  \label{fig:cumulative_returns}
\end{figure}

Market entry month dominates both forward and risk-adjusted returns, accounting for \RfExtRetMonthImpLo/--\RfExtRetMonthImpHi/\% of the explained variance for realised returns ($R^{2} = \RfExtRetRsqLo/$--$\RfExtRetRsqHi/$) and \RfExtAlphaMonthImpLo/--\RfExtAlphaMonthImpHi/\% for $\alpha$. \Cref{fig:cumulative_returns} confirms this pattern under repeated monthly optimisation: the cumulative excess return over the market benchmark trends progressively more negative for every strategy across the sample period. Portfolio $\beta$ is the second strongest signal, modest for realised returns (\RfExtRetBetaImpLo/--\RfExtRetBetaImpHi/\%) and more pronounced for $\alpha$ (\RfExtAlphaBetaImpLo/--\RfExtAlphaBetaImpHi/\%), while portfolio value, size $N$, and asset composition contribute marginally.

We furthermore provide Random Forest, SHAP, $\ell_1$-distance–$\alpha$, and per-snapshot robustness checks in Appendices~\ref{sec:appendix-pct-beat-market}, \ref{sec:appendix_distance_alpha}, and~\ref{sec:appendix_rf_shap}.

\subsection*{Takeaways}

The market benchmark outperforms all optimisation strategies, including \gls{MPT}, in median return.
The dominant predictor of performance is market entry timing: the age of an account explains \RfExtRetMonthImpLo/--\RfExtRetMonthImpHi/\% of return variance and \RfExtAlphaMonthImpLo/--\RfExtAlphaMonthImpHi/\% of $\alpha$ variance, far exceeding the contribution of allocation choice, portfolio size, or asset composition.
\emph{Overall, when an account enters the market matters far more than how it allocates.}

\section{Discussion}
\label{sec:discussion}

\subsection{Key Findings}

We present the first large-scale empirical study comparing actual on-chain portfolios to their theoretically optimal allocations under \gls{MPT}. We propose a portfolio reconstruction method, apply it to Ethereum, and compute portfolios for \TotalAccountsRounded/ accounts across \CoingeckoCleanedTokensN/ ERC-20 tokens.

First, we analyse on-chain portfolios and find that wealth is highly skewed: the top $1\%$ of accounts control \TopOneWealthTwentyfive/\% of token wealth in 2025, and most accounts (\PortSizeOneTokenPct/\%) hold a single asset. These results are consistent with prior work on narrow sets of assets (e.g., governance tokens~\cite{barbereau2022defi, nadler2022decentralized, ovezik2025sok}) and extend that evidence to a substantially larger and more heterogeneous set of tokens. High concentration thus persists across token categories. Relative to traditional finance, single-asset dominance on Ethereum is markedly more pronounced than the median US retail holding of three to four stocks~\cite{Kumar2008}. We also find little evidence of the naive $1/N$ heuristic~\cite{benartzi2001naive}, under which allocations are divided equally across available options. On-chain allocation behaviour therefore departs from a pattern that is common in traditional retirement and investment settings.

Second, we measure how far portfolios sit from the constrained efficient frontier. Small portfolios lie closest to the minimum-variance optimum, but this proximity is structural rather than deliberate: portfolios near the optimum are predominantly two-token portfolios that mechanically approximate the minimum-variance allocation. Apparent optimality erodes as portfolios grow; for five or more tokens, mean deviations from the frontier reach \LOneLargeMVRMean/\%--\LOneLargeMSRMean/\%, depending on the optimisation strategy. This extends prior evidence that low-dimensional allocations leave little room for optimisation to add value~\cite{demiguel2009optimal}. We further formalise the relationship between portfolio size and mean distance to the theoretical optima through a parsimonious three-parameter power-decay model.

Third, we quantify how deviations from the efficient frontier translate into realised performance relative to the market. Passive market-capitalisation weighting outperforms portfolios constructed under any of the three investigated \gls{MPT} optimisation strategies. The dominant predictor of differences in realised returns is \emph{when} an account enters the market, not \emph{how} it allocates: entry month alone explains \RfExtRetMonthImpLo/--\RfExtRetMonthImpHi/\% of the variance in returns and \RfExtAlphaMonthImpLo/--\RfExtAlphaMonthImpHi/\% of the variance in risk-adjusted returns, far exceeding the contribution of portfolio size or allocation choice. Moreover, even under repeated monthly optimisation, all strategies accumulate increasingly negative excess returns relative to the market benchmark; this underperformance is therefore not confined to a single market regime. These results stand in contrast to established findings from traditional finance, which attribute roughly $90\%$ of the time-series variation in portfolio returns to asset allocation policy~\cite{brinson1986determinants, ibbotson2000does}. A plausible explanation is offered by Liu et al.~\cite{liu2022common}, who show that a small number of common factors (in particular a cryptocurrency market factor) capture the variation in returns across cryptoassets, leaving limited scope for diversification benefits across tokens.

Taken together, these findings challenge the descriptive adequacy of \gls{MPT} in the cryptoasset domain. Decentralised markets are natural empirical testbeds for \gls{MPT}; actual allocations, however, bear little resemblance to mean-variance optimal portfolios. Pervasive under-diversification suggests that risk-return optimisation is not a salient feature at the wallet level. The dominance of entry timing over allocation choice further indicates that returns are governed primarily by a common market factor~\cite{liu2022common}, a regime in which the diversification benefits \gls{MPT} promises are structurally limited. Mean-variance optimisation thus appears neither descriptive of observed behaviour nor prescriptively useful in this setting, even if \gls{MPT} remains a normative benchmark.

Our findings have practical implications.
Our empirical analysis suggests that rebalancing toward \gls{MPT} optima does not reliably improve risk-adjusted performance, while passive market-cap weighting is observed to offer a more robust default. These findings could be of use to whoever wishes to import the mean-variance framework to the blockchain setting (e.g., \emph{retail investors or wealth managers}): doing so wholesale risks projecting an illusion of optimisation onto a return regime in which a dominant market factor leaves little for diversification gains.
For \emph{platform and protocol designers}, the persistence of single-asset portfolios among smaller accounts suggests that frictions (e.g., gas fees, limited attention) shape observed behaviour more than the celebrated Modern Portfolio Theory (MPT); interventions targeting transaction costs or portfolio-construction interfaces may therefore have greater practical impact than tools encouraging frontier-based optimisation. For \emph{regulators and supervisors}, the extreme wealth concentration we document raises concerns about systemic exposure that allocation-focused frameworks do not address.

\subsection{Limitations and Future Work}
\label{sec:limitations}

Our analysis is subject to several limitations,
each of which points to directions for future work. First, we cover only ERC-20 token portfolios on Ethereum, restricted to tokens listed on Coingecko. This subset excludes scams, abandoned projects, and illiquid assets, but is biased toward established tokens. Native ETH enters the dataset indirectly via its wrapped representation (WETH)~\cite{ethereum2026what}, the form of ETH usable in DeFi protocols. Assets held on custodial exchanges are not observable on-chain and are therefore excluded. This scope is analogous to prior work on traditional portfolios, which likewise focuses on specific markets and instruments (e.g.,~\cite{Kumar2008,huberman2001familiarity}). Extending the reconstruction to other EVM-compatible blockchains and to native-token flows can broaden coverage along this axis.

Second, we treat each Ethereum address as an independent portfolio. Linking addresses to common entities remains an open problem in blockchain analytics~\cite{yaish2026inequality,chemaya2025quantifying,liu2025sybil}; off-chain heuristics such as address clustering cannot guarantee full accuracy~\cite{meiklejohn2013fistful,victor2020address}. Entity-level portfolios may therefore differ in size and composition from the address-level portfolios we observe, and yield a complementary view at the actor level.

Third, our study evaluates portfolios against \gls{MPT} benchmarks, but does not analyse the underlying decision-making process (e.g., trading motives, herding, disposition effects). While most accounts underperform the market benchmark, characterising the minority of outperformers could reveal whether systematic strategies or informational advantages exist on-chain. The publicly available reconstruction pipeline and dataset (upon publication) provide a foundation for such analyses.

\section{Conclusion}
\label{sec:conclusion}

We reconstructed token portfolios for \TotalAccountsRounded/ Ethereum accounts across \CoingeckoCleanedTokensN/ ERC-20 tokens over the full chain history (2015--2025) and evaluated them against \gls{MPT} benchmarks.
Both the actual on-chain portfolios and their \gls{MPT}-optimal counterparts underperform the market benchmark in realised returns.
On-chain account portfolios are pervasively under-diversified, sitting close to the constrained efficient frontier by construction.
Realised returns are more closely tied to \emph{when} accounts enter the market rather than \emph{how} they allocate.
\gls{MPT} thus appears neither descriptive nor prescriptively useful in this setting, even as it retains its value as a normative benchmark.
The public reconstruction pipeline and dataset (upon publication) enable entity-level analyses, cross-chain extension, and outperformer characterisation.

\bibliography{main}

\ifdefined\IsShortPaper
\section*{Supplemental Material}\label{sec:supplement}
The full appendices A, B, C, D, and E (including glossary, robustness analyses, and additional figures and tables) are available in the extended version accompanying this submission and at \url{\datalink}.

\makeatletter
\def\@currentlabel{Supplement}
\makeatother
\label{app:risk-free-rate}
\label{sec:account_wealth}
\label{sec:appendix-pct-beat-market}
\label{sec:appendix_characteristics}
\label{sec:appendix_distance}
\label{sec:appendix_distance_alpha}
\label{sec:appendix_rf_shap}
\label{sec:computational_details}
\label{sec:l1_distance}
\label{sec:mean_reversion_details}
\else
\appendix

\newpage

\section{Glossary}
\label{sec:glossary}
Following is a list of the terms, notations, and acronyms used in the paper.
\setglossarystyle{long}
\printnoidxglossary[type=main]
\printnoidxglossary[type={acronym}]

\section{Supplementary Material: Portfolio Characteristics}
\label{sec:appendix_characteristics}

This appendix provides supplementary analysis for the portfolio characteristics discussed in \cref{sec:descriptive_statistics}.
It contains additional detail on account evolution and growth trajectories (\cref{sec:account_wealth}),
the decomposition of top holders by account type (\cref{sec:appendix_top1_ca_eoa}),
address-level attribution of the wealthiest accounts (\cref{sec:appendix_tag_coverage}),
and standard inequality measures confirming the wealth concentration documented in the main text (\cref{sec:appendix_inequality}).

\subsection{Accounts and Wealth}
\label{sec:account_wealth}

This subsection details account growth trajectories, the distribution of wealth across bins and account types, and the concentration of value among top holders, supplementing the descriptive statistics in \cref{sec:descriptive_statistics}.

\subsubsection{Account evolution.}

We analyse the evolution of accounts that have acquired at least one asset of our token selection.
We further distinguish between \gls{EOA}s and \gls{CA}s in the line plot of \cref{fig:number_of_accounts}.

\begin{figure}[htbp]
  \centering
  \includegraphics[width=1\linewidth]{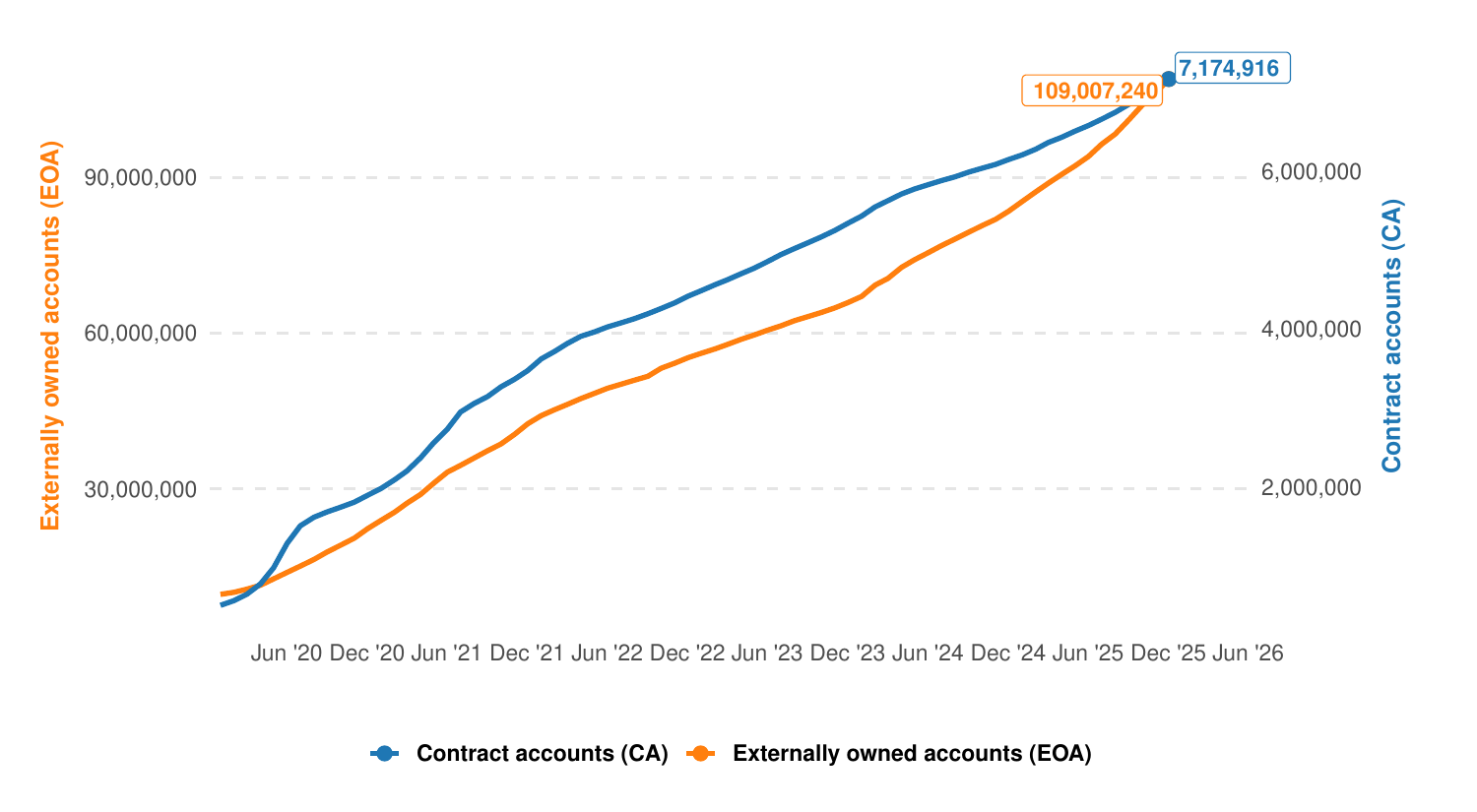}
  \caption{Number of accounts: the temporal evolution of accounts that have acquired at least one asset.}
  \label{fig:number_of_accounts}
\end{figure}

The two account types exhibit substantially different growth trajectories.
By the end of 2025, \gls{EOA}s reach \EOAAccountsEnd/ addresses against \CAAccountsEnd/ \glspl{CA},
so that contract accounts represent only \CAShareEndPct/ of all asset-holding accounts.
The \gls{EOA} population grows at a \gls{CAGR} of \EOACAGR/ over 2020--2025,
adding roughly \EOAAvgIncreaseYear/ addresses per year.
\Glspl{CA} grow at a comparable \gls{CAGR} of \CACAGR/,
corresponding to approximately \CAAvgIncreaseYear/ new contracts per year,
reflecting the steady deployment of \gls{DeFi} protocols and token infrastructure.

\subsubsection{Wealth distribution of accounts.}

We classify accounts into wealth bins
\[
(0,1],\, (1,100],\, (100,1\text{K}],\, \dots,\, (100\text{K}, \infty) \quad [\text{USD}],
\]
restricting the statistics to active addresses (non-zero wealth);
this binning scheme reflects the heavy-tailed nature of on-chain wealth
distributions.
\cref{tab:wealth_bucket_avg} reports the time-averaged share of accounts
in each bin separately for \glspl{CA} and \glspl{EOA}, while
\cref{fig:wealth_accounts_low,fig:wealth_accounts_high} show the temporal
evolution for the low- and high-wealth bins respectively.

\begin{table}[H]
\centering
\small
\caption{Wealth bin distribution: time-averaged share of accounts in each wealth bin across the observation window (non-zero USD value accounts only). Shares are percentages.}
\label{tab:wealth_bucket_avg}
\begin{tabular*}{\textwidth}{l @{\extracolsep{\fill}} rrrrrr}
\toprule
Type & 0--1 & 1--100 & 100--1K & 1K--10K & 10K--100K & $>$100K \\
\midrule
CA  & \WealthBucketCAZeroOneP/  & \WealthBucketCAOneHundredP/  & \WealthBucketCAHundredOneKP/  & \WealthBucketCAOneKTenKP/  & \WealthBucketCATenKHundredKP/  & \WealthBucketCAOverHundredKP/ \\
EOA & \WealthBucketEOAZeroOneP/ & \WealthBucketEOAOneHundredP/ & \WealthBucketEOAHundredOneKP/ & \WealthBucketEOAOneKTenKP/ & \WealthBucketEOATenKHundredKP/ & \WealthBucketEOAOverHundredKP/ \\
\bottomrule
\end{tabular*}
\end{table}

Across both account types, the majority of accounts consistently hold
wealth below \$100, but the within-type composition differs markedly:
for \glspl{CA} the lowest-wealth bin ($\leq \$1$) holds the largest
share ($\WealthBucketCAZeroOneP/\%$), whereas for \glspl{EOA} the
\$1--\$100 bin is the largest ($\WealthBucketEOAOneHundredP/\%$).
The share of low-wealth accounts decreases from 2020 to 2021 and rises
thereafter, while the share of high-wealth accounts peaks around
mid-2021, consistent with the market run-up and contraction described
above. The two account types reach their respective minima at different
times: the \gls{CA} minimum around mid-2021 (``DeFi Summer''), whereas
the \gls{EOA} minimum around late 2021 aligns with the broader market
peak, together producing the two wealth peaks visible in
\cref{fig:wealth_composition}.

\begin{figure}[htbp]
  \centering
  \includegraphics[width=1\linewidth]{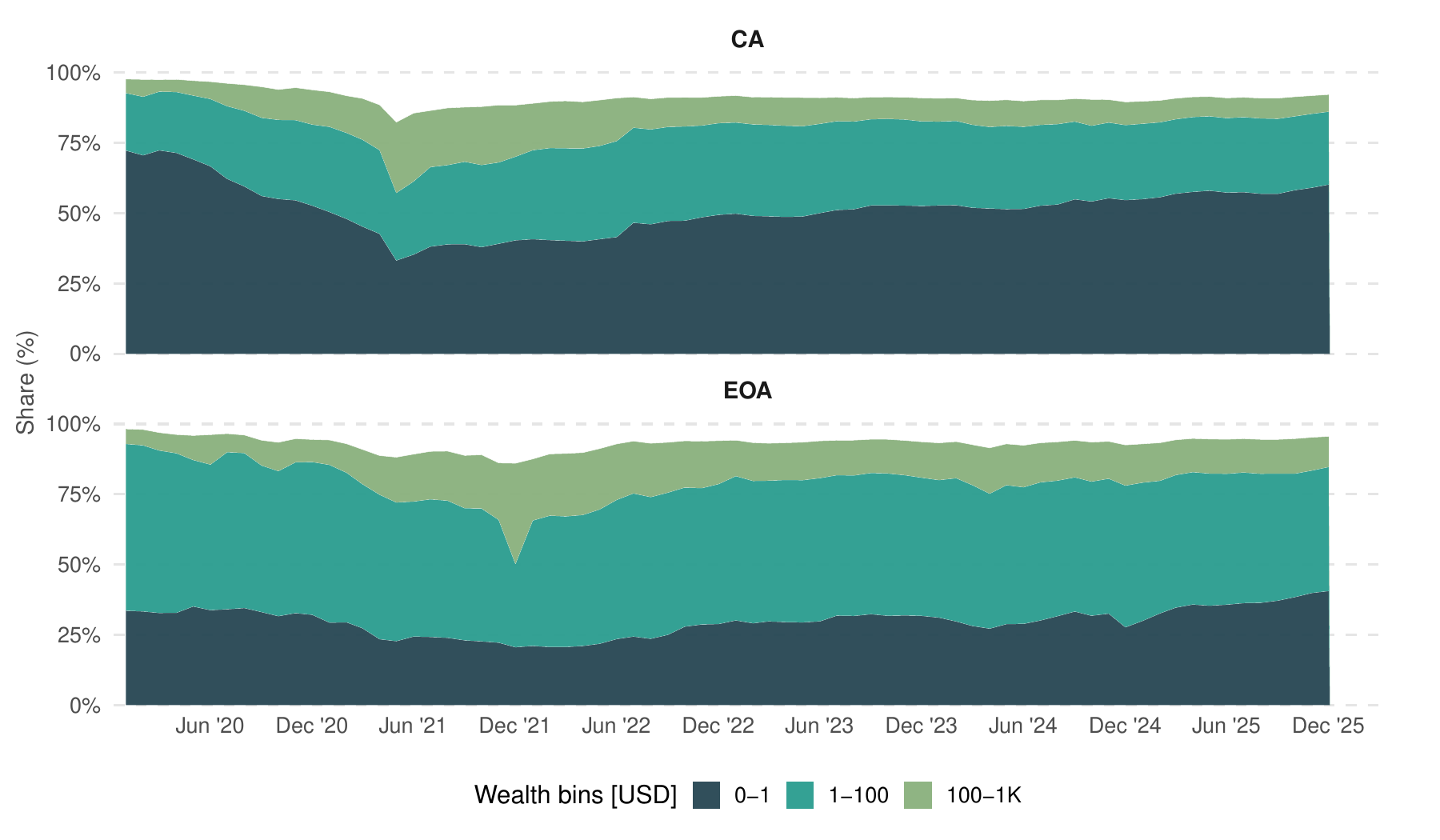}
  \caption{Distribution of accounts across \emph{low-wealth} bins over time (share of accounts with non-zero wealth). The top subplot shows \gls{CA}, the bottom subplot \gls{EOA}.}
  \label{fig:wealth_accounts_low}
\end{figure}

\begin{figure}[htbp]
  \centering
  \includegraphics[width=1\linewidth]{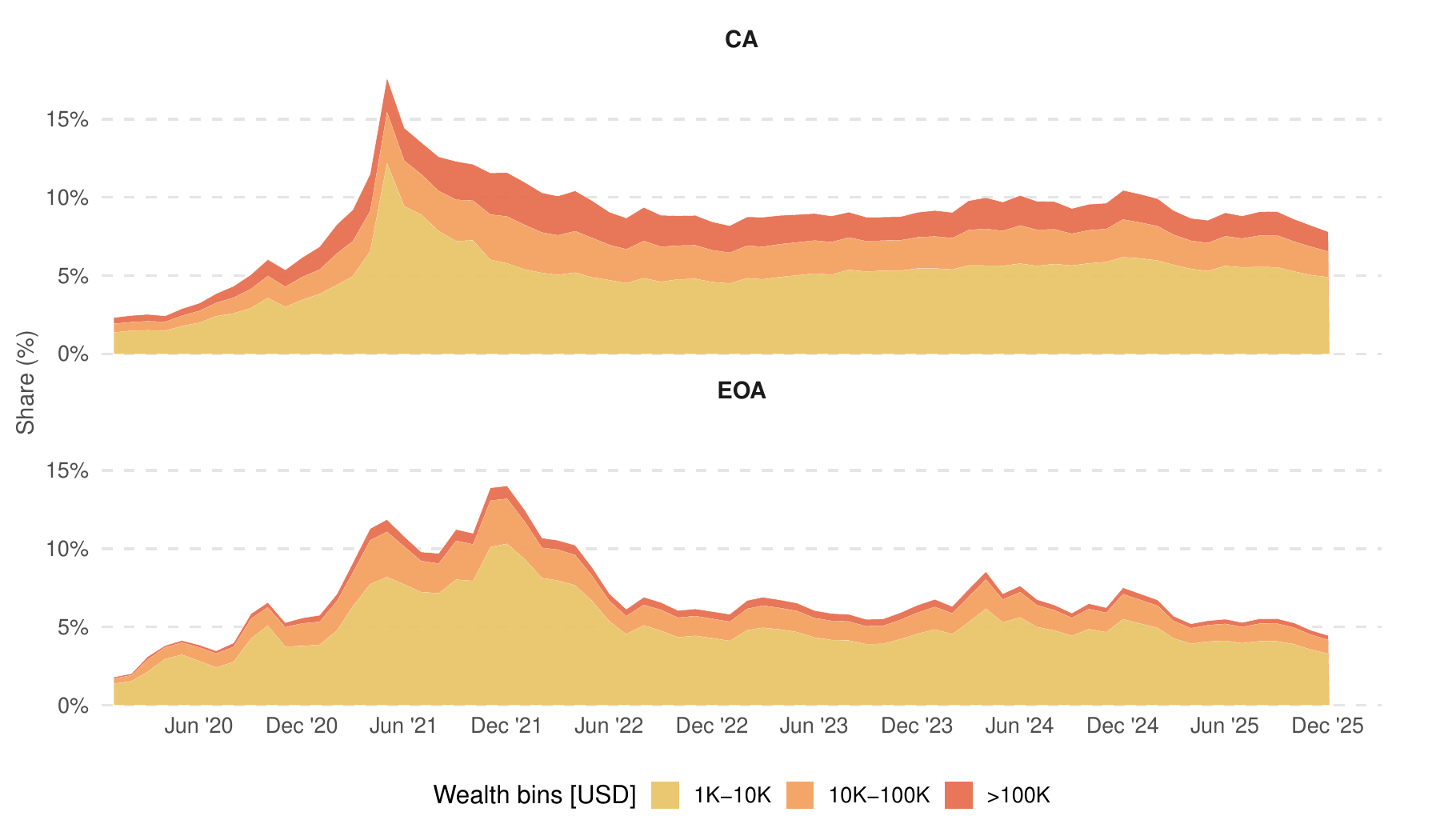}
  \caption{High-wealth bin distribution over time: share of accounts across high-wealth bins, with \gls{CA} (top) and \gls{EOA} (bottom). The peak in high-wealth \gls{CA} accounts around mid-2021 coincides with DeFi Summer, when protocols experienced rapid \gls{TVL} growth through liquidity mining.}
  \label{fig:wealth_accounts_high}
\end{figure}

\subsubsection{Wealth of top accounts.}

\cref{tab:topshare_yearly_avg_all} summarises the yearly concentration statistics for all accounts with non-zero wealth, extending the aggregate figures reported in \cref{sec:wealth_concentration}.

\begin{table}[H]
\centering
\small
\caption{Top-holder concentration by year: yearly averages of top-holder wealth shares for all accounts. Counts are in millions; shares in percent. The Top $1\%$ (min--max) column captures within-year variation across monthly snapshots.}
\label{tab:topshare_yearly_avg_all}
\begin{tabular*}{\textwidth}{l @{\extracolsep{\fill}} rrrrr}
\toprule
Year & \shortstack[r]{Avg accounts\\(M)} & \shortstack[r]{Top $1\%$\\(avg)} & \shortstack[r]{Top $1\%$\\(min--max)} & \shortstack[r]{Top $5\%$\\(avg)} & \shortstack[r]{Top $10\%$\\(avg)} \\
\midrule
2020 & 5.003  & 97.25 & 96.40--98.80 & 99.37 & 99.75 \\
2021 & 10.025 & 97.66 & 96.73--98.12 & 99.40 & 99.73 \\
2022 & 14.277 & 98.16 & 97.91--98.36 & 99.49 & 99.73 \\
2023 & 14.747 & 97.79 & 97.66--97.91 & 99.51 & 99.78 \\
2024 & 18.646 & 98.39 & 97.97--98.68 & 99.57 & 99.80 \\
2025 & 21.058 & 99.03 & 98.69--99.28 & 99.74 & 99.87 \\
\bottomrule
\end{tabular*}
\end{table}

\subsection{Account Type Breakdown of Top Holders}
\label{sec:appendix_top1_ca_eoa}

To understand whether the extreme wealth concentration documented in \cref{sec:wealth_concentration} is driven by protocol-controlled
accounts or individual holders, we decompose the top $1\%$ and top $0.1\%$ by account type.

Within the top $1\%$, contract accounts represent only $\TopOneCAPctWallets/\%$ of accounts by count but hold $\TopOneCAPctValue/\%$ of the top-$1\%$ wealth on average. \Glspl{EOA} account for the remaining $\TopOneEOAPctWallets/\%$ of accounts and $\TopOneEOAPctValue/\%$ of value. The \gls{CA} share is broadly consistent with the ecosystem-wide \gls{CA} wealth share of $\WealthShareCAMeanP/\%$ reported in \cref{sec:evolution_of_wealth}, indicating that the top-$1\%$ is not disproportionately composed of smart contracts relative to the overall population.

Narrowing the focus to the top $0.1\%$ sharpens the picture. \Glspl{CA} now constitute $\TopPointOneCAPctWallets/\%$ of addresses, nearly three times their share in the top $1\%$,
while their value share remains stable at $\TopPointOneCAPctValue/\%$. This indicates that the very largest holders include a substantially higher proportion of smart contracts, likely corresponding to \gls{DeFi} protocol treasuries, exchange hot wallets, and token vesting
contracts, while the majority of wealth even at this level is still held by individual \gls{EOA} addresses.

\subsection{Address Attribution of Top-0.1\% Holders}
\label{sec:appendix_tag_coverage}

To further characterise the entities behind the wealthiest addresses, we match the top $0.1\%$ of accounts at each monthly snapshot against address tags provided by iknaio~\cite{iknaio_tags}, queried via the GraphSense REST API~\cite{graphsense_rest}. The dataset assigns category labels and semantic concepts to known Ethereum addresses.

On average, $\TagCoveragePctMean/\%$ of top-$0.1\%$ addresses carry at least one tag (range: $\TagCoveragePctMin/$--$\TagCoveragePctMax/\%$ across snapshots), while the remaining $\TagUnlabeledPct/\%$ are unlabelled.
The limited coverage reflects the pseudonymous nature of Ethereum:
the majority of high-value addresses have no publicly attributed identity.
Among the $\approx 24\%$ with attribution, two complementary views emerge.

\textbf{By category.}
Of the addresses carrying a GraphSense \emph{category} label ($9.6\%$ of all
top-$0.1\%$ block--address pairs), the largest group is \emph{user} ($\TagCatUserPct/\%$ of categorised), followed by \emph{defi\_dex\_pair} ($\TagCatDexPairPct/\%$),
\emph{service} ($\TagCatServicePct/\%$), \emph{exchange} ($\TagCatExchangePct/\%$), and \emph{defi\_token} ($\TagCatDefiTokenPct/\%$). Together, exchange- and DeFi-related categories account for over $40\%$ of categorised addresses.

\textbf{By concept.}
The broader \emph{concept} taxonomy covers $21.4\%$ of top-$0.1\%$ pairs.
The dominant concept is \emph{defi} ($\TagConDefiPct/\%$ of concept-tagged
addresses), followed by \emph{defi\_custody} ($\TagConCustodyPct/\%$),
\emph{exchange} ($\TagConExchangePct/\%$), and \emph{defi\_dex}
($\TagConDexPct/\%$).
This confirms that the top of the wealth distribution is heavily populated by \gls{DeFi} protocol contracts (liquidity pools, custody vaults, token treasuries) and centralised exchange wallets, consistent with the disproportionate \gls{CA} value share reported in \cref{sec:appendix_top1_ca_eoa}.

\subsection{Wealth Inequality and Concentration}
\label{sec:appendix_inequality}

Given that the large majority of accounts hold less than \$100 on average,
we further investigate the wealth distribution to quantify the degree of concentration.
We use the Herfindahl-Hirschman Index (HHI) to measure the concentration of wealth among accounts
and \gls{gini}s to analyse the concentration and inequality of token portfolios.

Before turning to these indices, we briefly characterise the token-level holder base, which provides context for interpreting concentration measures.
Over time, the typical token holder base shrinks: the average number of holders per token decreases from \HoldersAvgYtwenty\ (median: \HoldersMedYtwenty) in 2020 to \HoldersAvgYtwentyfive\ (median: \HoldersMedYtwentyfive) in 2025.
This decline is driven by a compositional shift.
\cref{fig:holders_distribution_yearly} summarises the annual distribution of tokens grouped by their holder count.
The share of tokens with more than 10\,000 holders drops from \HoldersPctGtTenKYtwenty\ to \HoldersPctGtTenKYtwentyfive, while smaller tokens (100--1\,000 holders) grow from \HoldersPctHundredOneKYtwenty\ to \HoldersPctHundredOneKYtwentyfive.
Mid-sized tokens (1\,000--10\,000 holders) remain the dominant group
throughout, ranging from \HoldersPctOneKTenKYtwenty\ to \HoldersPctOneKTenKYtwentyfour.

\begin{figure}[htbp]
  \centering
  \includegraphics[width=1\linewidth]{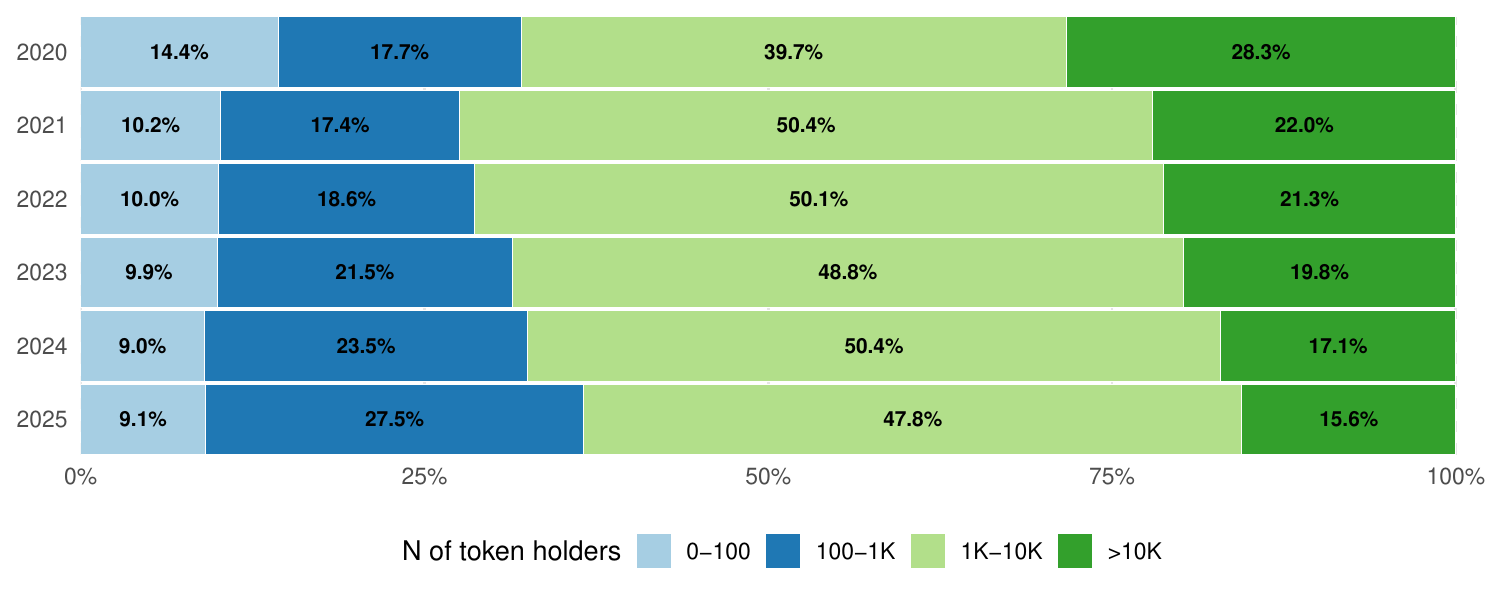}
  \caption{Token holder distribution: share of \gls{ERC-20} tokens grouped by number of holders per token (2020--2025).}
  \label{fig:holders_distribution_yearly}
\end{figure}

To complement the top-holder analysis in \cref{sec:wealth_concentration},
we compute two standard inequality measures for each token at every monthly snapshot and report their temporal evolution. We restrict the analysis to tokens with more than 100 holders at the time of the snapshot to focus on mature assets.

The \emph{Gini coefficient} ranges from $0$ (all accounts hold equal wealth) to $1$ (one account holds all wealth). The \emph{Herfindahl--Hirschman Index} (HHI) is calculated as
$\text{HHI} = \sum_{a=1}^{N} s_a^2$, where $s_a$ is the wealth share of account $a$ and $N$ is the total number of accounts; higher values indicate greater concentration.

\cref{fig:ineq_gini_hhi_appendix} displays the average and median
of both indices across all qualifying tokens over time. \Gls{gini} coefficients remain uniformly close to~$1$ throughout the observation period, with the cross-token average ranging from $0.954$ to $0.978$ and the median from $0.975$ to $0.990$.
HHI values indicate moderate to high concentration, with the average ranging from $0.221$ to $0.297$
and the median from $0.139$ to $0.220$. Both measures are stable over time, confirming that the extreme wealth concentration documented by the top-holder shares in the main text is a persistent structural feature of the \gls{ERC-20} token ecosystem.

\begin{figure}[htbp]
  \centering
  \includegraphics[width=1\linewidth]{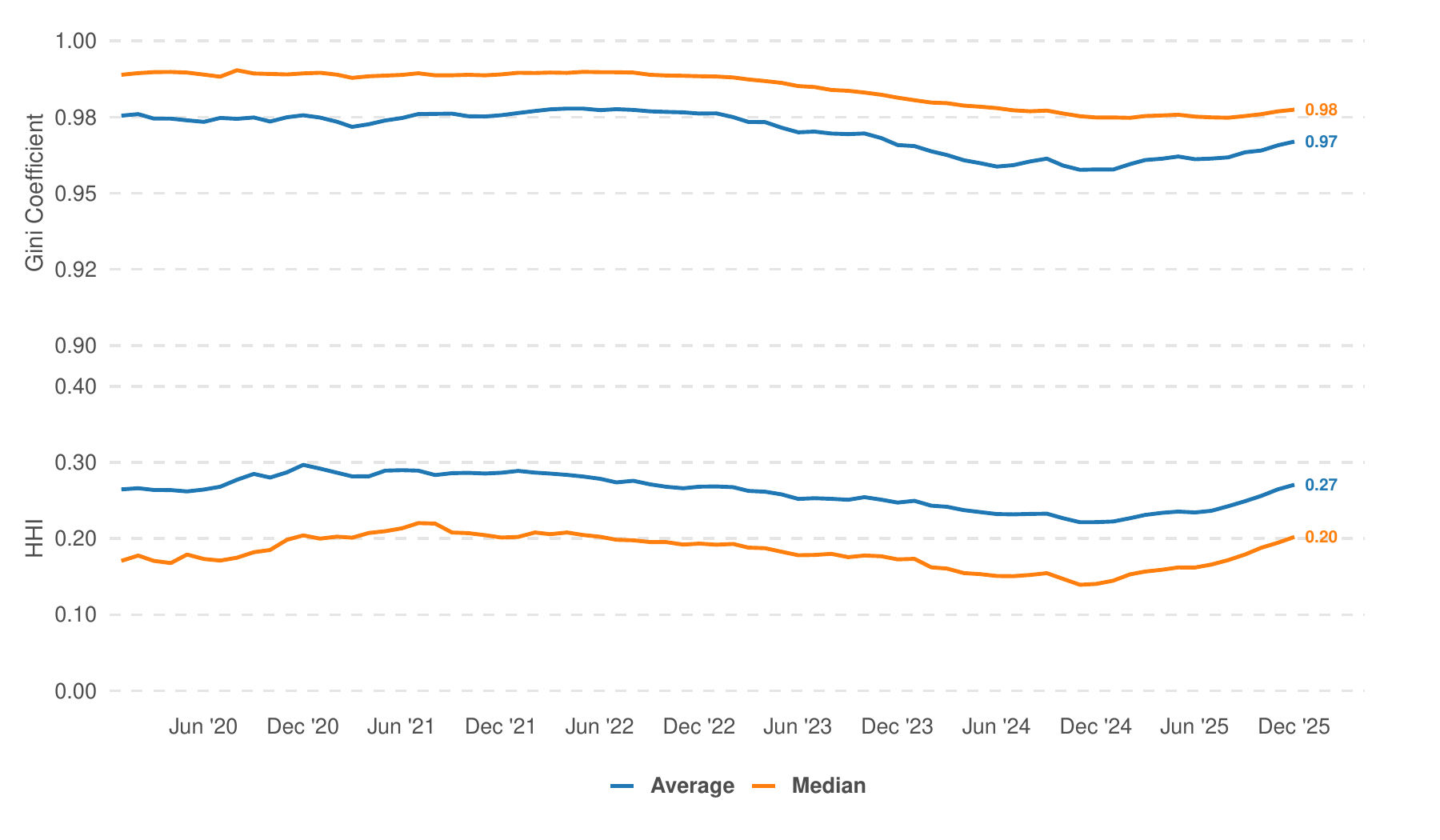}
  \caption{Inequality measures over time: average and median \gls{gini} coefficient (top) and \gls{HHI} (bottom) across all tokens with more than 100 holders, computed at each monthly snapshot (2020--2025).}
  \label{fig:ineq_gini_hhi_appendix}
\end{figure}

\section{Supplementary Material: Distance to the Efficient Frontier}
\label{sec:appendix_distance}

This appendix supplements the distance analysis in \cref{sec:mpt_optimal_portfolio}.
It covers the formal definition of the weight $\ell_1$ distance $d_s$ and the mean-reversion regularisation applied to return estimates (\cref{sec:l1_distance}),
the calibration and robustness of the risk-free rate $r_f$ (\cref{app:risk-free-rate}),
the determinants of $d_s$ and cross-strategy robustness checks (\cref{sec:l1_determinants_details}),
the relationship between $d_s$ and portfolio size with its fitted power-decay model (\cref{sec:distance_by_size}),
the comparison with naive allocation benchmarks (\cref{sec:naive_benchmark_details}),
and the optimisation volume and temporal evolution of $d_s$ (\cref{sec:mpt_optimisation_details}).

\subsection{Weight \texorpdfstring{$\ell_1$}{l1} Distance and Mean-Reversion Regularisation}
\label{sec:l1_distance}

This subsection provides the formal definition of the $\ell_1$ weight distance used throughout the main analysis (\cref{sec:mpt_optimal_portfolio}), together with the mean-reversion regularisation applied to return estimates in the optimisation pipeline.

\paragraph*{Formal definition.}
Let
\[
w_{(0)}=\bigl(w_{(0),1},\dots,w_{(0),N}\bigr),\qquad
w_{(s)}=\bigl(w_{(s),1},\dots,w_{(s),N}\bigr),
\]
with $\sum_{i=1}^N w_{(0),i} = \sum_{i=1}^N w_{(s),i} = 1$ and $w_i \ge 0$ (long-only).
We use $\ell_1$ because it has a direct \emph{rebalancing interpretation}: for fully-invested portfolios,
$\tfrac{1}{2}\lVert w_{(0)}-w_{(s)}\rVert_1$ equals the minimum one-way turnover (fraction of value traded) required to move from the current to the optimised weights.
We therefore define the distance with the $\tfrac{1}{2}$ factor included (see \cref{eq:l1_weight_distance}).

Using $|u-v| = u+v-2\min\{u,v\}$ for $u,v \ge 0$ and the fact that both portfolios are fully invested, we obtain
\begin{equation}
d_s
= \tfrac{1}{2}\left\lVert w_{(0)} - w_{(s)} \right\rVert_{1}
\;=\;
1 - \sum_{i=1}^{N} \min\!\left\{\, w_{(0),i},\, w_{(s),i} \,\right\}.
\label{eq:l1_weight_distance_min_form}
\end{equation}
For long-only portfolios, the distance is bounded by
\begin{equation}
0 \;\le\; d_s \;\le\; 1,
\label{eq:l1_weight_distance_bounds}
\end{equation}
where $0$ indicates identical allocations and values close to $1$ indicate nearly disjoint allocations (mass invested in different assets).

\paragraph*{Mean-reversion regularisation.}
\label{sec:mean_reversion_details}
Sample mean returns over the $T^{(-)}$-day lookback are noisy in cryptoasset markets, where extreme short-term moves are common and tend to be partially corrected over subsequent windows. We exploit this mean-reversion property as a regularisation mechanism: by shrinking each asset's sample mean $\mu_i$ toward a cross-sectional prior, we reduce sensitivity to transient price spikes and short-lived momentum effects.
Formally, for each asset $i$ we replace the raw sample mean return $\mu_i$ (\cref{eq:asset-mean}) by the shrunk estimate
\begin{equation}
\mu_i^{\mathrm{shrunk}}
=
\lambda\,\bar{\mu} \;+\; (1-\lambda)\,\mu_i,
\label{eq:mean_reversion}
\end{equation}
where $\bar{\mu} = \tfrac{1}{N}\sum_{j=1}^{N}\mu_j$ is the cross-asset mean and $\lambda \in [0,1]$ is the shrinkage strength. In our configuration we set $\lambda=0.5$, corresponding to a moderate pull toward the prior. The resulting portfolio expected return is then computed as usual by
\[
\mu^{(a,t)} \;=\; \sum_{i=1}^{N} w^{(a,t)}_i \mu_i^{\mathrm{shrunk}},
\]
which reduces overconfident forecasts while still preserving meaningful cross-sectional differences in recent performance. In practice, this produces smoother optimised weights and more stable allocations across snapshots.

\subsection{Calibration and Robustness of the Risk-Free Rate}
\label{app:risk-free-rate}

The \gls{MSR} (\cref{def:reference_portfolios}) is known as the \emph{tangency portfolio}, constructed by drawing a line from the risk-free rate $r_f$ tangent to the constrained efficient frontier. It is therefore the only place in our optimisation pipeline where $r_f$ enters. The \gls{MVR} and \gls{MRV} optima depend only on the covariance structure and the moments of the actual portfolio, and the \gls{CAPM} alphas are evaluated at $r_f = 0$ to keep the cross-sectional comparison invariant under a uniform shift in the riskless return. We adopt $r_f = \RfDefault/\%$ per annum throughout the \gls{MSR} optimisation on three independent grounds.

\paragraph*{Reference to the literature.}
The closest published precedent on portfolio optimisation with crypto assets~\cite{platanakis2019portfolio} proxies the risk-free rate with the U.S.\ 3-month Treasury bill yield. Their convention has carried over to subsequent crypto-portfolio work, and our setting matches it directly: in the 2023--2024 window, which contains the bulk of our account-time observations, the FRED series \texttt{TB3MS} averaged $\approx$\,\RfDefault/\%~\cite{fred_tb3ms}.

\paragraph*{Empirical match to on-chain stablecoin yields.}
Account holders in our sample interact natively with on-chain markets rather than U.S.\ Treasuries, so we cross-check the T-bill anchor against its closest on-chain analogue: the variable-rate supply APY for USDC and USDT on the major non-custodial money markets active on Ethereum during the sample window. We aggregate the supply APYs reported by DefiLlama's yields API across Aave v2 and v3, Compound v2 and v3, Morpho Blue, and Spark, restricting to pools deployed on Ethereum with TVL above one million USD and APY in $[0, 50]\%$ to exclude inactive or anomalous pools, and average the resulting series within each calendar month from January 2021 onwards. \Cref{tab:stablecoin-apy} reports the unweighted monthly statistics across all months in which lending data are available.

\begin{table}[htbp]
\centering
\small
\caption{On-chain stablecoin supply APY across the major non-custodial money markets on Ethereum (Aave v2/v3, Compound v2/v3, Morpho Blue, Spark), aggregated from DefiLlama's yields API. Unweighted monthly statistics, in \%.}
\label{tab:stablecoin-apy}
\begin{tabular*}{\textwidth}{l @{\extracolsep{\fill}} rrr}
\toprule
Asset & Avg APY & Median monthly APY & Months observed \\
\midrule
USDC                  & \StablecoinUSDCAvgAPY/   & \StablecoinUSDCMedianAPY/   & \StablecoinUSDCMonths/   \\
USDT                  & \StablecoinUSDTAvgAPY/   & \StablecoinUSDTMedianAPY/   & \StablecoinUSDTMonths/   \\
\midrule
USDC and USDT pooled  & \StablecoinPooledAvgAPY/ & \StablecoinPooledMedianAPY/ & \StablecoinPooledMonths/ \\
\bottomrule
\end{tabular*}
\end{table}

The pooled mean of \StablecoinPooledAvgAPY/\% essentially matches the T-bill anchor of \RfDefault/\%. We retain the T-bill convention of~\cite{platanakis2019portfolio} for comparability across the literature, but note that the on-chain alternative produces essentially the same number.

\paragraph*{Robustness to the value of $r_f$.}
\label{app:rf-robustness}
The asymmetry between $r_f = \RfDefault/\%$ in the \gls{MSR} optimisation and $r_f = 0$ in the \gls{CAPM} alpha evaluation is a deliberate modelling choice that we now show is empirically harmless on the \gls{MSR} side. We re-run the full optimisation pipeline with $r_f = 0$ on a stratified random sample of \RfRobustWalletsPerBlock/ accounts per monthly snapshot (\RfRobustTotalWallets/ accounts across all \RfRobustBlocks/ snapshots) and compare the resulting \gls{MSR} weights and distance metrics, account-by-account, against the production results obtained with $r_f = \RfDefault/\%$. \Cref{tab:rf-robustness} summarises the per-account differences.

\begin{table}[htbp]
\centering
\small
\caption{Robustness of the \gls{MSR} results to the value of $r_f$. Per-account quantities computed across \RfRobustTotalWallets/ matched accounts, normalised to the 0--100 scale used throughout the paper.}
\label{tab:rf-robustness}
\begin{tabular*}{\textwidth}{l @{\extracolsep{\fill}} rrr}
\toprule
Quantity & Median & Mean & 95th pct (abs) \\
\midrule
distance between \gls{MSR} weights ($r_f = 0$ vs.\ $r_f = \RfDefault/\%$) & \RfRobustWeightLOneMedianPct/  & \RfRobustWeightLOneMeanPct/  & \RfRobustWeightLOneTopFivePct/  \\
\addlinespace
$\Delta d_{\text{MSR}}$ ($w_{(0)}$ vs.\ \gls{MSR})                                 & \RfRobustGapDiffMedianPP/      & \RfRobustGapDiffMeanPP/      & \RfRobustGapDiffTopFivePP/      \\
$\Delta d$ (\gls{MSR} vs.\ equal-weight)                                           & \RfRobustVsEqualDiffMedianPP/  & \RfRobustVsEqualDiffMeanPP/  & \RfRobustVsEqualDiffTopFivePP/  \\
$\Delta d$ (\gls{MSR} vs.\ market-cap-weight)                                      & \RfRobustVsMcapDiffMedianPP/   & \RfRobustVsMcapDiffMeanPP/   & \RfRobustVsMcapDiffTopFivePP/   \\
\bottomrule
\end{tabular*}
\end{table}

The median account sees no change in any reported quantity. The mean changes in the three distance metrics are well under three percentage points in absolute value. At the 95th percentile only the \gls{MSR} weights themselves shift meaningfully ($\RfRobustWeightLOneTopFivePct/\%$), reflecting that the tangency portfolio slides along the same constrained efficient frontier when $r_f$ changes. The distance $d_{\text{MSR}}$ from the actual allocation $w_{(0)}$ to that frontier barely moves. Crucially, the change in the headline distance $d_{\text{MSR}}$ has a mean of \RfRobustGapDiffMeanPP/\,pp and a 95th-percentile absolute change of \RfRobustGapDiffTopFivePP/\,pp on a 0--100 scale.

\subsection{Determinants and Robustness of Distance}
\label{sec:l1_determinants_details}

We investigate which observable account characteristics determine distance using the feature set \texttt{num\_tokens}, \texttt{log\_value\_usd}, and \texttt{month} on a 10-million-row sample.

\paragraph*{Regression and feature importance.}
\textit{OLS.} Linear models explain little variance for \gls{MRV} ($R^2 = \LOneOlsBrRsq/$) and \gls{MSR} ($R^2 = \LOneOlsMsRsq/$), but substantially more for \gls{MVR} ($R^2 = \LOneOlsSrRsq/$). In all three, \texttt{num\_tokens} carries the largest coefficient ($p < 0.001$). An extended specification with polynomial and interaction terms raises $R^2$ to $\LOneOlsExtBrRsq/$, $\LOneOlsExtMsRsq/$, and $\LOneOlsExtSrRsq/$ respectively. The dominant additional predictor is $\log(\texttt{num\_tokens})$, confirming the logarithmic relationship between portfolio complexity and distance visible in \cref{fig:tokens_vs_distance}.

\textit{Random Forest.} A 300-tree model on a held-out set achieves $R^2 = \LOneRfBrRsq/$ (MAE $= \LOneRfBrMae/$ pp) for \gls{MRV}, $R^2 = \LOneRfSrRsq/$ (MAE $= \LOneRfSrMae/$ pp) for \gls{MVR}, and $R^2 = \LOneRfMsRsq/$ (MAE $= \LOneRfMsMae/$ pp) for \gls{MSR}. Feature importances, shown in \cref{fig:rf_importance}, diverge sharply. For \gls{MVR}, portfolio size dominates with $\LOneRfSrTokensImp/\%$ of total importance. For the other two, importance is spread more evenly across \texttt{log\_value\_usd} (${\sim}\LOneRfBrMsValueImp/\%$), \texttt{num\_tokens} ($\LOneRfBrMsTokensImpLo/$--$\LOneRfBrMsTokensImpHi/\%$), and \texttt{month} ($\LOneRfBrMsMonthImpLo/$--$\LOneRfBrMsMonthImpHi/\%$). This confirms the qualitative split visible in \cref{fig:tokens_vs_distance}: \gls{MVR} distance is almost entirely determined by portfolio size, whereas \gls{MRV} and \gls{MSR} distances additionally depend on wealth and market regime.

\begin{figure}[htbp]
    \centering
    \includegraphics[width=1\linewidth]{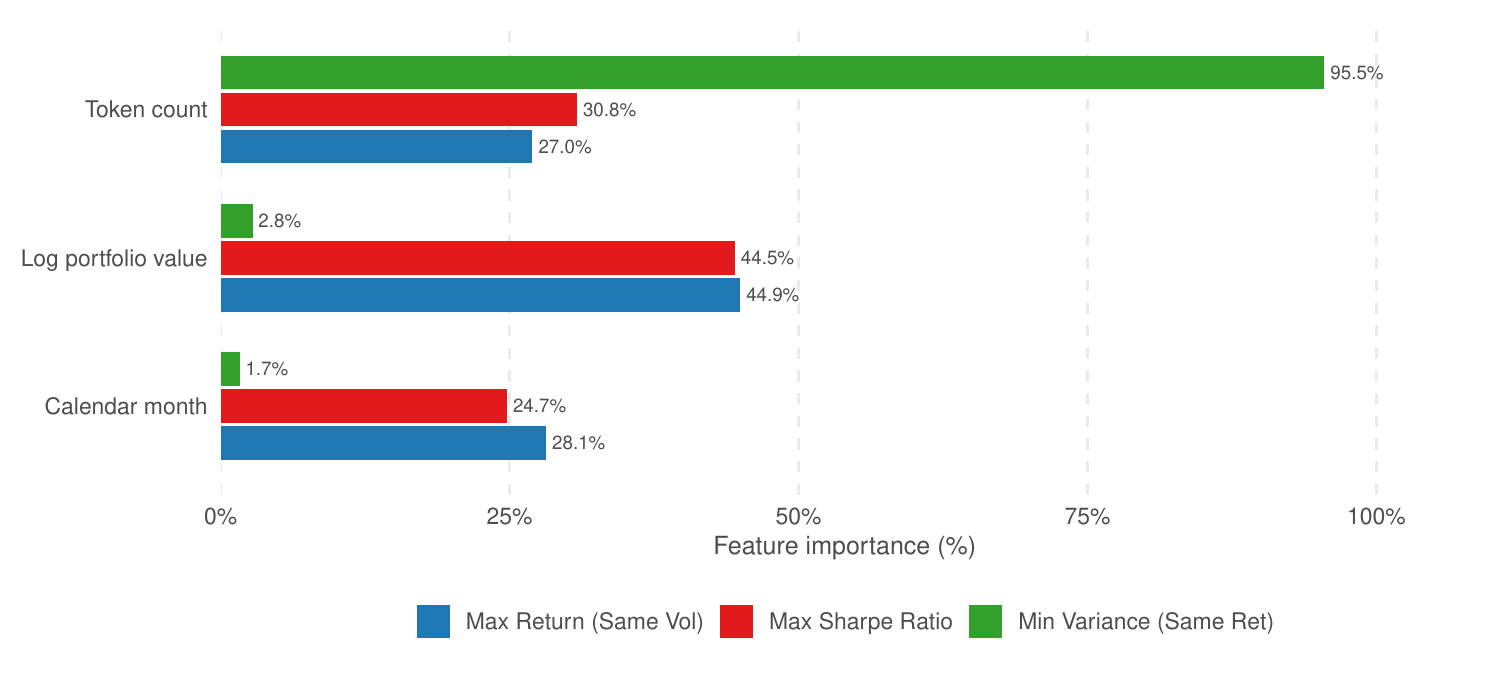}
    \caption{Random Forest feature importances: permutation-based importances for predicting distance under each optimisation strategy. Portfolio size accounts for $\LOneRfSrTokensImp/\%$ of importance under \gls{MVR} but only $\LOneRfBrMsTokensImpLo/$--$\LOneRfBrMsTokensImpHi/\%$ under the other two objectives.}
    \label{fig:rf_importance}
\end{figure}

\textit{Spearman correlations} on a 50M-row sample confirm the non-linearity: $\rho = \LOneSpearmanSrTokens/$ between \texttt{num\_tokens} and \gls{MVR} distance, far exceeding the Pearson $r$ of $\LOnePearsonSrTokens/$. For \gls{MRV} and \gls{MSR} the Spearman $\rho$ is $\LOneSpearmanBrTokens/$ and $\LOneSpearmanMsTokens/$, positive but less dominant.

\textit{Quantile regression} reveals that at $\tau = 0.25$, only \texttt{num\_tokens} is significant for \gls{MRV} and \gls{MVR} (low-distance accounts hold exactly two tokens regardless of value), while at $\tau = 0.75$ the \texttt{log\_value\_usd} coefficient also becomes significant, indicating that wealth matters primarily in the high-distance tail.

\paragraph*{Near-optimal accounts.}
\label{sec:near_optimal_profile}
We examine accounts whose distance falls within $1\%$ of the optimum. Across all three objectives, near-optimal accounts are overwhelmingly simple: $\LOneNearOptTwoTokenPctLo/$--$\LOneNearOptTwoTokenPctHi/\%$ hold exactly two tokens, and cumulative three-or-fewer exceeds $99\%$ in every case. Median portfolio value is modest (\$$\LOneNearOptMedianValueLo/$--\$$\LOneNearOptMedianValueHi/$). The key distinction is the \textit{size} of the near-optimal pool: $\LOneNearOptSrN/$M observations ($\LOneNearOptSrPct/\%$) are near-optimal under \gls{MVR}, $\LOneNearOptBrN/$M ($\LOneNearOptBrPct/\%$) under \gls{MRV}, and only $\LOneNearOptMsN/$M ($\LOneNearOptMsPct/\%$) under \gls{MSR}, confirming that minimum-variance optimality is structurally easy to approximate, while Sharpe optimality is difficult even for the simplest allocations.

\paragraph*{Per-strategy distance by portfolio size.}
\cref{fig:tokens_vs_distance_detail} disaggregates the relationship between portfolio size and distance for each optimisation strategy individually, showing the median (line), mean (dots), interquartile range (shaded band), and $\pm 1$ standard-deviation envelope.
The standard-deviation bands reveal substantial cross-sectional heterogeneity at every portfolio size, particularly for \gls{MSR}, where the upper band exceeds $80\%$ even for portfolios with $N > 10$.
For \gls{MVR}, the standard deviation narrows rapidly with portfolio size, confirming that larger portfolios converge tightly toward the minimum-variance optimum.

\begin{figure}[htbp]
    \centering
    \includegraphics[width=1\linewidth]{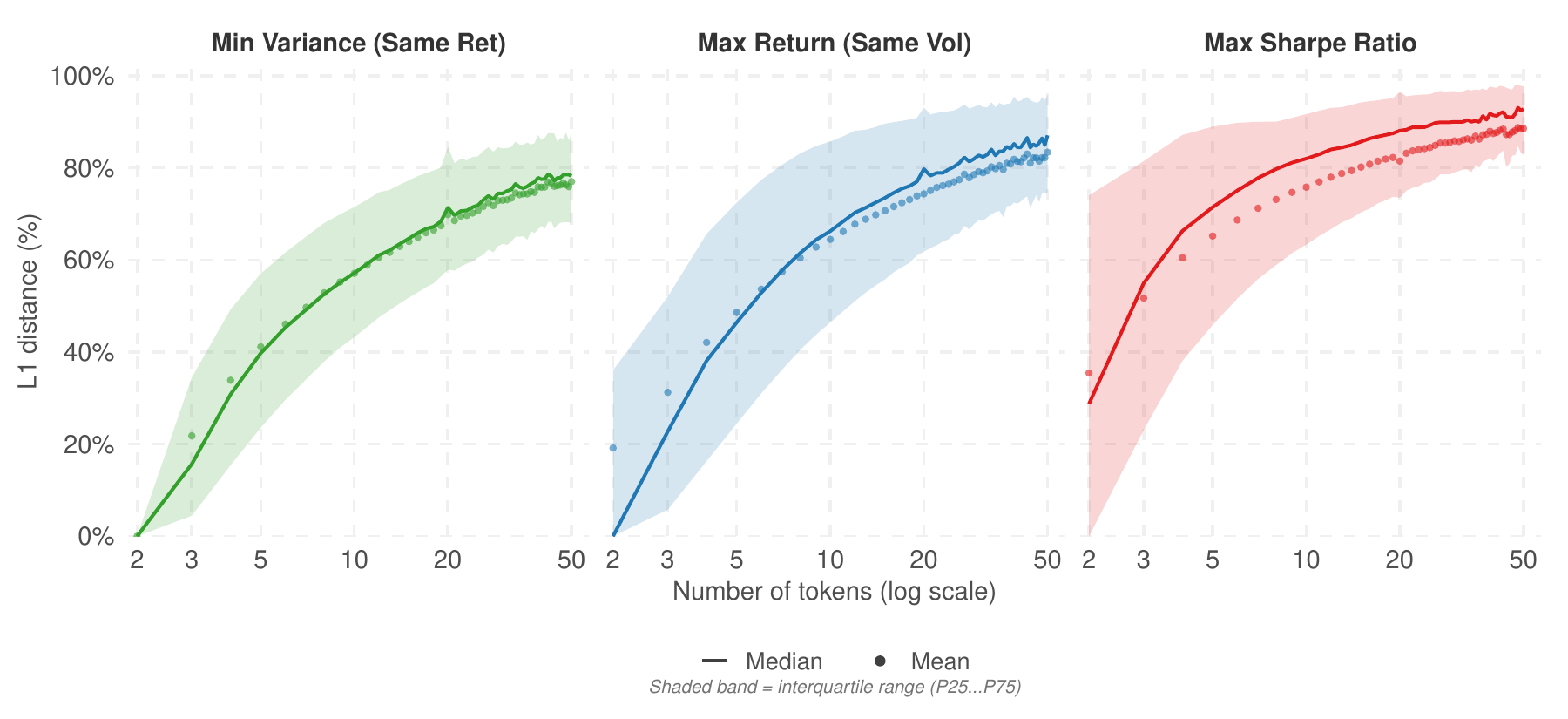}
    \caption{Per-strategy distance by portfolio size: median distance (line), mean (dots), interquartile range (shaded band), and $\pm 1$ standard-deviation envelope as a function of portfolio size $N$ (log-scaled axis). Based on a 5M-row sample.}
    \label{fig:tokens_vs_distance_detail}
\end{figure}

\paragraph*{Cross-strategy correlation.}
\label{sec:robustness_and_correlation}
The three distance measures are only moderately correlated with one another: $r = \LOneCorrBrSr/$ between \gls{MRV} and \gls{MVR}, $r = \LOneCorrBrMs/$ between \gls{MRV} and \gls{MSR}, and $r = \LOneCorrSrMs/$ between \gls{MVR} and \gls{MSR}. Being optimally allocated for one objective thus provides limited information about optimality for another, so portfolio ``quality'' in \gls{DeFi} is not a unidimensional concept.

Joint optimality is exceedingly rare. Only \LOneJointAllThreeN/ account-observations ($\LOneJointAllThreePct/\%$) achieve exact zero distance under all three objectives simultaneously. Pairwise exact-zero overlap is also thin: $\LOneJointBrSrPct/\%$ for \gls{MRV} $\cap$ \gls{MVR}, $\LOneJointBrMsPct/\%$ for \gls{MRV} $\cap$ \gls{MSR}, and $\LOneJointSrMsPct/\%$ for \gls{MVR} $\cap$ \gls{MSR}.

\paragraph*{Robustness to portfolio size.}
The preceding results are heavily shaped by the two-asset majority. To test whether the findings generalise to genuinely diversified portfolios, we repeat the full analysis on the subset of account-observations with $N \geq 5$. The near-optimal population essentially vanishes: the share within $1\%$ of the optimum drops from $\LOneSrCombinedLeOnePct/\%$ to $\LOneFivePlusSrLeOnePct/\%$ for \gls{MVR}, from $\LOneBrCombinedLeOnePct/\%$ to $\LOneFivePlusBrLeOnePct/\%$ for \gls{MRV}, and from $\LOneMsCombinedLeOnePct/\%$ to $\LOneFivePlusMsLeOnePct/\%$ for \gls{MSR}, confirming that the high near-optimality rates in the full dataset were an artefact of portfolio simplicity rather than genuine optimisation. The distribution also shifts markedly rightward: under \gls{MSR}, $\LOneFivePlusMsHighBinPct/\%$ of $N \geq 5$ accounts fall in the $(80,100]\%$ bucket, compared to $\LOneMsHighBinPct/\%$ in the full data.

Predictive models lose most of their explanatory power. Random Forest $R^2$ for \gls{MVR} drops from $\LOneRfSrRsq/$ to $\LOneFivePlusRfSrRsq/$, and the Spearman $\rho$ between \texttt{num\_tokens} and \gls{MVR} distance falls from $\LOneSpearmanSrTokens/$ to $\LOneFivePlusSpearmanSrTokens/$, indicating that the previously dominant portfolio-size effect was largely capturing the gap between two-asset and multi-asset accounts. Feature importances reshuffle accordingly: for \gls{MSR}, \texttt{month} becomes the leading predictor ($\LOneFivePlusMsMonthImp/\%$), suggesting that shifting market regimes make the Sharpe target a moving goalpost for diversified portfolios. Cross-strategy correlations also increase (most notably, $r$ between \gls{MRV} and \gls{MSR} rises from $\LOneCorrBrMs/$ to $\LOneFivePlusCorrBrMs/$), implying that among diversified accounts, suboptimality across objectives converges. Overall, these results reinforce the paper's central finding: observable portfolio characteristics explain suboptimality primarily through the coarse channel of portfolio size, while \textit{which} tokens are held (unobserved in our feature set) is the dominant predictive feature for any non-trivial allocation.

\subsection{Distance by Portfolio Size}
\label{sec:distance_by_size}
\Cref{fig:l1_mean_token_bucket} disaggregates the binary split from \cref{fig:l1_mean_token_group}
into eight portfolio-size buckets.
Across all three strategies, mean distance rises steeply as portfolio size increases from $N=2$ to $N=5$
and then plateaus for larger portfolios.
For \gls{MVR}, $N=2$ accounts have a mean distance near zero,
while $N=5$ accounts already approach the level observed for the largest portfolios ($N \geq 21$).
The pattern is qualitatively similar for \gls{MRV} and \gls{MSR}, though the absolute distances are higher
and the plateau sets in at a slightly higher level.
This saturation motivates the choice of $N=5$ as the stratification threshold in the main analysis: below $N=5$, distance is a mechanical artefact of portfolio simplicity, while above $N=5$, additional tokens contribute little further deviation.

\begin{figure}[htbp]
    \centering
    \includegraphics[width=1\linewidth]{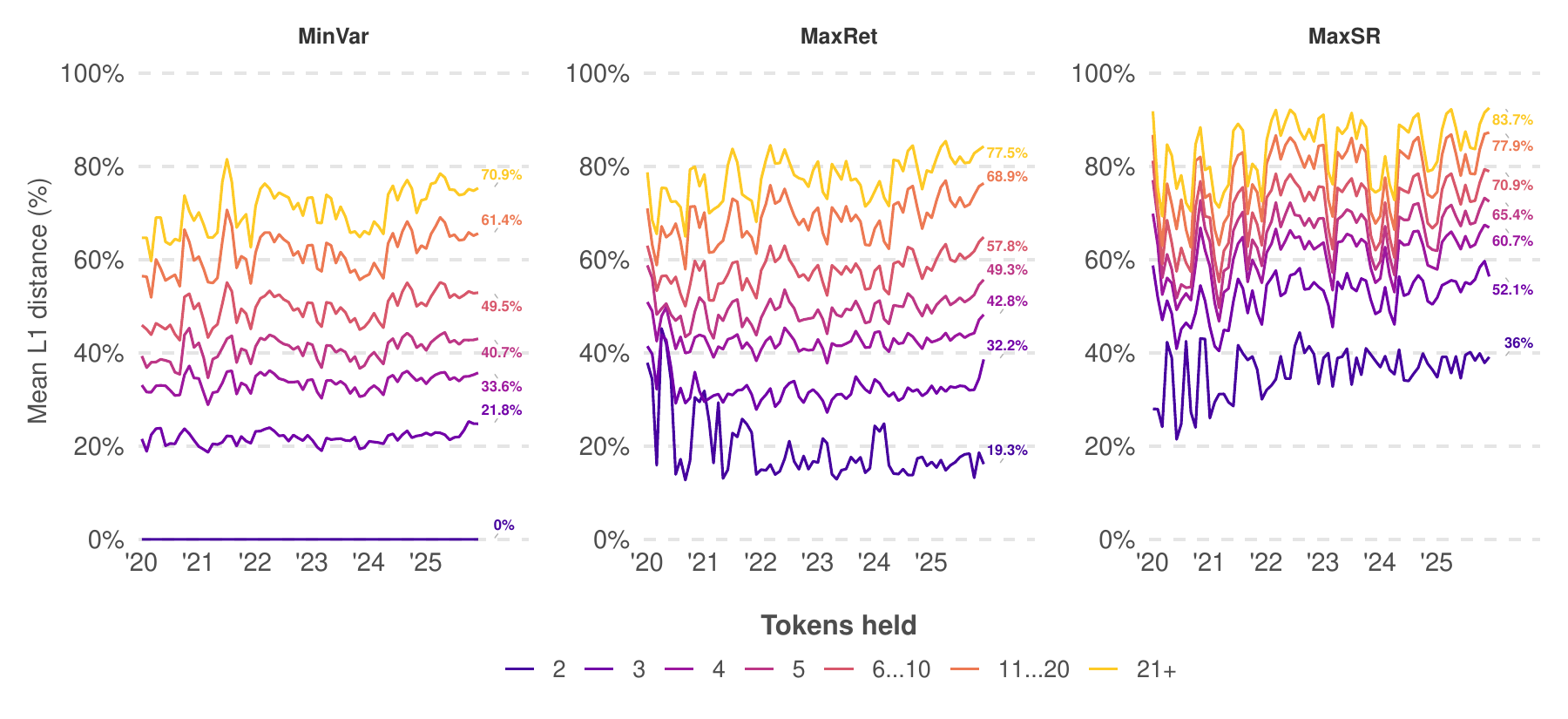}
    \caption{Mean distance over time by granular portfolio-size bucket and optimisation strategy.
    Legend entries report the time-series average for each bucket.
    Distance rises steeply from $N=2$ to $N=5$ and plateaus thereafter.}
    \label{fig:l1_mean_token_bucket}
\end{figure}

\paragraph*{Threshold sensitivity.}
To verify that the binary split is robust to the choice of cutoff,
we repeat the paired Wilcoxon signed-rank test from \cref{sec:measuring_distance}
for thresholds $N \in \{3, 4, 5, 6, 7\}$.
\Cref{tab:threshold_sweep} reports the results.
All fifteen tests reject the null at $p < 1.7 \times 10^{-13}$
(the minimum attainable for \TotalBlocks/ paired blocks),
confirming that the small-versus-large contrast is not an artefact of the specific cutoff.
At $N=5$, the threshold used in the main text, Cohen's $d$ is $\LOneCohenMRV/$ for \gls{MRV}, $\LOneCohenMVR/$ for \gls{MVR}, and $\LOneCohenMSR/$ for \gls{MSR}, the last two reflecting the near-zero distances achievable by small portfolios under minimum-variance optimisation.
Cohen's $d$ increases monotonically with $N$ because raising the threshold
purifies the ``below'' group (dominated by one- and two-token accounts that are structurally close to optimal),
thereby widening the gap.
The marginal gain in $d$ is largest between $N=3$ and $N=5$ and diminishes thereafter,
consistent with the saturation visible in \cref{fig:l1_mean_token_bucket}.

\begin{table}[htbp]
\centering
\small
\caption{Paired Wilcoxon signed-rank tests for the distance gap
between portfolios with fewer than $N$ tokens and those with $N$ or more,
at varying thresholds. All $p$-values equal $1.66 \times 10^{-13}$ (Wilcoxon floor for \TotalBlocks/ blocks).}
\label{tab:threshold_sweep}
\begin{tabular*}{\textwidth}{@{\extracolsep{\fill}} l c r r r r}
\toprule
Strategy & $N$ & Mean${<}N$ (\%) & Mean${\geq}N$ (\%) & $\Delta$ (pp) & Cohen's $d$ \\
\midrule
\gls{MRV} & 3 & 19.3 & 42.4 & 23.0 &  3.39 \\
          & 4 & 22.3 & 51.4 & 29.1 &  4.54 \\
          & 5 & \LOneSmallMRVMean/ & \LOneLargeMRVMean/ & \LOneDeltaMRV/ & \LOneCohenMRV/ \\
          & 6 & 25.3 & 60.7 & 35.4 &  5.30 \\
          & 7 & 26.1 & 63.6 & 37.5 &  5.59 \\
\midrule
\gls{MVR} & 3 &  0.0 & 32.8 & 32.8 & 12.62 \\
          & 4 &  4.9 & 42.7 & 37.8 & 13.84 \\
          & 5 & \LOneSmallMVRMean/ & \LOneLargeMVRMean/ & \LOneDeltaMVR/ & \LOneCohenMVR/ \\
          & 6 &  9.0 & 52.6 & 43.6 & 14.18 \\
          & 7 & 10.0 & 55.7 & 45.8 & 14.50 \\
\midrule
\gls{MSR} & 3 & 36.0 & 59.7 & 23.7 &  4.46 \\
          & 4 & 39.7 & 66.5 & 26.9 &  5.22 \\
          & 5 & \LOneSmallMSRMean/ & \LOneLargeMSRMean/ & \LOneDeltaMSR/ & \LOneCohenMSR/ \\
          & 6 & 42.6 & 72.8 & 30.2 &  5.79 \\
          & 7 & 43.3 & 74.6 & 31.3 &  6.02 \\
\bottomrule
\end{tabular*}
\end{table}

\paragraph*{Power-decay fit.}
\label{sec:l1_fit_details}
We model the relationship between portfolio size and mean distance using a three-parameter power-decay specification
$\hat{d}_s(n) = \delta_s^{\infty}\,(1 - \psi_s \cdot n^{-\gamma_s})$,
estimated by weighted nonlinear least squares (Levenberg--Marquardt) on 49 portfolio-size bins ($n \in [2,50]$, minimum 30 observations per bin), with the constraint $\delta_s^{\infty} \leq 100\%$.
Each bin is weighted by $\sqrt{N_n}$, where $N_n$ is the number of account-observations at that portfolio size, so that well-populated bins (e.g.\ $n{=}2$ with 38M observations) dominate the fit over sparse tail bins.
The model is monotonically increasing and bounded: it equals $\delta_s^{\infty}(1 - \psi_s \cdot 2^{-\gamma_s})$ at the minimum portfolio size and asymptotes to $\delta_s^{\infty}$ as $n \to \infty$.
The asymptote $\delta_s^{\infty}$ need not equal $100\%$: \gls{MRV} hits the upper bound ($\delta^{\infty} = \FitDinfMRV/\%$), confirming that sufficiently diversified wallets can share zero weight overlap with the return-maximising portfolio.

In contrast, \gls{MVR} and \gls{MSR} plateau at $\delta^{\infty} = \FitDinfMVR/\%$ and $\delta^{\infty} = \FitDinfMSR/\%$ respectively, indicating that even infinitely diversified wallets retain some structural overlap with the minimum-variance and Sharpe-optimal allocations.
The exponent $\gamma_s$ controls the saturation speed.
For \gls{MSR}, $\gamma = \FitGammaMSR/$ is the largest, meaning that distance rises steeply with just a few additional tokens and largely plateaus beyond $n \approx 10$.
For \gls{MRV}, $\gamma = \FitGammaMRV/$ is the smallest, indicating a more gradual climb that continues to increase noticeably up to $n \approx 20$.
\gls{MVR} sits in between ($\gamma = \FitGammaMVR/$), consistent with the intermediate position of minimum-variance distance throughout our analysis.

\cref{fig:l1_fit_vs_actual} overlays the fitted curves on the empirical means. The fits track the data closely across the entire range, with the largest deviations appearing in the sparse tail beyond 30 tokens where individual bin estimates are noisier.

\begin{figure}[htbp]
    \centering
    \includegraphics[width=1\linewidth]{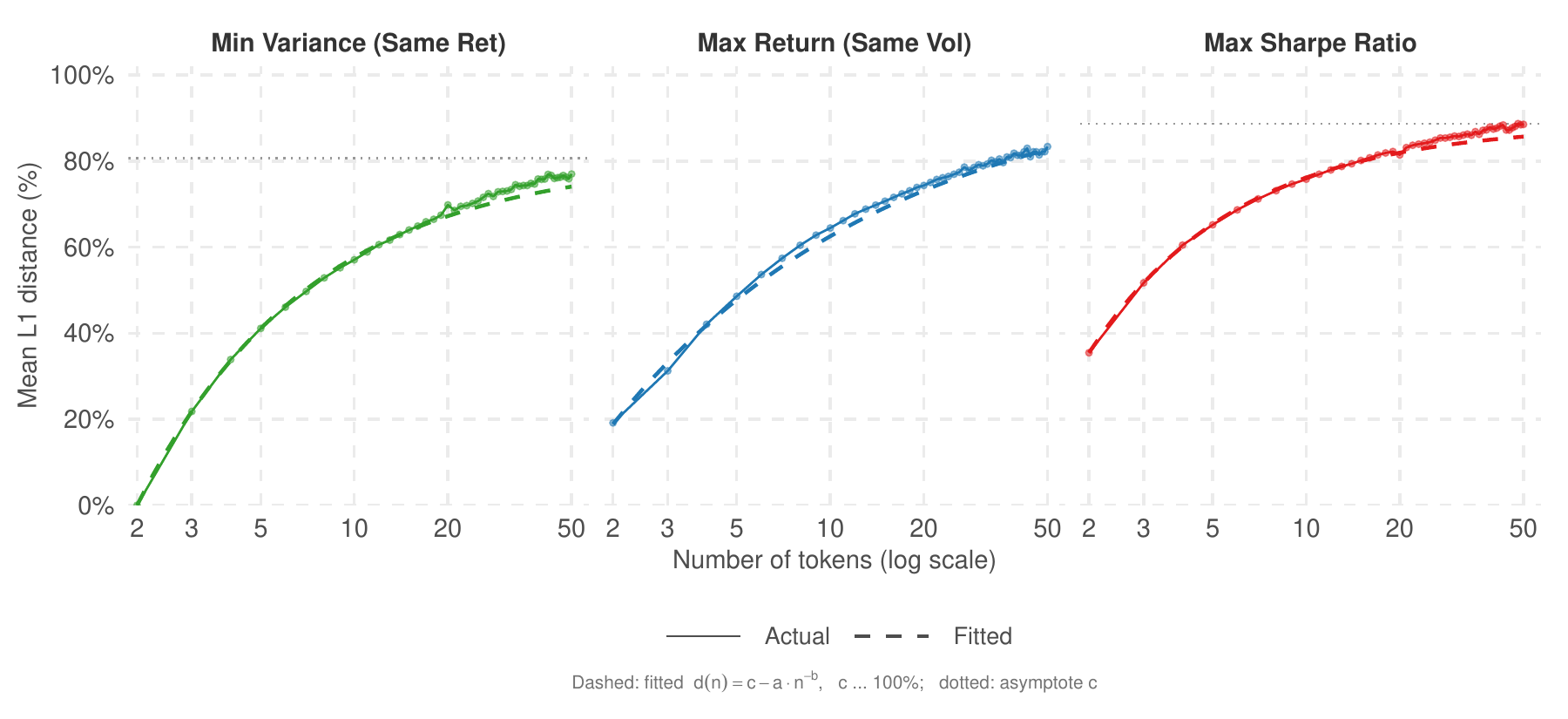}
    \caption{Fitted versus empirical distance: empirical mean distance (solid) versus fitted power-decay model $\hat{d}_s(n) = \delta_s^{\infty}\,(1 - \psi_s \cdot n^{-\gamma_s})$ (dashed) for each strategy. Dotted lines mark the asymptote $\delta_s^{\infty}$ where $\delta_s^{\infty} < 100\%$.}
    \label{fig:l1_fit_vs_actual}
\end{figure}

\subsection{Naive Benchmark Comparison}
\label{sec:naive_benchmark_details}

\cref{tab:l1_naive} reports the distance between the actual allocation $w_{(0)}$ and two naive benchmarks.
Both naive strategies are substantially farther from $w_{(0)}$ than the closest \gls{MPT} strategy (\gls{MVR}), confirming that on-chain accounts do not simply replicate either equal-weight or market-cap-weight allocations.
The mean--median relationship differs across benchmarks: for equal-weight the median exceeds the mean, indicating a left-skewed distribution with a concentration of accounts near the $\LOneNaiveEqualMedian/\%$ mark, whereas for market-cap-weight the mean exceeds the median, reflecting a right tail of accounts that deviate substantially from cap-weighting.

\begin{table}[htbp]
\centering
\small
\caption{Distance to naive benchmarks: distance between the actual allocation $w_{(0)}$ and naive allocation benchmarks, expressed as percentages. The closest \gls{MPT} strategy (\gls{MVR}) is included for reference.}
\label{tab:l1_naive}
\begin{tabular*}{\textwidth}{l @{\extracolsep{\fill}} cc}
\toprule
Benchmark & Mean $d_s$ (\%) & Median $d_s$ (\%) \\
\midrule
Equal-weight ($1/N$)       & \LOneNaiveEqualMean/  & \LOneNaiveEqualMedian/ \\
Market-cap-weight           & \LOneNaiveMcapMean/   & \LOneNaiveMcapMedian/ \\
\midrule
\textit{\gls{MVR} (reference)} & \textit{\LOneMeanSaferRisk/} & \textit{0.0} \\
\bottomrule
\end{tabular*}
\end{table}

\cref{tab:mpt_vs_naive} and \cref{fig:mpt_vs_naive} decompose the structural effect of each \gls{MPT} strategy on distance to the naive benchmarks.
Three patterns stand out.
First, \gls{MVR} is the most conservative optimiser: it leaves the majority of accounts unchanged against both equal-weight (\LOneDeltaSrEqualUnchangedPct/\%) and cap-weight (\LOneDeltaSrMcapUnchangedPct/\%), and when it does move accounts, it predominantly moves them closer to equal-weight (closer-to-farther ratio $\approx$ \LOneDeltaSrEqualRatio/:1).
Second, \gls{MRV} shows a split personality: it moves accounts closer to equal-weight (mean $\Delta d = \LOneDeltaBrEqualMean/$) but farther from cap-weight (mean $\Delta d = \LOneDeltaBrMcapMean/$), indicating that return-maximising portfolios tend to diversify away from concentrated large-cap holdings.
Third, \gls{MSR} pushes the majority of accounts farther from both benchmarks, with over half moving away from equal-weight (\LOneDeltaMsEqualFartherPct/\%) and cap-weight (\LOneDeltaMsMcapFartherPct/\%), reflecting the highly specific asset weightings required to maximise the Sharpe ratio.

\begin{table}[htbp]
\centering
\small
\caption{Directional effect of \gls{MPT} optimisation on naive-benchmark distance: $\Delta d < 0$ indicates the optimised portfolio moved closer; $\Delta d > 0$ farther. Accounts with $|\Delta d| < 0.001$ are classified as unchanged.}
\label{tab:mpt_vs_naive}
\begin{tabular*}{\textwidth}{l @{\extracolsep{\fill}} lcccc}
\toprule
Strategy & Naive Benchmark & Mean $\Delta d$ & Closer (\%) & Farther (\%) & Unchanged (\%) \\
\midrule
\gls{MRV} & Equal-weight & $\LOneDeltaBrEqualMean/$ & \LOneDeltaBrEqualCloserPct/ & \LOneDeltaBrEqualFartherPct/ & \LOneDeltaBrEqualUnchangedPct/ \\
          & MCap-weight  & $\LOneDeltaBrMcapMean/$  & \LOneDeltaBrMcapCloserPct/  & \LOneDeltaBrMcapFartherPct/  & \LOneDeltaBrMcapUnchangedPct/ \\
\midrule
\gls{MVR} & Equal-weight & $\LOneDeltaSrEqualMean/$ & \LOneDeltaSrEqualCloserPct/ & \LOneDeltaSrEqualFartherPct/ & \LOneDeltaSrEqualUnchangedPct/ \\
          & MCap-weight  & $\LOneDeltaSrMcapMean/$  & \LOneDeltaSrMcapCloserPct/  & \LOneDeltaSrMcapFartherPct/  & \LOneDeltaSrMcapUnchangedPct/ \\
\midrule
\gls{MSR} & Equal-weight & $\LOneDeltaMsEqualMean/$ & \LOneDeltaMsEqualCloserPct/ & \LOneDeltaMsEqualFartherPct/ & \LOneDeltaMsEqualUnchangedPct/ \\
          & MCap-weight  & $\LOneDeltaMsMcapMean/$  & \LOneDeltaMsMcapCloserPct/  & \LOneDeltaMsMcapFartherPct/  & \LOneDeltaMsMcapUnchangedPct/ \\
\bottomrule
\end{tabular*}
\end{table}

\begin{figure}[H]
    \centering
    \includegraphics[width=1\linewidth]{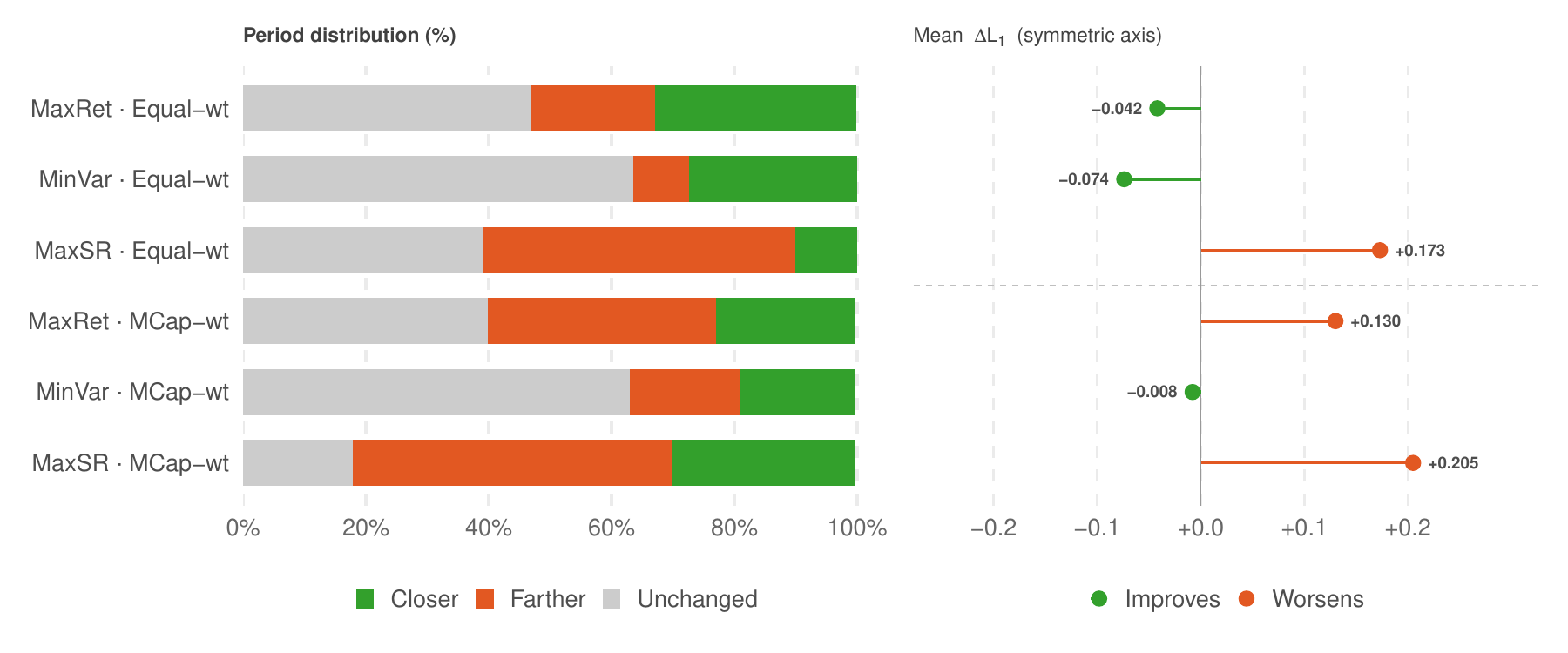}
    \caption{\gls{MPT} optimisation versus naive benchmarks: left, proportion of accounts moved closer, farther, or left unchanged; right, mean $\Delta d$ (negative = toward the benchmark; positive = away).}
    \label{fig:mpt_vs_naive}
\end{figure}

\subsection{Optimisation Volume and Temporal Evolution}
\label{sec:mpt_optimisation_details}

This subsection reports the optimisation volume and temporal distance statistics underlying the main analysis in \cref{sec:mpt_optimal_portfolio}.

\paragraph*{Aggregate temporal evolution of distance.}
\label{sec:appendix_l1_time}

\Cref{fig:l1_time_std} reports the aggregate mean and median distance,
together with $\pm 1$ standard-deviation bands, for each optimisation strategy over the full sample period.
All three strategies exhibit a mild upward drift in mean distance over time,
consistent with the gradual increase in portfolio complexity documented in the main analysis.
The persistent and substantial gap between mean and median under \gls{MVR} reflects the distributional asymmetry of that strategy's distance: a large mass of accounts with near-zero distance
(pooled $\leq 1\%$ share: \LOneSrCombinedLeOnePct/\%) coexists with a thin but heavy right tail,
pulling the mean well above the median.
Under \gls{MRV} and \gls{MSR} the mean--median gap is narrower, indicating a less skewed distribution,
consistent with their higher pooled mean distances of \LOneMeanBetterReturn/\% and \LOneMeanMaxSharpe/\% against medians of \LOneMedianBetterReturn/\% and \LOneMedianMaxSharpe/\%, respectively.
The broad standard-deviation bands across all strategies confirm the high cross-sectional heterogeneity
that motivates the portfolio-size stratification presented in the main text
(\cref{fig:l1_mean_token_group}): once accounts are split by portfolio size,
the \gls{MVR} mean ranges from \LOneLargeMVRMean/\% ($N \geq 5$) to \LOneSmallMVRMean/\% ($N < 5$), a spread of more than 40 percentage points that the aggregate view entirely obscures.

\begin{figure}[htbp]
    \centering
    \includegraphics[width=1\linewidth]{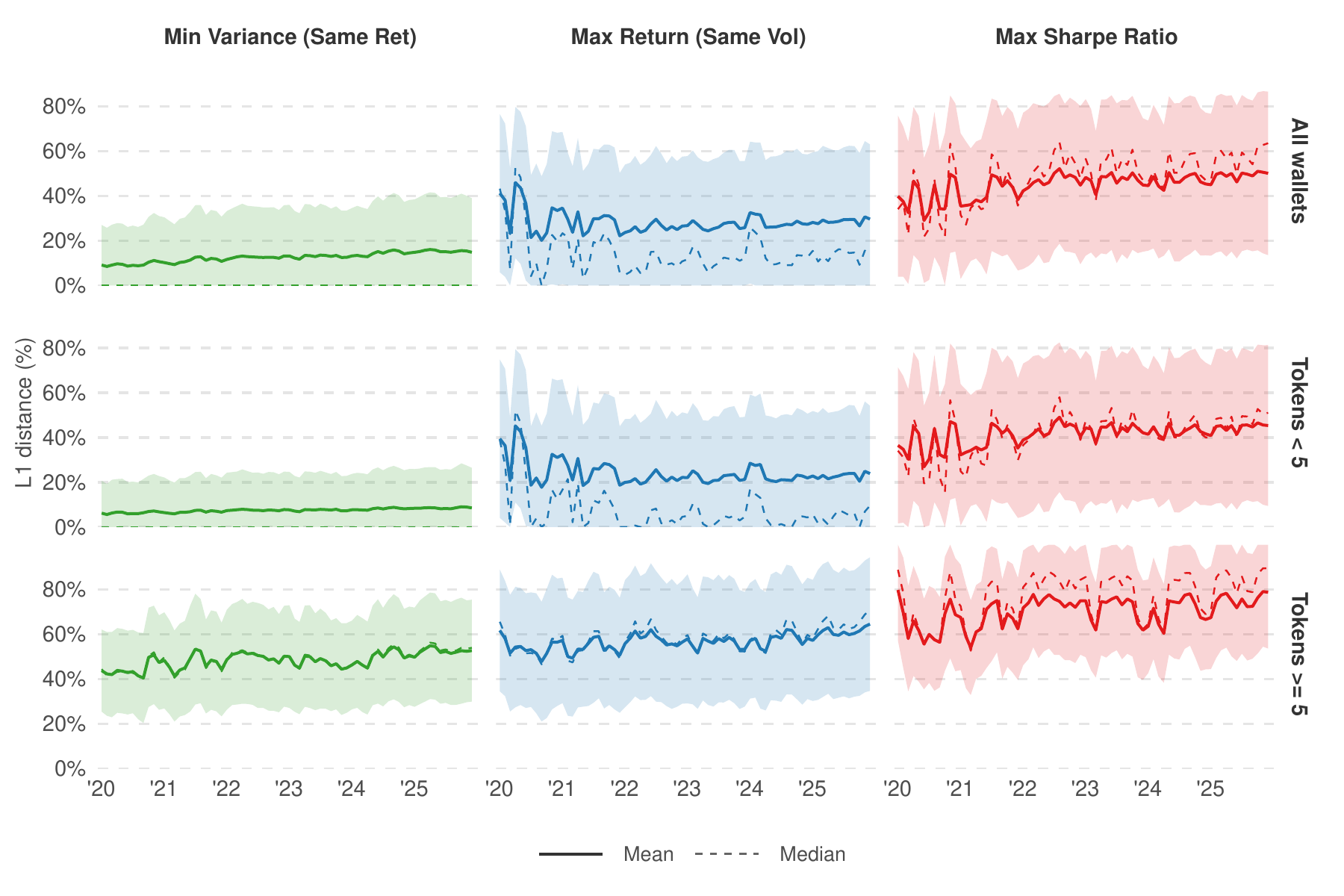}
    \caption{Aggregate distance over time: monthly mean and median with $\pm 1$ standard-deviation bands for each strategy, pooled across all accounts ($N = \LOneObservationsN/$, \LOneMonthsN/ monthly snapshots).}
    \label{fig:l1_time_std}
\end{figure}

\section{Supplementary Material: Realised Portfolio Performance}
\label{sec:appendix_performance}

This appendix supplements the realised performance analysis in \cref{sec:returns_and_capm}.
It covers the frequency of outperforming the market and the cross-sectional magnitude of beat-market margins (\cref{sec:appendix-pct-beat-market}),
the analysis of portfolio $\beta$ and market exposure (\cref{sec:appendix_beta}),
the association between frontier distance and risk-adjusted return (\cref{sec:appendix_distance_alpha}),
and the Random Forest and SHAP analysis of return and alpha predictability (\cref{sec:appendix_rf_shap}).

\subsection{Frequency of Outperforming the Market}
\label{sec:appendix-pct-beat-market}

\Cref{tab:ret_alpha_summary} in \cref{sec:forward_returns} reports realised return and alpha statistics aggregated over all account observations. As a complementary view, we report the within-snapshot share of accounts that beat the \gls{wETH} and \gls{WBTC} market index and summarise these per-snapshot percentages across the \CAPMBlocksN/ monthly snapshots. For each strategy and each monthly snapshot we compute (i) the share of accounts whose realised \CAPMHorizonDays/-day return exceeds the market and (ii) the share whose \gls{CAPM} alpha is strictly positive. Both numerator and denominator are restricted to accounts where the relevant return and beta values are present, so each strategy is evaluated on its own valid subpopulation. The baseline filter from \cref{sec:forward_returns} is applied throughout.

\Cref{tab:pct_beat_market} reports the resulting distributions. Three patterns stand out. First, cross-strategy variation is narrow: the per-snapshot median ranges from \PctBeatRetEqualBlockMedian/\% to \PctBeatRetBaselineBlockMedian/\% by raw return and from \PctPosAlphaBetterBlockMedian/\% to \PctPosAlphaBaselineBlockMedian/\% by alpha, a spread of \PctBeatRetMedianStrategySpread/ and \PctPosAlphaMedianStrategySpread/ percentage points respectively. Second, cross-snapshot variation is an order of magnitude larger: the baseline rate alone spans \PctBeatRetBaselineBlockMin/\% to \PctBeatRetBaselineBlockMax/\% across snapshots, and pairwise Pearson correlations of the per-snapshot series across strategies all exceed \PctBeatCrossStratCorrLo/, indicating that strategies move almost in lockstep over the calendar. Third, in \PctBlocksBelowFifty/\% of snapshots fewer than half of baseline accounts beat the market index, reflecting the right-skewed cross section of \gls{ERC-20} returns where a long tail of underperforming tokens drags most accounts below the \gls{wETH}--\gls{WBTC} blend in the typical month. Together these patterns reinforce the conclusion of \cref{sec:forward_returns}: the choice of allocation strategy has at most a marginal effect on whether an account beats the market in any given month, while the prevailing market regime explains the bulk of the variation.

The frequency view summarised in \cref{tab:pct_beat_market} captures only whether each account beat the market in a given month, not by how much. As a complementary magnitude view, for each monthly snapshot we compute the within-snapshot percentiles of the beat-market margin across accounts and report the median of each percentile across the \CAPMBlocksN/ monthly snapshots. Two distributional metrics are reported in parallel: the raw excess return $r_s - r_{(m)}$ and the cross-sectional \gls{CAPM} alpha $r_s - \beta_s\, r_{(m)}$. The per-strategy distributions are visualised in \cref{fig:beat_market_quantile_box_excess,fig:beat_market_quantile_box_alpha}, and selected percentiles are reported in \cref{tab:beat_market_quantiles}.

\begin{table}[htbp]
\centering
\small
\caption{Per-snapshot frequency of beating the \gls{wETH}--\gls{WBTC} market index, summarised across the \CAPMBlocksN/ monthly snapshots. Each cell is the unweighted median or mean across snapshots of the within-snapshot share of accounts whose realised return exceeds the market (left pair) or whose \gls{CAPM} $\alpha$ is strictly positive (right pair).}
\label{tab:pct_beat_market}
\begin{tabular*}{\textwidth}{@{\extracolsep{\fill}} l rr rr @{}}
\toprule
& \multicolumn{2}{c}{Beats market by return (\%)} & \multicolumn{2}{c}{Positive \gls{CAPM} $\alpha$ (\%)} \\
\cmidrule(lr){2-3} \cmidrule(lr){4-5}
Strategy        & Median & Mean & Median & Mean \\
\midrule
Baseline        & \PctBeatRetBaselineBlockMedian/  & \PctBeatRetBaselineBlockMean/  & \PctPosAlphaBaselineBlockMedian/  & \PctPosAlphaBaselineBlockMean/  \\
MaxRet          & \PctBeatRetBetterBlockMedian/    & \PctBeatRetBetterBlockMean/    & \PctPosAlphaBetterBlockMedian/    & \PctPosAlphaBetterBlockMean/    \\
MinVar          & \PctBeatRetSaferBlockMedian/     & \PctBeatRetSaferBlockMean/     & \PctPosAlphaSaferBlockMedian/     & \PctPosAlphaSaferBlockMean/     \\
MaxSR           & \PctBeatRetMaxSharpeBlockMedian/ & \PctBeatRetMaxSharpeBlockMean/ & \PctPosAlphaMaxSharpeBlockMedian/ & \PctPosAlphaMaxSharpeBlockMean/ \\
Equal-weight    & \PctBeatRetEqualBlockMedian/     & \PctBeatRetEqualBlockMean/     & \PctPosAlphaEqualBlockMedian/     & \PctPosAlphaEqualBlockMean/     \\
MCap-weight     & \PctBeatRetMcapBlockMedian/      & \PctBeatRetMcapBlockMean/      & \PctPosAlphaMcapBlockMedian/      & \PctPosAlphaMcapBlockMean/      \\
\bottomrule
\end{tabular*}
\end{table}

Three patterns are immediately visible. First, every strategy produces a strongly right-skewed cross-sectional distribution: the upper tail extends much further from zero than the lower tail. For the baseline portfolio the top one percent of accounts beat the market by \BeatExcRetBaselineTopOne/\% in a typical month while the bottom one percent return \BeatExcRetBaselineBotOneAbs/\% below it, an asymmetry of \BeatExcRetBaselineSkew/ percentage points ($Q_{99} + Q_{01}$, with positive values indicating right skew). The asymmetry reproduces qualitatively across all six strategies and reflects the well-documented right skew of token-level returns on Ethereum, which dominates the cross section even after aggregation to the account level. Second, the centre of mass lies below zero for every strategy: median excess return ranges from \BeatExcRetMcapMed/\% for MCap-weight to \BeatExcRetBetterMed/\% for MaxRet, restating the conclusion of \cref{tab:pct_beat_market} as a location parameter rather than a binary frequency. Third, the cross-strategy contrasts that the figure makes most visible all live in the tails rather than at the median. \gls{MSR} has the deepest lower tail ($Q_{01} = \BeatExcRetMaxSharpeBotOne/\%$), reflecting the classical Markowitz problem of concentrated bets on noisy historical covariances. Equal-weight produces the narrowest distribution overall ($Q_{01} = \BeatExcRetEqualBotOne/\%$, $Q_{99} = \BeatExcRetEqualTopOne/\%$), trading upper-tail upside for lower-tail robustness. MCap-weight is the only strategy whose entire distribution shifts noticeably toward zero, consistent with its mechanical alignment with the market index.

\begin{figure}[htbp]
\centering
\includegraphics[width=1\textwidth]{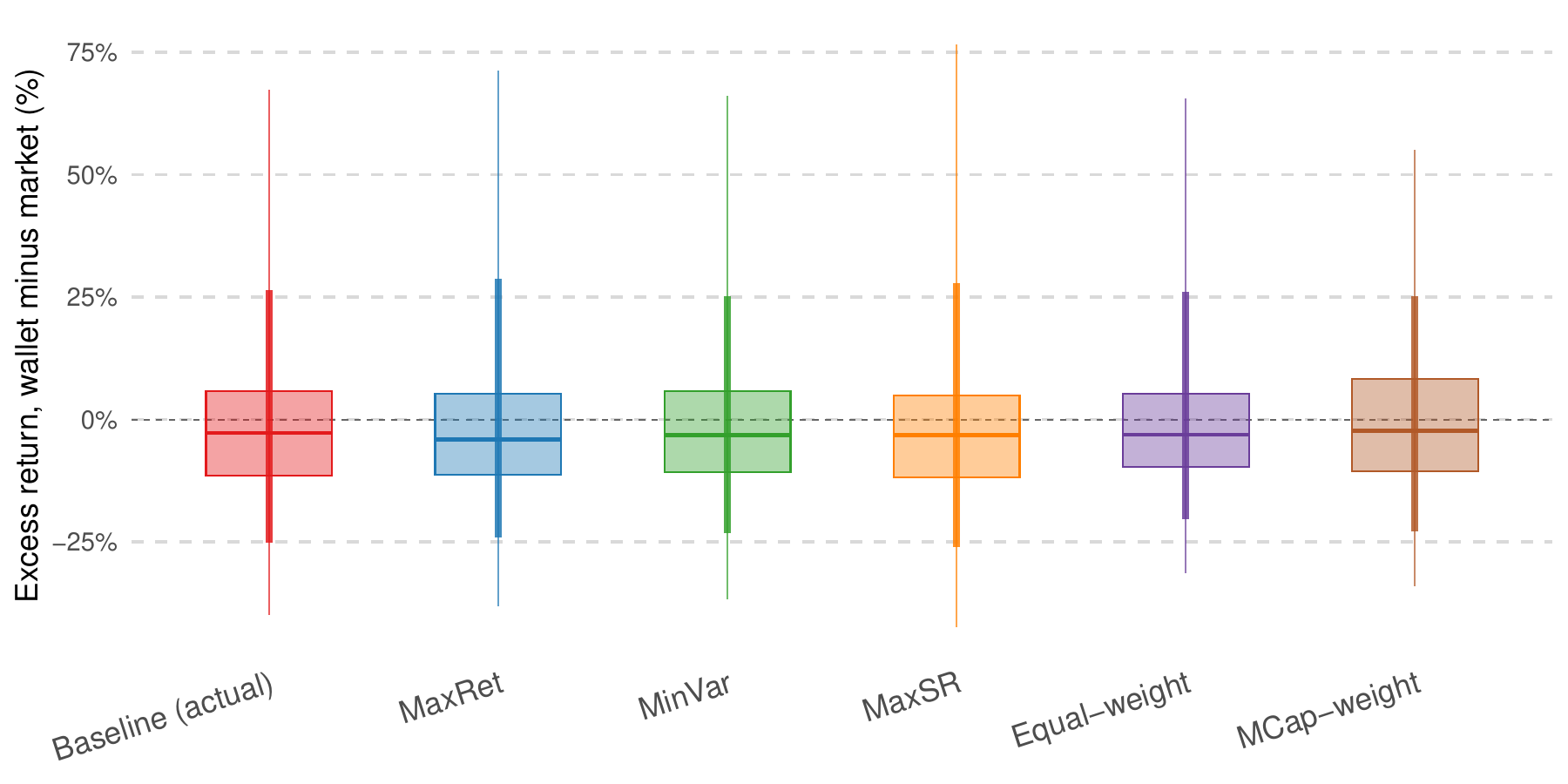}
\caption{Cross-sectional distribution of the per-account excess return $r_s - r_{(m)}$, by strategy. Each box summarises the within-snapshot percentiles across accounts (filled box $Q_{25}$ to $Q_{75}$, inner line $Q_{50}$, thick whisker $Q_{05}$ to $Q_{95}$, thin whisker $Q_{01}$ to $Q_{99}$), aggregated as the median of each percentile across the \CAPMBlocksN/ monthly snapshots. The dashed line marks the market.}
\label{fig:beat_market_quantile_box_excess}
\end{figure}

The \gls{CAPM} $\alpha$ distribution shown in \cref{fig:beat_market_quantile_box_alpha} mirrors the excess-return shape closely, with slightly more pronounced tails after the $\beta$ adjustment (baseline $Q_{01} = \BeatAlphaBaselineBotOne/\%$ versus \BeatExcRetBaselineBotOne/\% on raw excess return, and $Q_{99} = \BeatAlphaBaselineTopOne/\%$ versus \BeatExcRetBaselineTopOne/\%), indicating that accounts in both tails carry larger absolute $\beta$ exposure to market moves. The cross-strategy ranking is unchanged.

\begin{figure}[htbp]
\centering
\includegraphics[width=1\textwidth]{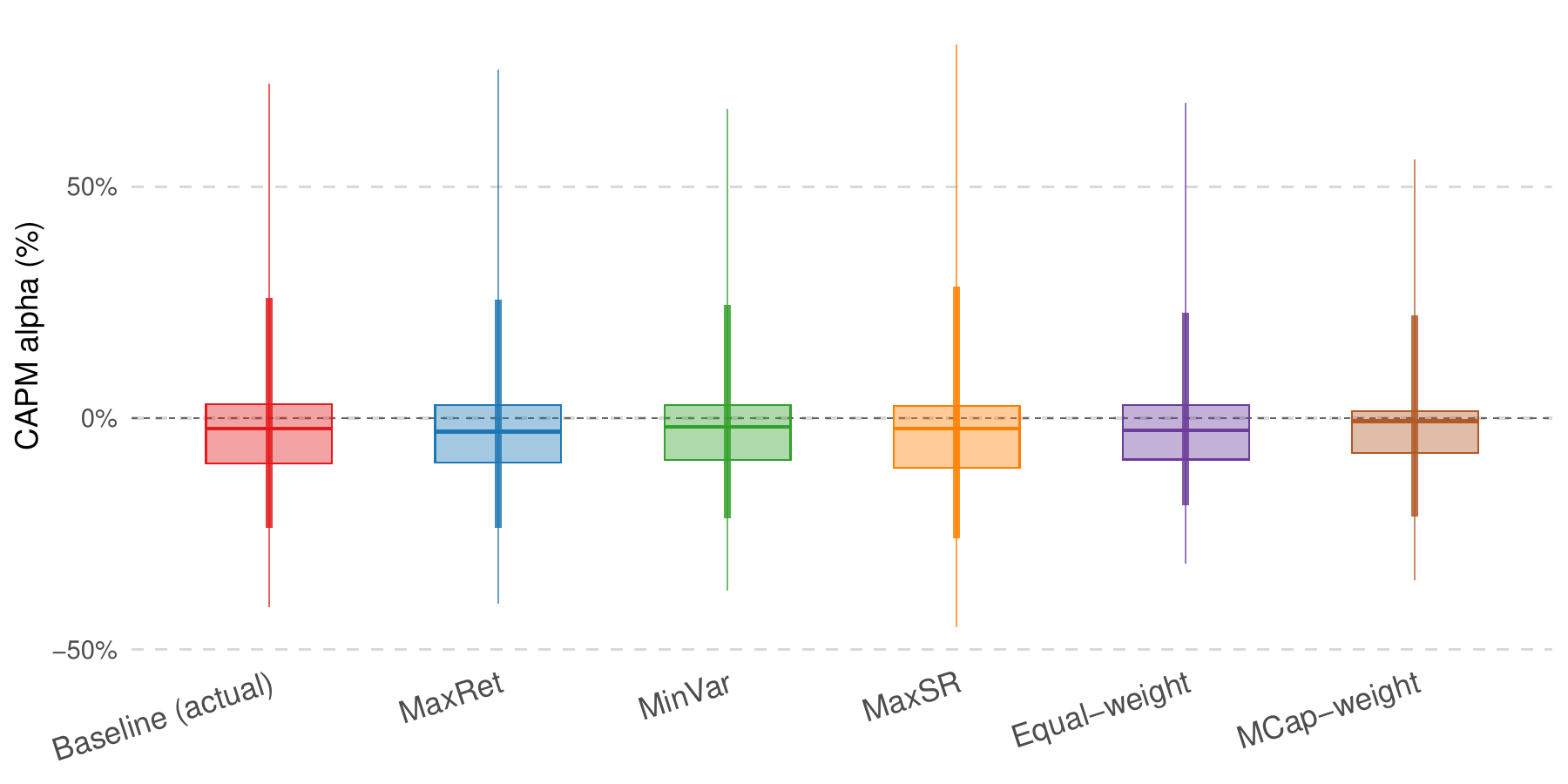}
\caption{Distribution of the per-account \gls{CAPM} $\alpha$
$r_s - \beta_s\, r_{(m)}$ by strategy, with quantiles aggregated as
the median across all 72 monthly snapshots. Boxes are
constructed as in \cref{fig:beat_market_quantile_box_excess}, with
the dashed line marking zero alpha.}
\label{fig:beat_market_quantile_box_alpha}
\end{figure}

\begin{table}[htbp]
\centering
\small
\caption{Distribution of the per-account beat-market margin, summarised across the \CAPMBlocksN/ monthly snapshots. For each monthly snapshot we compute the listed within-snapshot percentiles across accounts and report the median of each percentile across snapshots. The \emph{Skew} column is $Q_{99} + Q_{01}$. Positive values indicate right skew, and values are reported in percentage points.}
\label{tab:beat_market_quantiles}
\begin{tabular*}{\textwidth}{@{\extracolsep{\fill}} l rrrr rrrr @{}}
\toprule
& \multicolumn{4}{c}{Excess return $r_s - r_{(m)}$ (\%)} & \multicolumn{4}{c}{\gls{CAPM} $\alpha$ $r_s - \beta_s\, r_{(m)}$ (\%)} \\
\cmidrule(lr){2-5} \cmidrule(lr){6-9}
Strategy     & $Q_{01}$ & $Q_{50}$ & $Q_{99}$ & Skew & $Q_{01}$ & $Q_{50}$ & $Q_{99}$ & Skew \\
\midrule
Baseline     & \BeatExcRetBaselineBotOne/  & \BeatExcRetBaselineMed/  & \BeatExcRetBaselineTopOne/  & \BeatExcRetBaselineSkew/  & \BeatAlphaBaselineBotOne/  & \BeatAlphaBaselineMed/  & \BeatAlphaBaselineTopOne/  & \BeatAlphaBaselineSkew/  \\
MaxRet       & \BeatExcRetBetterBotOne/    & \BeatExcRetBetterMed/    & \BeatExcRetBetterTopOne/    & \BeatExcRetBetterSkew/    & \BeatAlphaBetterBotOne/    & \BeatAlphaBetterMed/    & \BeatAlphaBetterTopOne/    & \BeatAlphaBetterSkew/    \\
MinVar       & \BeatExcRetSaferBotOne/     & \BeatExcRetSaferMed/     & \BeatExcRetSaferTopOne/     & \BeatExcRetSaferSkew/     & \BeatAlphaSaferBotOne/     & \BeatAlphaSaferMed/     & \BeatAlphaSaferTopOne/     & \BeatAlphaSaferSkew/     \\
MaxSR        & \BeatExcRetMaxSharpeBotOne/ & \BeatExcRetMaxSharpeMed/ & \BeatExcRetMaxSharpeTopOne/ & \BeatExcRetMaxSharpeSkew/ & \BeatAlphaMaxSharpeBotOne/ & \BeatAlphaMaxSharpeMed/ & \BeatAlphaMaxSharpeTopOne/ & \BeatAlphaMaxSharpeSkew/ \\
Equal-weight & \BeatExcRetEqualBotOne/     & \BeatExcRetEqualMed/     & \BeatExcRetEqualTopOne/     & \BeatExcRetEqualSkew/     & \BeatAlphaEqualBotOne/     & \BeatAlphaEqualMed/     & \BeatAlphaEqualTopOne/     & \BeatAlphaEqualSkew/     \\
MCap-weight  & \BeatExcRetMcapBotOne/      & \BeatExcRetMcapMed/      & \BeatExcRetMcapTopOne/      & \BeatExcRetMcapSkew/      & \BeatAlphaMcapBotOne/      & \BeatAlphaMcapMed/      & \BeatAlphaMcapTopOne/      & \BeatAlphaMcapSkew/      \\
\bottomrule
\end{tabular*}
\end{table}

As a complementary temporal view, \cref{fig:returns_panel_delta_market} contrasts the between-strategy return differences with the magnitude of market-wide movements across snapshots, underscoring that allocation choice is a second-order effect relative to the market cycle.

\begin{figure}[htbp]
  \centering
  \includegraphics[width=1\linewidth]{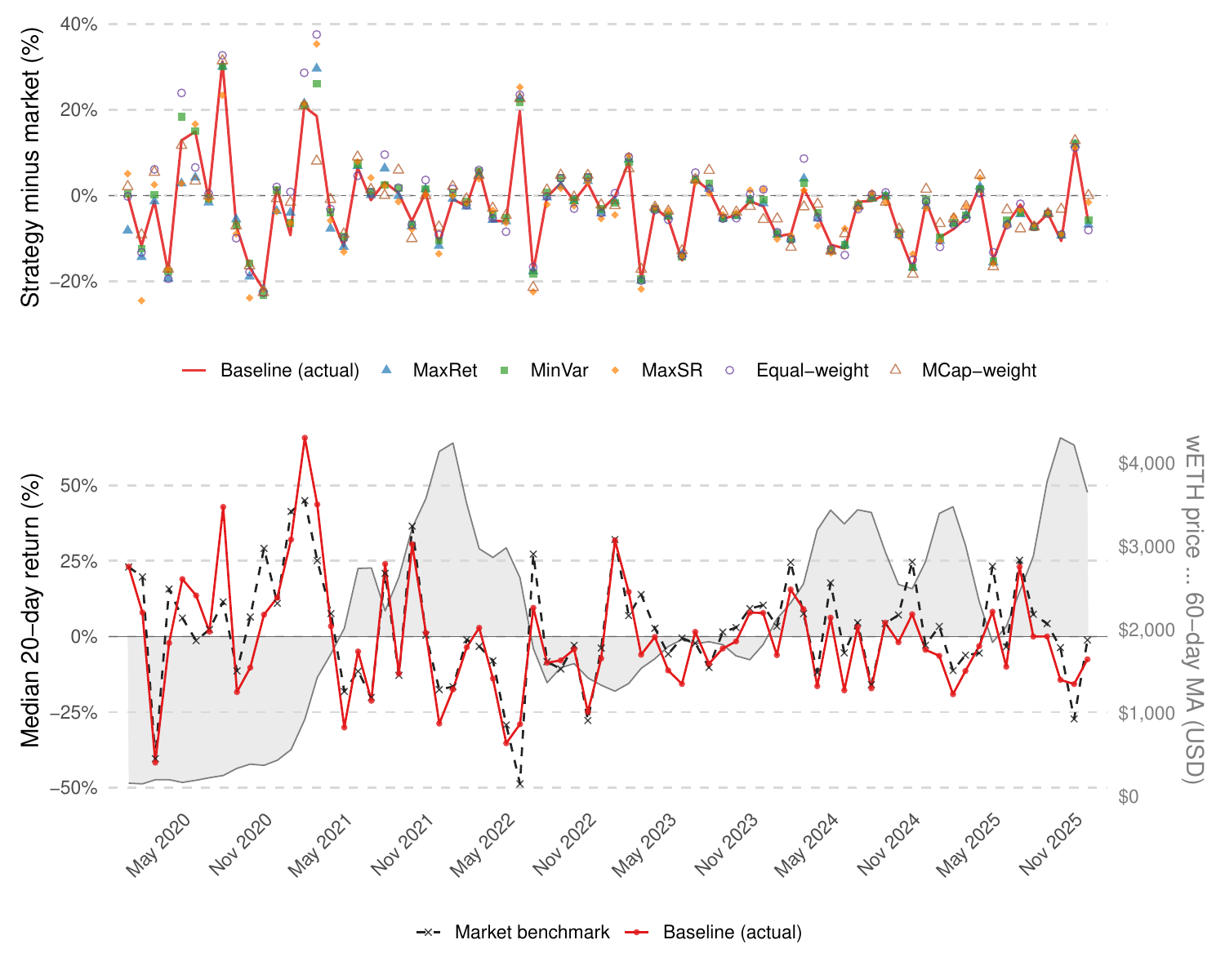}
  \caption{Returns relative to the market. Top: median $T^{(+)}$-day return minus the market benchmark per snapshot, showing that allocation choice has little effect relative to market-wide movements. Bottom: median baseline return and market benchmark overlaid on the \gls{wETH} 60-day moving average price. The between-strategy spread rarely exceeds 10\,pp, while the market swing spans over 100\,pp.}
  \label{fig:returns_panel_delta_market}
\end{figure}

\subsection{Portfolio $\beta$ and Market Exposure}
\label{sec:appendix_beta}

Portfolio $\beta$, which captures sensitivity to the market, has a median of \CAPMBetaBaselineMedian/ across accounts (IQR \CAPMBetaBaselinePtwentyfive/--\CAPMBetaBaselinePseventyfive/) for the baseline allocation.
Every optimisation strategy reduces $\beta$ on average, market-cap-weighting most strongly ($\Delta\beta = \CAPMDeltaBetaMcapMean/$).
By construction, $\alpha$ is orthogonal to $\beta$ in \gls{CAPM} (empirical $\alpha \sim \beta$: $\rho = \CAPMSpearmanBetaAlpha/$), so reducing market exposure provides no mechanism for improving risk-adjusted returns, and $\alpha$ remains negative for every strategy.

\subsection{Frontier Distance and Risk-Adjusted Return}
\label{sec:appendix_distance_alpha}

In \cref{sec:mpt_optimal_portfolio} we showed that portfolio size is the dominant determinant of $\ell_1$-distance from the efficient frontier. If \gls{MPT} optimality matters in practice, we would expect a negative association between distance ($d$) and risk-adjusted returns~($\alpha$). We find neither: Spearman rank correlations between $\ell_1$-distance and baseline $\alpha$ are small and, if anything, positive ($\rho = \CAPMSpearmanSrAlpha/$ to $\CAPMSpearmanMsAlpha/$), meaning farther accounts earn slightly higher risk-adjusted returns. The raw return is higher for zero-distance accounts ($\RetDistZeroMedianBaseline/\%$) than for the 80--100\,\% bucket ($\RetDistHighMedianBaseline/\%$), but this gap is structural rather than a payoff to optimisation: as \cref{sec:mpt_optimal_portfolio,sec:portfolio_characteristics} showed, near-optimal accounts are almost exclusively two-token portfolios that mechanically land close to the frontier.

\subsection{Random Forest and SHAP Details}
\label{sec:appendix_rf_shap}

To quantify the relative importance of account characteristics for predicting performance, we train Random Forest regressors (300 trees, depth~18, 10\,M sample) on two target variables---realised $T^{(+)}$-day return (\%) and cross-sectional $\alpha$---for each of the six strategies.
We consider two feature sets: a \emph{baseline} model with three features (entry month, $\log$ portfolio value $V$, portfolio size $N$) and an \emph{extended} model of six features that adds portfolio $\beta$, a binary ETH-wrapper indicator, and a binary BTC-wrapper indicator.

\paragraph*{Feature importance.}
\Cref{tab:rf_importance} reports the permutation-based feature importances and $R^{2}$ for both models on the baseline strategy.
The pattern is consistent across all six strategies (full results in the accompanying data release): month dominates, $\beta$ is a meaningful but secondary predictor, and asset composition is negligible.

\begin{table}[htbp]
\centering
\small
\caption{Random Forest feature importances: permutation-based importances and out-of-sample $R^{2}$ for predicting baseline returns and $\alpha$. The 6-feature model adds portfolio $\beta$, ETH-wrapper, and BTC-wrapper indicators.}
\label{tab:rf_importance}
\begin{tabular*}{\textwidth}{l @{\extracolsep{\fill}} cccc}
\toprule
 & \multicolumn{2}{c}{Return (\%)} & \multicolumn{2}{c}{$\alpha$} \\
\cmidrule(lr){2-3} \cmidrule(lr){4-5}
Feature & 3-feat & 6-feat & 3-feat & 6-feat \\
\midrule
Entry month       & 73\% & 70\% & 55\% & 55\% \\
Portfolio $\beta$ & ---  & 22\% & ---  & 26\% \\
Log portfolio value & 26\% & 8\%  & 16\% & 17\% \\
Portfolio size $N$ & 1\%  & $<$1\% & 1\%  & $<$1\% \\
Holds ETH wrapper & ---  & $<$1\% & ---  & 1\% \\
Holds BTC wrapper & ---  & $<$1\% & ---  & $<$1\% \\
\midrule
$R^{2}$ (test) & 0.55 & 0.65 & 0.36 & 0.48 \\
\bottomrule
\end{tabular*}
\end{table}

\paragraph*{SHAP analysis.}
We compute SHAP~\cite{lundberg2017unified} values via TreeExplainer~\cite{lundberg2020local} on a 30\,K subsample to assess how each feature shifts individual predictions.
\Cref{fig:shap_summary} shows the SHAP summary for baseline returns.
Snapshot month dominates, shifting predicted returns by $\pm\ShapMonthPp/$\,pp on average. The portfolio $\beta$ contributes $\pm\ShapBetaPp/$\,pp, and all remaining features contribute less than $\pm 1$\,pp.
\begin{figure}[H]
  \centering
  \includegraphics[width=1\linewidth]{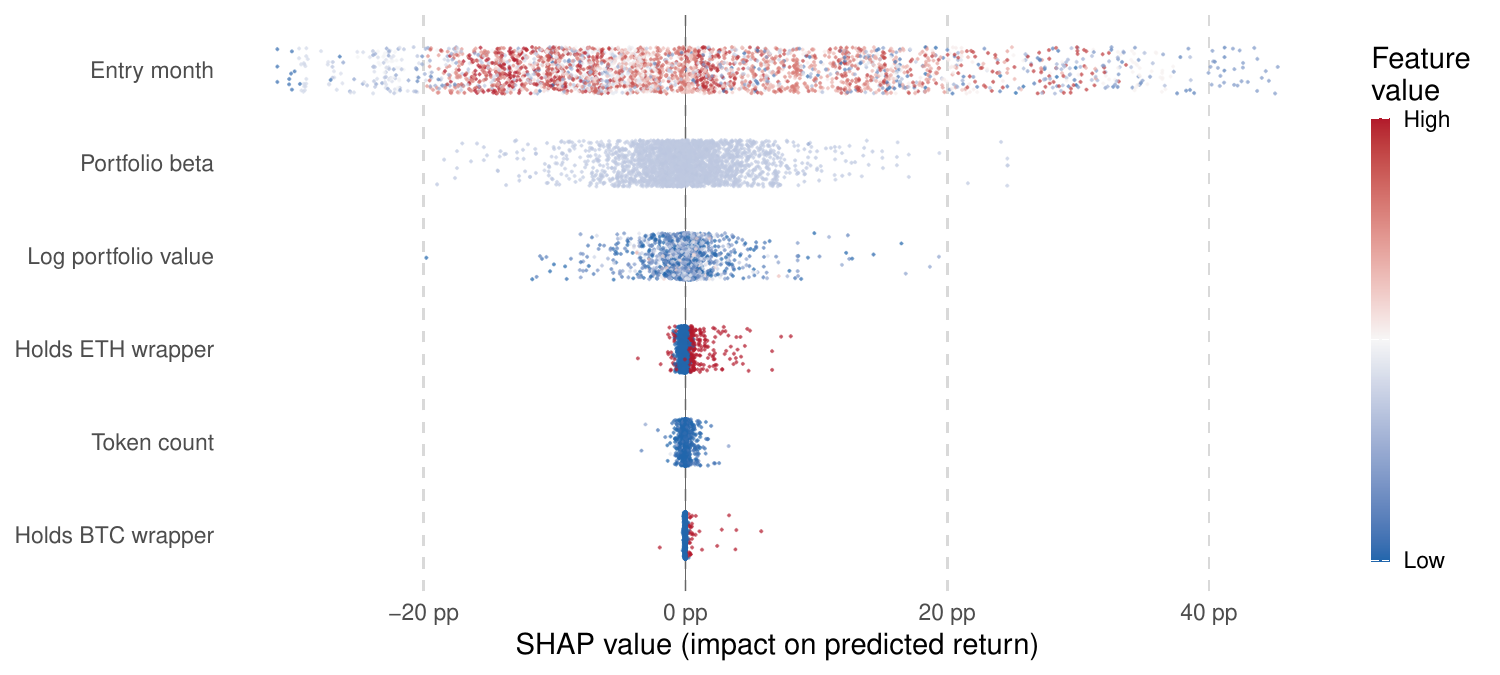}
  \caption{SHAP summary for baseline returns. Each dot represents one account-snapshot, the $x$-axis shows feature contribution to predicted return (pp), and colour encodes feature value (blue\,=\,low, red\,=\,high). Snapshot month dominates ($\pm$\ShapMonthPp/\,pp on average).}
  \label{fig:shap_summary}
\end{figure}

The SHAP dependence plot for snapshot month (\cref{fig:shap_month}) shows predicted returns swinging by tens of percentage points across the sample period. The per-month median (red line) tracks the broader crypto market cycle. The model is not learning which accounts allocate well, rather, which months were good or bad to be in the market.
\begin{figure}[H]
  \centering
  \includegraphics[width=1\linewidth]{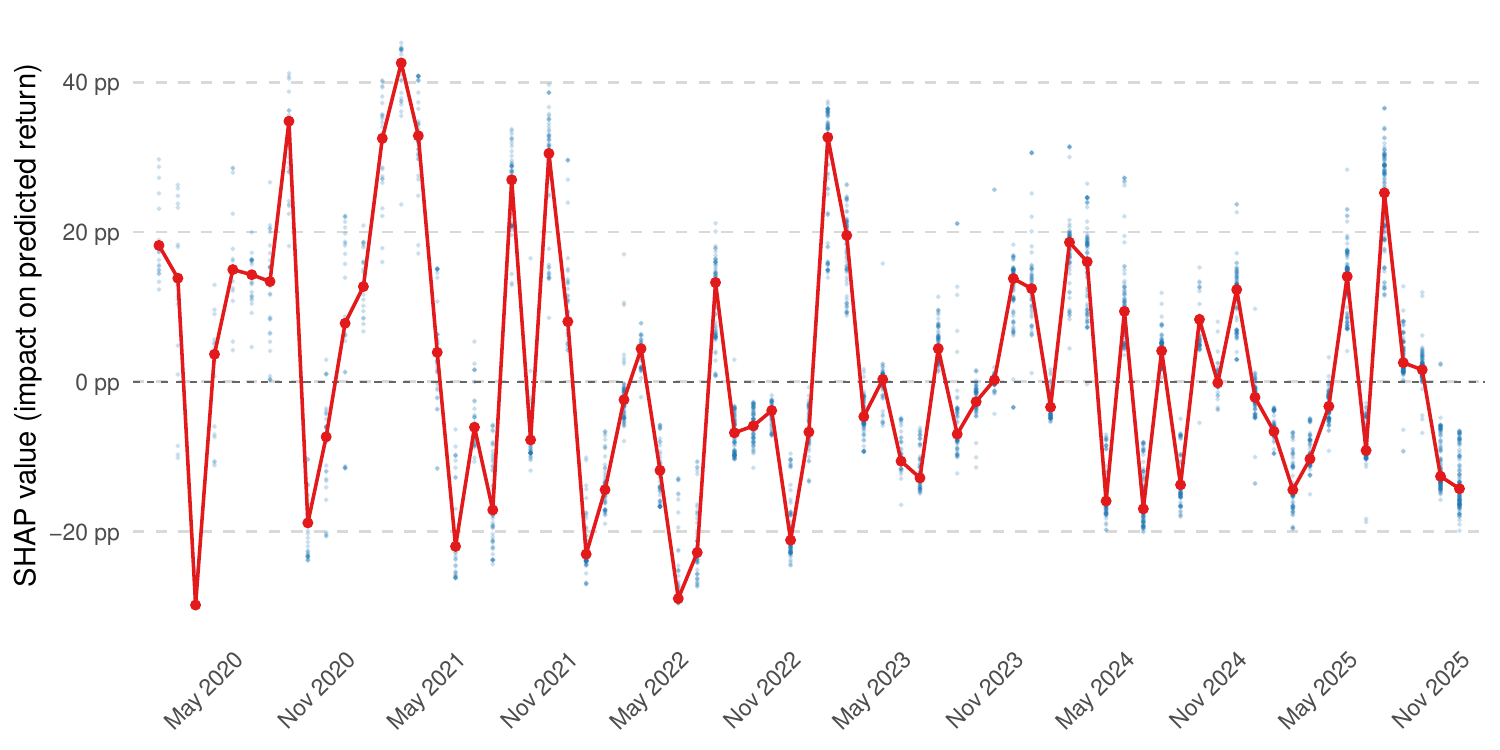}
    \caption{SHAP dependence for snapshot month. Each dot is one account-snapshot, the $x$-axis is the calendar month of the snapshot, and the $y$-axis the SHAP value (pp). The red line marks the per-month median.}
  \label{fig:shap_month}
\end{figure}

\paragraph*{3-feature vs.\ 6-feature comparison.}
Adding $\beta$ and asset-composition indicators improves $R^{2}$ by \RfDeltaRsqRetLo/--\RfDeltaRsqRetHi/ for returns and \RfDeltaRsqAlphaLo/--\RfDeltaRsqAlphaHi/ for $\alpha$ across all strategies (\cref{tab:rf_3v6}).
The improvement is driven almost entirely by $\beta$: the two binary asset indicators contribute less than 1\% of importance in every model.
This confirms that $\beta$ carries genuine predictive content beyond timing and wealth, but timing remains the dominant factor in both specifications.

\begin{table}[htbp]
\centering
\small
\caption{Model comparison across strategies: out-of-sample $R^{2}$ for 3-feature versus 6-feature Random Forest models.}
\label{tab:rf_3v6}
\begin{tabular*}{\textwidth}{l @{\extracolsep{\fill}} cccccc}
\toprule
 & \multicolumn{3}{c}{Return (\%)} & \multicolumn{3}{c}{$\alpha$} \\
\cmidrule(lr){2-4} \cmidrule(lr){5-7}
Strategy & 3-feat & 6-feat & $\Delta$ & 3-feat & 6-feat & $\Delta$ \\
\midrule
Baseline      & 0.55 & 0.65 & +0.10 & 0.36 & 0.48 & +0.12 \\
MaxRet        & 0.51 & 0.60 & +0.09 & 0.32 & 0.42 & +0.10 \\
MinVar        & 0.56 & 0.65 & +0.09 & 0.37 & 0.47 & +0.10 \\
MaxSR         & 0.49 & 0.56 & +0.07 & 0.29 & 0.38 & +0.09 \\
Equal-weight  & 0.58 & 0.65 & +0.07 & 0.39 & 0.47 & +0.08 \\
MCap-weight   & 0.52 & 0.62 & +0.10 & 0.35 & 0.46 & +0.11 \\
\bottomrule
\end{tabular*}
\end{table}

\section{Computational Details}
\label{sec:computational_details}

This appendix describes the computational infrastructure and software stack used to produce the results presented in this paper, supplementing the methodological overview in \cref{sec:portfolio_reconstruction}. It covers the execution environment (\cref{subsec:execution_environment}), the three-stage computational pipeline (\cref{subsec:computational_pipeline}), and the software stack (\cref{subsec:software_stack}).

\subsection{Execution Environment}
\label{subsec:execution_environment}

\ifdefined\anonymousAck
All large-scale stages (portfolio reconstruction from raw \gls{ERC-20} logs and the downstream \gls{MPT} optimisation/backtesting) were executed on a public-sector academic HPC cluster (specific infrastructure anonymised for double-blind review).
Compute nodes are AMD Zen~3 systems equipped with two AMD EPYC 7713 processors (128 physical cores per node), typically with 512\,GB RAM and a local 2\,TB NVMe disk.

\textbf{Storage model.} Large parquet datasets are stored on a tiered IBM Spectrum Scale (GPFS) system with on the order of hundreds of TB of flash and a few PB of HDD capacity; frequently accessed files are automatically promoted to the flash tier.
\else
All large-scale stages (portfolio reconstruction from raw \gls{ERC-20} logs and the downstream \gls{MPT} optimisation/backtesting) were executed on the Austrian Scientific Computing (ASC) infrastructure\footnote{\url{https://asc.ac.at/}},
using the Vienna Scientific Cluster VSC-5\footnote{\url{https://asc.ac.at/systems/vsc-5/}}.

All experiments were run on the default VSC-5 CPU partition \texttt{zen3\_0512}\footnote{\url{https://docs.asc.ac.at/systems/vsc5.html} \;\;and\;\; \url{https://docs.asc.ac.at/running_jobs/vsc5_queues.html}},
which provides AMD Zen~3 nodes equipped with two AMD EPYC 7713 processors (128 physical cores per node), typically with 512\,GB RAM and a local 2\,TB NVMe disk.

\textbf{Storage model.} We store all large parquet datasets on the ASC project filesystem \texttt{\$DATA} (e.g., \texttt{/gpfs/data/}),
which is implemented as a tiered IBM Spectrum Scale (GPFS) system with 500\,TB of flash and approximately 5\,PB of HDD capacity\footnote{\url{https://docs.asc.ac.at/storage/where_store_data.html}}.
Frequently accessed files are automatically promoted to the flash tier, enabling efficient large-scale scans while keeping long-term storage feasible.
\fi

\subsection{Computational Pipeline}
\label{subsec:computational_pipeline}

The pipeline consists of three stages: building a data lake from raw Ethereum blockchain data, reconstructing per-account portfolio snapshots, and running \gls{MPT} optimisation with forward-looking backtesting.

\subsubsection{Data lake infrastructure.}
As a foundation to the reconstruction procedure, we utilise a data lake solution to store the raw Ethereum data.
The \textit{logs} and \textit{traces} are the primary inputs to our data lake, organised in partitions of \num{10000} Ethereum blocks in an efficiently accessible \textit{Parquet} format.
Parquet files use efficient compression and allow fast access through a scannable column-oriented storage format.
This makes the format well-suited for parallel processing and provides a systematic infrastructure for processing the data.

\subsubsection{Portfolio reconstruction.}
 
The most storage-intensive phase involves extracting and parsing the complete Ethereum log history
from block~\#0 to block~\#23{,}943{,}069 (Jul-30-2015 to Dec-05-2025, approximately ten years of data).
We scan the complete Ethereum log history and extract transfer events
for tokens matching our allow-list, yielding \ReconTransferEvents/ transfer events across \ReconTokensDiscovered/ discovered tokens,
stored as per-token ledger Parquet files totalling \ReconParquetSizeGB/\,GB.
The transfer volume is heavily concentrated:
the single largest token (\gls{wETH}) accounts for \ReconTopTokenTransfers/\,M transfers
(\ReconTopTokenSharePct/\% of the total),
while the median token has only \ReconMedianTokenTransfers/ transfers.
After validation filtering (verifying \gls{ERC-20} compliance, price-data availability,
sufficient trading volume, and plausible fully-diluted valuation) \ReconTokensAfterFilter/
tokens pass all criteria and enter the reconstruction stage.
 
The MapReduce approach (described in \cref{sec:portfolio_reconstruction})
proceeds in two MapReduce passes.
The first pass scans the raw log partitions, decodes Transfer, Deposit, and Withdrawal events,
and writes per-token shard files. If a batch exceeds its timeout, the scheduler automatically splits it in half and retries both halves, avoiding manual intervention on slow partitions.
The second pass merges shards into a single sorted ledger per token, storing each transfer as two rows (a negative entry for the sender and a positive entry for the recipient) so that account balances at any block height can be recovered by cumulative summation.
 
Portfolio reconstruction itself reads these ledgers,
aggregates balances at each of the \ReconSnapshotBlocks/ monthly snapshots via DuckDB,
enriches holdings with decimals and USD prices, and writes the final snapshot Parquet.
All balances for a given token--account pair are computed in a single SQL pass
using conditional sums (\texttt{SUM(CASE WHEN block <= b THEN val END)}) per snapshot,
rather than issuing \ReconSnapshotBlocks/ separate queries.
We configure the pipeline with \ReconPipelineWorkers/ parallel workers
and a batch size of \ReconBatchSize/ accounts.
Due to the dataset scale, reconstruction is split into two runs
(2020--2022 and 2023--2025).
The first run produces \ReconOutputRowsFirst/\,M rows, the second \ReconOutputRowsSecond/\,M rows,
for a combined total of \ReconOutputRows/\,M account--snapshot observations
(\ReconOutputSizeGB/\,GB compressed Parquet).

\paragraph*{Proxy contract resolution.}\label{sec:proxy_resolution}
A token contract may be deployed as a proxy that delegates execution to an underlying implementation. We resolve such proxies before applying the ERC-20 selector checks of \cref{sec:data_collection}, covering EIP-1167 minimal proxies, EIP-1967 Transparent/UUPS proxies, beacon proxies, ZeppelinOS ``unstructured storage'' proxies, and proxies that expose an \texttt{implementation()} method.

\subsubsection{MPT optimisation and backtesting.}
 
The optimisation stage takes the reconstructed holdings snapshots and, for each of the
\MptBlocks/ monthly snapshots and every account with at least two tokens,
solves three constrained mean--variance problems:
maximise return at the account's current volatility (\gls{MRV}),
minimise variance at the account's current expected return (\gls{MVR}),
and maximise the Sharpe ratio (\gls{MSR}).
Two naive benchmarks (equal-weight and market-cap-weight) are computed alongside,
yielding \MptStrategies/ strategy variants per account.
For each strategy, the pipeline evaluates realised \MptHorizonDays/-day forward returns
and computes \gls{CAPM} alpha and beta against a market benchmark.

The input dataset contains \MptObservationsM/\,M account--snapshot observations.
Per-snapshot account counts range from \MptMinWalletsPerBlock/ to \MptMaxWalletsPerBlock/
(median: \MptMedianWalletsPerBlock/), reflecting the growth of the Ethereum account base
over the 2020--2025 observation window.
Due to dataset scale, optimisation jobs are split by snapshot ranges and executed
on \MptCores/-core nodes with \MptWorkers/ parallel workers per job.
Price histories for all tokens in each snapshot's universe are pre-loaded into shared memory
and distributed to workers via process-level initialisation,
avoiding redundant I/O across the millions of per-wallet optimisations.
The final output comprises \MptObservations/ rows across \MptBlocks/ partitioned Parquet files
totalling \MptOutputSizeGB/\,GB.

\subsection{Software Stack}
\label{subsec:software_stack}

This subsection describes the key libraries and tools underlying the pipeline described in \cref{subsec:computational_pipeline}.

We implement the pipeline in Python 3.11 using PyArrow as the foundational columnar backend and Polars for efficient batch transformations.
Parquet datasets are accessed via streaming interfaces to avoid materialising full tables in memory.
For large-scale aggregation and lightweight analytical queries, we additionally employ DuckDB as an embedded analytical database (with explicitly bounded memory settings for stability).
This stack provides a practical balance between scalability (HPC-ready I/O throughput), reproducibility (deterministic parquet-based artifacts), and modularity (separable reconstruction and optimisation stages).

For transparency and reproducibility, we provide the full pipeline implementation at \url{\datalink}.

\fi

\end{document}